\documentclass[twocolumn,twocolappendix]{aastex63}

\usepackage{natbib,aas_macros,amsmath}
\usepackage{multirow,color}
\usepackage{amsmath,amssymb}
\usepackage{natbib}
\usepackage{longtable}
\usepackage{lineno}

\def\farcs{%
 \mbox{%
  \kern  0.13ex.%
  \kern -0.95ex\arcsec%
  \kern -0.1ex%
 }%
}%

\newcommand{\oiii}{[O\,{\sc iii}]}

\newcommand{\nii}{[N\,{\sc ii}]}

\newcommand{\muv}{$M_{\rm UV}$}

\newcommand{\hst}{{\it HST}}
\newcommand{\spitzer}{{\it Spitzer}}
\newcommand{\jwst}{{\it JWST}}

\newcommand{\targ}{GLIMPSE-16043}
\newcommand{\targb}{JOF-21739}
\newcommand{\zackmodel}{Yggdrasil}

\submitjournal{ApJ}
\shorttitle{
An Ultra-faint Pop~III Candidate at $z=6.5$ Discovered in GLIMPSE}
\shortauthors{Fujimoto \& Naidu et al.}

\begin{document}

\title{
GLIMPSE: An ultra-faint \boldmath $\simeq 10^{5}\,M_{\odot}$ Pop~III Galaxy Candidate \\ 
and First Constraints on the Pop~III UV Luminosity Function at $z\simeq$ 6--7
}

\correspondingauthor{Seiji Fujimoto \& Rohan Naidu}
\email{fujimoto@utexas.edu, rnaidu@mit.edu}
\author[0000-0001-7201-5066]{Seiji Fujimoto$^{\dagger}$}
\altaffiliation{Hubble Fellow}
\affiliation{Department of Astronomy, The University of Texas at Austin, Austin, TX 78712, USA}

\author[0000-0003-3997-5705]{Rohan P.~Naidu$^{*}$}
\altaffiliation{These authors contributed equally to this work.}
\affiliation{
MIT Kavli Institute for Astrophysics and Space Research, 70 Vassar Street, Cambridge, MA 02139, USA}

\author[0000-0002-0302-2577]{John Chisholm}
\affiliation{Department of Astronomy, The University of Texas at Austin, Austin, TX 78712, USA}

\author[0000-0002-7570-0824]{Hakim Atek}
\affiliation{Institut d'Astrophysique de Paris, CNRS, Sorbonne Universit\'e, 98bis Boulevard Arago, 75014, Paris, France}
\author[0000-0003-4564-2771]{Ryan Endsley}
\affiliation{Department of Astronomy, The University of Texas at Austin, Austin, TX 78712, USA}
\author[0000-0002-5588-9156]{Vasily Kokorev}
\affiliation{Department of Astronomy, The University of Texas at Austin, Austin, TX 78712, USA}
\author[0000-0001-6278-032X]{Lukas J. Furtak}
\affiliation{Department of Physics, Ben-Gurion University of the Negev, P.O. Box 653, Be'er-Sheva 84105, Israel}
\author[0000-0002-9651-5716]{Richard Pan}
\affiliation{Department of Physics \& Astronomy, Tufts University, MA 02155, USA}

\author[0000-0002-4966-7450]{Boyuan Liu}
\affiliation{Institut f\"{u}r Theoretische Astrophysik, Zentrum f\"{u}r Astronomie, Universit\"{a}t Heidelberg, D-69120 Heidelberg, Germany}

\author[0000-0003-0212-2979]{Volker Bromm}
\affiliation{Department of Astronomy, The University of Texas at Austin, Austin, TX 78712, USA}

\author[0000-0003-2237-0777]{Alessandra Venditti}
\affiliation{Department of Astronomy, The University of Texas at Austin, Austin, TX 78712, USA}

\author[0000-0002-8365-0337]{Eli Visbal}
\affiliation{Department of Physics and Astronomy and Ritter Astrophysical Research Center, University of Toledo, 2801 W Bancroft Street, Toledo, OH 43606, USA}
\author[0000-0002-8013-5970]{Richard Sarmento}
\affiliation{School of Earth and Space Exploration, Arizona State University, P.O. Box 871404, Tempe, AZ, 85287-1404, USA}

\author[0000-0001-8928-4465]{Andrea Weibel}
\affiliation{Department of Astronomy, University of Geneva, Chemin Pegasi 51, 1290 Versoix, Switzerland}

\author[0000-0001-5851-6649]{Pascal A.\ Oesch}
\affiliation{Department of Astronomy, University of Geneva, Chemin Pegasi 51, 1290 Versoix, Switzerland}
\affiliation{Cosmic Dawn Center (DAWN), Copenhagen, Denmark}
\affiliation{Niels Bohr Institute, University of Copenhagen, Jagtvej 128, K{\o}benhavn N, DK-2200, Denmark}

\author[0000-0003-2680-005X]{Gabriel Brammer}
\affiliation{Cosmic Dawn Center (DAWN), Niels Bohr Institute, University of Copenhagen, Jagtvej 128, K{\o}benhavn N, DK-2200, Denmark}

\author[0000-0001-7144-7182]{Daniel Schaerer}
\affiliation{Department of Astronomy, University of Geneva, Chemin Pegasi 51, 1290 Versoix, Switzerland}

\author[0000-0002-8192-8091]{Angela Adamo}
\affiliation{Department of Astronomy, Oskar Klein center, Stockholm University, AlbaNova University center, SE-106 91 Stockholm, Sweden}

\author[0000-0002-4153-053X]{Danielle A. Berg}
\affiliation{Department of Astronomy, The University of Texas at Austin, Austin, TX 78712, USA}

\author[0000-0001-5063-8254]{Rachel Bezanson}
\affiliation{Department of Physics and Astronomy and PITT PACC, University of Pittsburgh, Pittsburgh, PA 15260, USA}

\author[0000-0002-4989-2471]{Rychard Bouwens}
\affiliation{Leiden Observatory, Leiden University, NL-2300 RA Leiden, Netherlands}

\author[0009-0009-9795-6167]{Iryna Chemerynska}
\affiliation{Institut d'Astrophysique de Paris, CNRS, Sorbonne Universit\'e, 98bis Boulevard Arago, 75014, Paris, France}

\author[0000-0001-7940-1816]{Adélaïde Claeyssens}
\affiliation{Department of Astronomy, Oskar Klein center, Stockholm University, AlbaNova University center, SE-106 91 Stockholm, Sweden}

\author[0000-0003-0348-2917]{Miroslava Dessauges-Zavadsky}
\affiliation{Department of Astronomy, University of Geneva, Chemin Pegasi 51, 1290 Versoix, Switzerland}

\author[0000-0002-2139-7145]{Anna Frebel}
\affiliation{MIT Kavli Institute for Astrophysics and Space Research, 70 Vassar Street, Cambridge, MA 02139, USA}

\author[0000-0002-3897-6856]{Damien Korber}
\affiliation{Department of Astronomy, University of Geneva, Chemin Pegasi 51, 1290 Versoix, Switzerland}

\author[0000-0002-2057-5376]{Ivo Labbe}
\affiliation{Centre for Astrophysics and Supercomputing, Swinburne University of Technology, Melbourne, VIC 3122, Australia}

\author[0000-0001-8442-1846]{Rui Marques-Chaves}
\affiliation{Department of Astronomy, University of Geneva, Chemin Pegasi 51, 1290 Versoix, Switzerland}

\author[0000-0003-2871-127X]{Jorryt Matthee}
\affiliation{Institute of Science and Technology Austria (ISTA), Am Campus 1, 3400 Klosterneuburg, Austria}

\author[0000-0001-5538-2614]{Kristen B.~W.\ McQuinn}
\affiliation{Space Telescope Science Institute, 3700 San Martin Dr., Baltimore, MD 21218, USA}
\affiliation{Department of Physics \& Astronomy, Rutgers, The State University of New Jersey, Piscataway, NJ 08854, USA}

\author[0000-0002-8984-0465]{Julian B.~Mu\~noz}
\affiliation{Department of Astronomy, The University of Texas at Austin, Austin, TX 78712, USA}

\author[0000-0002-5554-8896]{Priyamvada Natarajan}
\affiliation{Department of Astronomy, Yale University, 219 Prospect Street, New Haven, CT 06511, USA}
\affiliation{Department of Physics, Yale University, 217 Prospect Street, New Haven, CT 06511, USA}
\affiliation{Black Hole Initiative at Harvard University, 20 Garden Street, Cambridge, MA 02138, USA}

\author[0000-0001-8419-3062]{Alberto Saldana-Lopez}
\affiliation{Department of Astronomy, Oskar Klein Centre, Stockholm University,106 91 Stockholm, Sweden}

\author[0000-0002-1714-1905]{Katherine A. Suess}
\affiliation{Department for Astrophysical \& Planetary Science, University of Colorado, Boulder, CO 80309, USA}

\author[0000-0002-3216-1322]{Marta Volonteri}
\affiliation{Institut d'Astrophysique de Paris, CNRS, Sorbonne Universit\'e, 98bis Boulevard Arago, 75014, Paris, France}

\author[0000-0002-0350-4488]{Adi Zitrin}
\affiliation{Department of Physics, Ben-Gurion University of the Negev, P.O. Box 653, Be'er-Sheva 84105, Israel}

\def\apj{ApJ}%
\def\apjl{ApJL}%
\def\apjs{ApJS}%

\def\rme{\rm e}
\def\rmstar{\rm star}
\def\rmFIR{\rm FIR}
\def\itHubble{\it Hubble}
\def\rmyr{\rm yr}

\begin{abstract}
Detecting the first generation of stars, Population III (Pop~III), has been a long-standing goal in astrophysics, yet they remain elusive even in the \jwst\ era. Here we present a novel NIRCam-based selection method for Pop~III galaxies, and carefully validate it through completeness and contamination simulations. We systematically search $\simeq$~500~arcmin$^2$ across \jwst\ legacy fields for Pop~III candidates, including GLIMPSE which, assisted by gravitational lensing, has produced \jwst's deepest NIRCam imaging thus far. We discover one promising Pop~III candidate (GLIMPSE-16043) at $z=6.50^{+0.03}_{-0.24}$, a moderately lensed galaxy ($\mu=2.9^{+0.1}_{-0.2}$) with an intrinsic UV magnitude of $M_{\rm\,UV}=-15.89^{+0.12}_{-0.14}$. It exhibits key Pop~III features: strong H$\alpha$ (rest-frame EW $2810\pm550$~\AA); Balmer jump; no dust (UV slope $\beta=-2.34\pm0.36$); and undetectable metal lines (e.g.,\,\oiii;\,\oiii/H$\beta<0.44$) implying a gas-phase metallicity of $Z_{\rm\,gas}/Z_{\odot}<0.5\%$. These properties indicate the presence of a nascent, metal-deficient young stellar population ($<5$~Myr) with a stellar mass of $\simeq10^{5}M_{\odot}$. Intriguingly, this source deviates significantly from the extrapolated UV-metallicity relation derived from recent \jwst\ observations at $z=4-10$, consistent with UV enhancement by a top-heavy Pop~III initial mass function or the presence of an extremely metal-poor AGN. We also derive the first observational constraints on the Pop~III UV luminosity function at $z\simeq6-7$. The volume density of GLIMPSE-16043 ($\approx10^{-4}$cMpc$^{-3}$) is in excellent agreement with theoretical predictions, independently reinforcing its plausibility. This study demonstrates the power of our novel NIRCam method to finally reveal distant galaxies even more pristine than the Milky Way's most metal-poor satellites, thereby promising to bring us closer to the first generation of stars than we have ever been before.
\end{abstract}
\keywords{ galaxies: formation --- galaxies: evolution --- galaxies: high-redshift --- galaxies: structure -- galaxies -- galaxies starburst -- ISM: dust}

\section{Introduction}
\label{sec:intro} 

Members of the first generation of stars that formed from primordial, i.e. metal-free, gas are referred to as Population~III (Pop~III) stars \citep[e.g.,][]{bromm2013,frebel2015,inayoshi2020,klessen2023}. Hypothesized since the 1960s \citep[e.g.,][]{page1966}, detecting any signatures of the presumably very massive and short-lived Pop~III stars remains one of the elusive, ultimate frontiers of observational astrophysics. In many ways, Pop III stars represent the beginning of our beginnings. By forging metals for the very first time, these stars enriched the sterile ocean of hydrogen and helium gas left behind by the Big Bang, setting in motion the chemical evolution of the universe that would one day lead to life on Earth. If not for their legacy of metals, the hot gas entrained in dark matter halos would struggle to cool and condense into stars and galaxies. And if not for the compact remnants they left behind, the behemoth supermassive black holes rippling spacetime today may have never been seeded. Few goals in astrophysics seem more fundamental than discovering these stars that are the very first chapter of our origin story.

Detecting light from the first Pop III stars forming at $z\approx15-20$ demands imaging depths of $\approx40$ mag, which is beyond the reach of all current and planned future facilities \citep[e.g.,][]{schauer2020}. However, recent simulations suggest that Pop~III star formation may have persisted for some time, down to lower redshifts, $z\simeq$ 6--7 \citep[e.g.,][]{jimenez2006, tornatore2007, pallottini2014b, venditti2023,venditti2024b,katz2023}. In these models, pristine pockets of gas that give rise to Pop~III star formation in small dark matter host halos survive unpolluted due to e.g., their remote locations on the edges of the then-emerging cosmic web and forming structures, inefficient feedback and metal mixing in their immediate surroundings, or their residence in low-mass dark matter halos that have not yet undergone their first episode of star formation. When considering an observational search for early galactic systems at high redshift that contain Pop~III stars, low-mass, faint ($L<1\%L^{*}$) galaxies are hence the prime targets. As we demonstrate in \S\ref{sec:depth}, a pure Pop~III-dominated source forming stars from gas in a primordial, metal-free dark matter halo is expected to have a stellar mass of $\approx10^{5} M_{\rm{\odot}}$ requiring a depth of $\approx31$ mag for detection.

What might the smoking gun signatures of Pop III stars in such systems be? Made from just H and He, massive Pop III stars embedded in primordial H {\sc ii} regions should produce spectra revealing strong hydrogen (e.g., H$\alpha$, H$\beta$) and helium lines \citep[e.g., He{\sc ii}$\lambda$1640, 4686;][]{tumlinson2001,schaerer2002}
accompanied by a complete absence of metal lines (e.g., \oiii$\lambda$5008).
For pristine gas or extremely metal-poor conditions, top-heavy initial mass functions (IMF) are expected, perhaps including extremely massive stars reaching $\approx100~M_{\rm{\odot}} - 1000~~M_{\rm{\odot}}$ \citep[e.g.,][]{hirano2014, chon2024}. The consequent hot and highly ionized ISM conditions of primordial gas are expected to result in a steep Balmer jump in the nebular continuum portion of the spectrum \citep[e.g.,][]{zackrisson2011}. 

Prior to \jwst\ \citep{gardner2023}, efforts to search for Pop III stars at $z\gtrsim6$ were limited to identifying restframe-UV signatures (Ly$\alpha$ and He {\sc ii}; e.g., \citealt{sobral2015}), typically in fairly luminous systems ($\gtrsim0.1 L^{*}$) in order to be accessible to ground-based spectroscopy. \jwst's sensitivity now allows us to push into the crucial low-mass dwarf galaxy regime where Pop III-dominated systems are most likely to occur, while also providing access to the full suite of restframe-UV to restframe-optical diagnostics. The question is therefore, what is the most efficient way of deploying \jwst\ to scan a large number of sources for rare Pop III signatures? 

Recent NIRSpec IFU results are tantalizing, identifying a candidate He{\sc ii} emitting clump at $z=10.6$ \citep{maiolino2023b} and ``LAP1", a highly magnified \oiii-faint source at $z=6.6$ \citep{vanzella2023b}. Intriguingly, these source detections lack stellar continuum in their spectra, raising the possibility that instead of metal-poor stars, these two systems may instead represent nebular emission from pristine gas clouds due to irradiation by nearby star-forming regions (or AGN). However, given the limited survey volume (field of view of $3\farcs0\times3\farcs0$), the IFU, and more broadly, slit-spectroscopy, offer a more viable approach for follow-up studies \citep[e.g.,][]{venditti2024b}. 

NIRCam grism spectroscopy captures every source in the field of view, offering a combination of superb statistics, high resolution (e.g., to split \oiii+H$\beta$ out to $z\approx9$), and spatially resolved spectra to identify metal-poor clumps \citep[e.g.,][]{oesch2023, kashino2023, matthee2023}. However, this comes at the cost of sensitivity, as a result of which the deepest grism surveys are yet to push below $\approx29$ mag. This implies that meaningful Pop III searches with the grism hinge on the discovery of extremely magnified arcs \citep[e.g.,][]{naidu2024}.

The situation at hand is rather surprising -- even after two full years of \textit{JWST} operations only a single source (LAP1) has been robustly confirmed below the ``metallicity floor" of the Milky Way's ultra-faint dwarf galaxy satellite population ([Fe/H]$ \sim-2.5$; e.g., \citealt{fu2023}). Intriguingly, these local satellite galaxies may bear signatures of Pop III stars that exploded at $z>6$ \citep[e.g.,][]{jeon2017}. Remarkably, then, the best examples of primordial stellar populations are still at $z\sim0$ and not at $z>6$. This is despite the fact that these surviving systems formed their near-pristine metal-poor stars at exactly the redshifts ($z\gtrsim6$) that are now well within our grasp with \textit{JWST} \citep[e.g.,][]{brown2014, weisz2014, savino2023}. At these early epochs, however, \jwst\ is yet to push to the low metallicity regimes and low stellar masses that the local dwarf satellite galaxies span. Ultra-deep observations and new selection techniques are required to provide a breakthrough if we ever want to capture the high-redshift ancestors of today's ancient ultra-faint satellite galaxies.

Indeed, an efficient approach that is yet to be systematically exploited involves capturing Pop III signatures through deep imaging surveys that deploy strategically chosen medium-bands and broad-bands. 
While the maximal rest-frame equivalent width (EW) of the rest-UV helium lines is $\simeq$~100--200~${\rm \AA}$ \citep[e.g.,][]{schaerer2002,nakajima2022b}, the strong rest-frame optical H$\alpha$ line reaches EWs of $\gtrsim$~2000--3000$ {\rm \AA}$ in young stellar populations \citep[e.g.,][]{endsley2023a}. This could significantly boost the flux even in broad-band filters. In fact, this flux excess in broad-band photometry has been used to constrain the EWs of the strong rest-frame optical H$\alpha$ and \oiii\ lines of $z\simeq$6--9 galaxies with \hst\ and \spitzer\ for several years \citep[e.g.,][]{labbe2013, smit2015, roberts-borsani2016, stefanon2022}. The same experiment may now be performed with \jwst/NIRCam out to $z\simeq9$ \citep[e.g.,][]{endsley2023a, simmonds2024, llerena2024} and even at $z>10$ with MIRI, albeit at much reduced efficiency \citep[e.g.,][]{helton2024}. This provides a new opportunity for finding Pop~III star systems, provided the classical line-flux excess analysis is optimized to capture the unique SED shapes of Pop III-dominated galaxies exhibiting strong H$\alpha$ and H$\beta$ but are lacking any \oiii\ lines. 

The redshift range of $z\simeq$ 6--7 presents an ideal window to identify Pop~III galaxies purely with NIRCam using the extensive rest-frame UV and optical information available. This is the highest redshift where H$\alpha$ is accessible. Furthermore, NIRCam's sensitivity peaks at $3.5-4\mu$m \citep{rigby2023}. This implies that the most stringent upper limit on the \oiii\ line at 5008\AA\ may be placed at $z\sim6.5$. Higher redshifts may be probed using NIRCam+MIRI \citep[e.g.,][]{zackrisson2011, trussler2023}, but MIRI's smaller field of view and much lower throughput provide significant overall limitations. 
Given the large volumes already surveyed and planned with NIRCam, a pure NIRCam search at $z\simeq$ 6--7 not only lays a promising foundation for the discovery of early Pop-III galaxies, but also enables the provision of stringent constraints on their volume density. Importantly, even non-detections provide valuable insights, offering strong constraints on Pop~III star formation predictions, enrichment processes in chemically unevolved dark matter halos, and feedback mechanisms in early galaxies. For example, even upper limits on the Pop III star rate at these redshifts would guide the \textit{Roman} mission in undertaking dedicated surveys for the putative Pair Instability Supernovae (PISNe), whose yields remain speculative at present \citep[e.g.,][]{moriya2022}. Similarly, forecasts for the number of events (e.g., black hole mergers) detected by future gravitational wave observatories (e.g., LISA) critically depend on the still observationally unconstrained Pop~III IMF. 

\begin{figure*}[t!]
\begin{center}
\includegraphics[trim=0cm 0.1cm 0cm 0cm, clip, angle=0,width=0.9\textwidth]{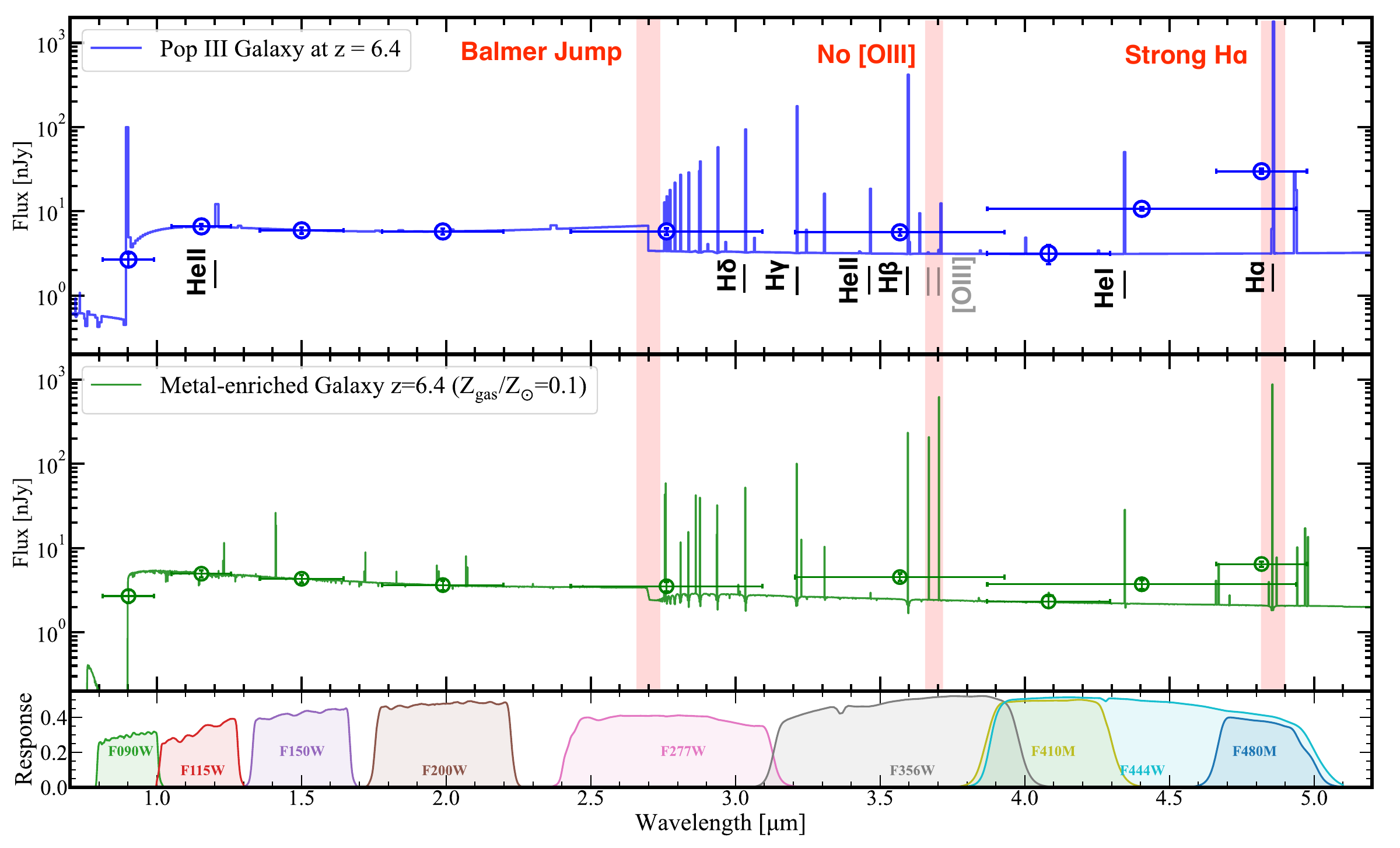}
\end{center}
\vspace{-0.4cm}
 \caption{
\textbf{Key Pop III galaxy features motivating our search strategy illustrated using the GLIMPSE filter-set.} 
\textbf{\textit{Top:}} \zackmodel\ Pop~III galaxy SED model \citep{zackrisson2011} with a moderately top-heavy IMF (blue curve). 
This model is characterized by $\log(M_{\rm star}/M_{\odot})=6$, $f_{\rm cov}=1.0$, and $t_{\rm age}=1.0$~Myr at $z=6.4$, with a characteristic mass of individual Pop~III stars of $M_{\star, \rm IMF} \simeq 10\,M_{\odot}$ (Pop~III.2).  
\textbf{\textit{Middle:}} 
A metal-enriched galaxy SED model (green curve) is also shown, generated using \texttt{BAGPIPES} with the following parameters: $Z_{\rm gas}/Z_{\odot}=0.1$, $n_{\rm e}=1000$~cm$^{-3}$, and $\log(U)=-2$.  
In both models, we masked the strong Ly$\alpha$ given its frequent damping by the neutral IGM at this redshift. 
\textbf{\textit{Bottom:}} 
The filter responses of the nine NIRCam filters used in GLIMPSE.  
In the top and middle panels, the expected NIRCam photometry is presented as open circles, where the y-axis errors denote the 1$\sigma$ uncertainty based on the GLIMPSE data depth.  
Red shading highlights several unique SED features of the Pop~III galaxy, including strong Balmer lines, the absence of the \oiii\ line, and the Balmer jump, that are effectively captured by the NIRCam photometry.  
\label{fig:sed_comp}}
\end{figure*}

In this paper, we present our novel technique to select the best candidate Pop~III galaxies using \jwst/NIRCam data alone, and the first systematic search of Pop~III galaxies in publicly available data from deep \jwst\ legacy surveys. Notably, our search includes the Cycle~2 large program GLIMPSE (PID~3293; PIs H. Atek \& J. Chisholm) that has produced \jwst's deepest NIRCam imaging dataset to date (accounting for lensing). It allows for the first time to push into the early low-mass dwarf galaxy regime where Pop III dominated galaxies are the likeliest to occur. By leveraging strong lensing produced by the well-studied massive galaxy cluster of Abell S1063 \citep{lutz2016}, GLIMPSE is designed to detect ultra-faint galaxies in the epoch of reionization. The GLIMPSE filter-set is comprised of seven wide and two medium bands, notably including F480M which enables clean isolation of the H$\alpha$ line at the highest redshift possible with NIRCam data.

The paper is structured as follows. Section~\ref{sec:method} outlines our selection method, including dedicated simulations for completeness and contamination. In Section~\ref{sec:obs}, we describe the observations and data processing of GLIMPSE and other public NIRCam datasets suitable for the Pop~III galaxy search. In Section~\ref{sec:result}, we present Pop~III candidates identified in the deep NIRCam data based on our new technique and discuss their physical properties and volume density. In Section~\ref{sec:caveats}, we explore an array of alternate scenarios regarding the origins of our Pop III galaxy candidates. We discuss future prospects in light of our findings in Section~\ref{sec:future} and summarize our results in Section~\ref{sec:summary}.
Throughout this paper, we assume a flat universe with $\Omega_{\rm m} = 0.3$, $\Omega_\Lambda = 0.7$, $\sigma_8 = 0.8$,
and $H_0 = 70$ km s$^{-1}$ Mpc$^{-1}$. We use magnitudes in the AB system \citep{oke1983}. 
We adopt an uncertainty floor of 5\% for the input photometry when running \texttt{EAZY} and SED fitting to capture systematic uncertainties \citep[e.g.,][]{weibel2024}.

\setlength{\tabcolsep}{5pt}
\begin{table*}
    \centering
    \caption{Summary of the models used in our selection methods, completeness/contamination simulations, and additional tests.}
    \vspace{-0.3cm}
    \label{tab:models}
    \begin{tabular}{lp{2.3cm}p{3.5cm}p{4.5cm}p{4cm}}
        \hline
        Type & Reference & IMF & Key descriptions & Usage \\
        \hline
        Pop~III & 
        \citet{zackrisson2011} & 
        Power-law ($\alpha$=2.35) 50--500$M_{\odot}$ (PopIII.1), Log-normal (center = $10M_{\odot}$, $\sigma=1M_{\odot}$) 1--500$M_{\odot}$ (PopIII.2), Kroupa 0.1--100$M_{\odot}$ (PopIII.Kroupa) \newline
        &
        $n_e$=100cm$^{-3}$, $f_{\rm cov}$ = 1.0, 
        $t_{\mathrm{age}}$ = [0.01, 0.1, 1.0, 3.6] Myr (PopIII.1), 
        [0.01, 1.0, 5, 10, 50] Myr (PopIII.2), 
        [0.01, 1.0, 5, 10, 50] Myr (PopIII.Kroupa)
        & 
        Color-based selection (\S\ref{sec:color}),  \newline
        SED-based selection (\S\ref{sec:sed}), \newline
        Completeness simulation (\S\ref{sec:comp}) \\
                              & 
         \citet{nakajima2022b} &
         Salpeter 1--100$M_{\odot}$ (PopIII.Sal1-100),  1--500$M_{\odot}$ (PopIII.Sal1-500),  50--500$M_{\odot}$ (PopIII.Sal50-500)
         &
         $n_e =$ 10$^3$cm$^{-3}$, $\log U$ = -1.5, $t_{\mathrm{age}}$ = 0 Myr 
        & 
        SED-based selection  (\S\ref{sec:sed}) \\ \hline
        Pop~II & 
        \texttt{BAGPIPES} \newline \citep{carnall2018}  
        &
        Kroupa
        &
        $Z_{\rm gas}/Z_\odot$=[0.01, 0.05, 0.10, 0.20] \newline 
        $\log U$=[-3, -2, -1], $A_{\rm V}$=0 \newline 
        & 
        Color-based selection (\S\ref{sec:color}) \\
                             & 
        \texttt{EAZY} \newline  \citep{brammer2008} &
        Salpeter 
        &
        \texttt{blue\_sfhz\_13}, a strong line emitter template \citep{carnall2023} \newline  &
        SED-based selection (\S\ref{sec:sed}) \\ 
                             & 
         SC SAM \newline  \citep{yung2022} & 
         Chabrier &
         Semi-analytically simulated 1,472,791 galaxies across a light cone in $\sim$800 arcmin$^{2}$ at $z=$0--10 &
        Contamination simulation (\S\ref{sec:sed}) \\   \hline
        Metal-poor AGN &
        \cite{inayoshi2022a, inayoshi2022b} & 
        \nodata &
        Seed BH mass $M_{\rm BH}=10^{6}\,M_{\odot}$ in a metal-poor galaxy ($Z_{\rm gas}/Z_{\odot}=0.01$). Broken power-law accretion disk, nebular emission, and dense accretion disk radiation (0.1--100 pc), with viewing angle $60^\circ$. \newline           
        &
        Additional SED test (\S\ref{sec:agn}, Appendix~\ref{sec:appendix_agn}) \\
         &
        \cite{nakajima2022b} & 
        \nodata &
        DCBH with thermal Big Bump ($T_{\rm bb}=5\times10^4$, $1\times10^5$, $2\times10^5$ K), power-law continuum ($\alpha=$-1.2,-1.6,-2.0), and nebular emission ($n_e=$10$^{3}$cm$^{-3}$, $Z_{\rm gas}/Z_\odot$=0, $\log U$=-1.5) in plane parallel.
        & 
        Additional SED test (\S\ref{sec:agn}, Appendix~\ref{sec:appendix_agn}) \\
        \hline
    \end{tabular}
\end{table*}

\section{Efficient Pop~III Search with NIRCam}
\label{sec:method}

\begin{figure*}[t!]
\begin{center}
\includegraphics[trim=0cm 0cm 0cm 0cm, clip, angle=0,width=1\textwidth]{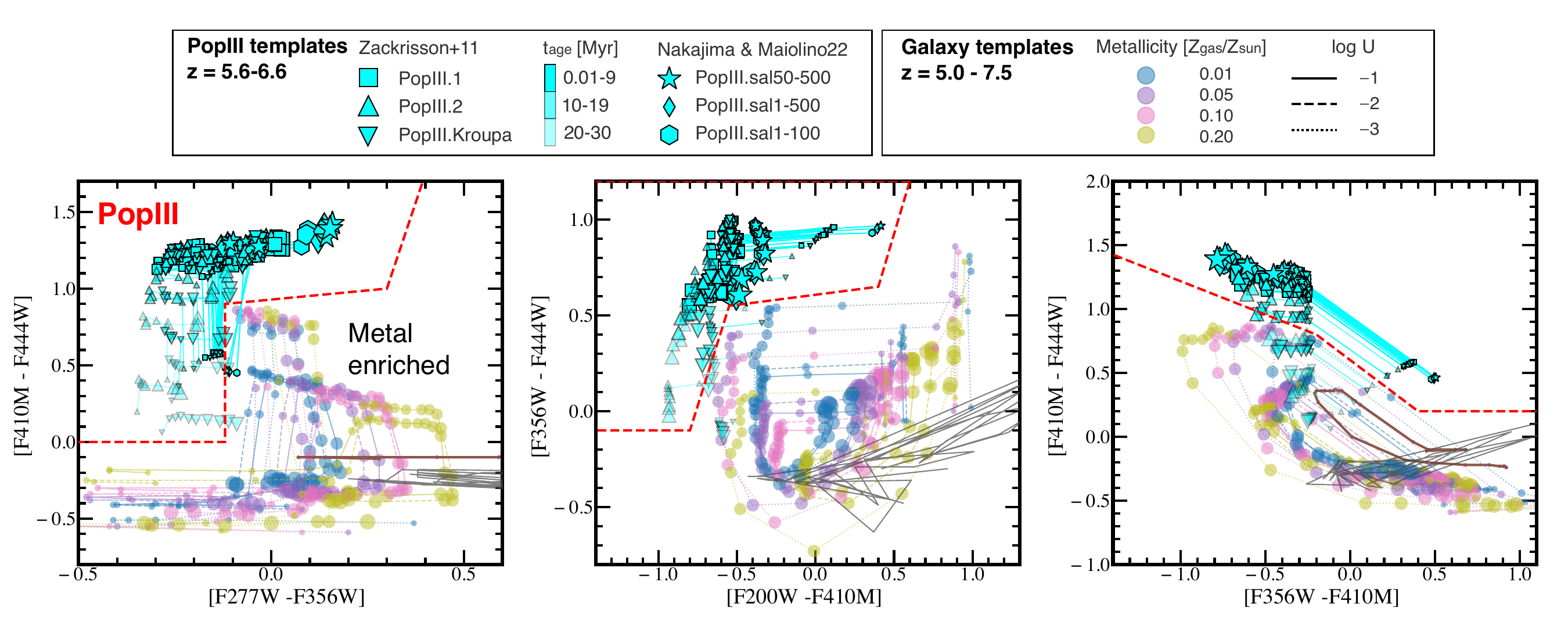}
\end{center}
\vspace{-0.2cm}
 \caption{
\textbf{NIRCam color-color diagrams for selecting $z \simeq$~6--7 Pop~III galaxies.} 
In all diagrams, the selection (dashed red line) is guided by the same set of features: the x-axis colors detect weak \oiii\ lines and/or the Balmer jump, while the y-axis colors capture the strong H$\alpha$ EW.  
The cyan symbols represent Pop~III galaxies at $z=5.6$--6.6, taken from the \zackmodel\ models \citep{zackrisson2011}, assuming a very top-heavy IMF (Pop~III.1), a moderately top-heavy IMF (Pop~III.2), and a standard Kroupa IMF (Pop~III.Kroupa) with $t_{\rm age}=0.01$--30~Myr. \citet{nakajima2022b} models are also shown, assuming a Salpeter IMF with mass ranges of 1--100~$M_{\odot}$ (Pop~III.Sal1-100), 50--500~$M_{\odot}$ (Pop~III.Sal50-500), and 1--500~$M_{\odot}$ (Pop~III.Sal1-500). The transparency of the symbols increases along three tiers of $t_{\rm age}$: 0.01--10~Myr, 10--20~Myr, and 20--30~Myr.  
The other colored symbols represent metal-enriched mock galaxies at $z=5.0$--7.5, generated using \texttt{BAGPIPES}. The blue, dark magenta, pink, and gold colors denote $Z_{\rm gas}/Z_{\odot}=0.01$, 0.05, 0.10, and 0.20, respectively. The solid, dashed, and dotted lines correspond to $\log(U)=-1$, $-2$, and $-3$. All metal-enriched galaxies are assumed to have $A_{\rm V}=0$. 
The brown curve represents a dusty star-forming galaxy at $z=5.0$--7.5 with $Z_{\rm gas}/Z_{\odot}=1.0$, $A_{\rm V}=2.0$, and $\log(U)=-2$ (its red color is off the scale in the middle panel).  
The grey curve indicates cool dwarfs with $T_{\rm eff}=$500--2500~K taken from Phoenix stellar template library in \texttt{EAZY}.  
In all symbols, the sizes, ranging from small to large, correspond to increasing redshift. 
Accounting for contaminants from a wide range of redshifts is critical in the selection process, and thus this wide range is displayed for the metal-enriched galaxies.
\label{fig:color}}
\end{figure*}

In this section we establish that a Pop III search using NIRCam is feasible, and that such a search may be executed with high completeness and negligible contamination. We carefully quantify the imaging depth and filter-set coverage required for a successful search. We begin with Figure~\ref{fig:sed_comp}, where we compare the spectral energy distributions (SEDs) of Pop~III and metal-enriched galaxies at $z=6.4$. Three key features of Pop~III SEDs are highlighted: the absence of \oiii, the strong H$\alpha$, and the Balmer jump\footnote{
Although the Balmer jump is not unique to Pop~III stellar populations, it becomes more pronounced (e.g., [F200W-F410M] $<-1.0$; see Figure~\ref{fig:color}) in very young, metal-free systems due to their hotter stars. Therefore, in this study we include the Balmer jump among the key features for distinguishing Pop~III galaxies in broad-band photometry.
}. 
These three features are strong enough to appear in the NIRCam photometry out to $z\sim7$ enabling us to develop an efficient selection method. 
We note that these SED features may diminish with age after the Pop~III starburst or in case the escape fraction of ionizing photons is relatively high. 
Our study specifically focuses on enabling the search for the young phase of Pop~III galaxies, rather than encompassing all their evolutionary stages.

To quantitatively analyze these features and develop a search strategy, we introduce two photometric approaches purely using NIRCam in the following subsections: i) color-color diagrams, ii) SED fitting. We also evaluate the completeness and contamination rates for each method, with the goal of identifying the most effective approach.

\subsection{Color-based Selection}
\label{sec:color}

In Figure~\ref{fig:color}, we present possible color-color spaces for distinguishing Pop~III galaxies from metal-enriched galaxies at $z\sim$~6--7. The filter combinations are optimized to focus on the distinctive characteristics of Pop~III objects described above -- the complete absence of \oiii, strong hydrogen lines, and Balmer jump. 

For Pop~III galaxies, we utilize the \zackmodel\ models presented in \citet{zackrisson2011}. 
The \zackmodel\ framework combines stellar and nebular emission contributions by employing synthetic spectra from stellar population models and photoionization calculations. 
Additionally, we incorporate the Cloudy-based Pop~III models introduced in \citet{nakajima2022b}, 
which explore a range of physical conditions for Pop~III stellar populations and their surrounding gas. 
We summarize the key parameters of these models in Table~\ref{tab:models} and provide their detail descriptions in Appendix~\ref{sec:appendix_sed}.  
 
For modeling metal-enriched (albeit still quite metal-poor, ``Pop II" galaxies), we employ mock galaxy SEDs generated using \texttt{BAGPIPES} \citep{carnall2018}, a Bayesian spectral fitting code designed for modeling galaxy emission across far-ultraviolet to microwave wavelengths, including stellar population synthesis models, nebular emission, and dust attenuation. 
The key parameters and detailed descriptions of the \texttt{BAGPIPES}-based mock SEDs are also summarized in Table~\ref{tab:models} and Appendix~\ref{sec:appendix_sed}. 

Figure~\ref{fig:color} presents the color tracks of Pop~III galaxies (blue symbols) at $z=5.6$--6.6 and metal-enriched galaxies (other colored symbols) at $z=5.0$--7.5. 
The wide redshift range displayed for the metal-enriched galaxies is helpful to ensure separation even with contaminants from different redshifts in the color-color spaces (see Section~\ref{sec:lowz} for discussion of even lower redshift interlopers). 
We find that Pop~III galaxies are distinctly separated from metal-enriched galaxies in specific regions of the color-color space owing to their unique spectral features highlighted in Figure~\ref{fig:sed_comp}. 

It is important to note that these features, especially the strong H$\alpha$ line reflected in the y-axis colors in all diagrams, are sensitive to the stellar age (\(t_{\rm age}\)).  
The colors of the Pop~III galaxies suggest that these distinctive features begin to diminish around \(t_{\rm age} \simeq 10\)--30~Myr (see faint cyan points in Fig. \ref{fig:color}). Importantly, the effectiveness of the color separation depends on the redshift range, as the technique relies on strong emission lines falling in specific NIRCam filters. Since Pop~III star formation is more likely to occur at higher redshifts, this study focuses on the \(z \simeq 6\)--7 range, where the H$\alpha$ line is still captured by the NIRCam filters. However, extending this technique to lower redshifts using different filter sets is an avenue worth exploring in future studies \citep[cf. $z\sim4$--5; ][]{nishigaki2023}. In Appendix~\ref{appendix:sed}, we also discuss the contamination rate when extending the Pop~III search to lower redshifts. 
We also note that the separation between populations seen in Figure~\ref{fig:color} is driven by F410M that is sensitive to the continuum level thereby constraining the high H$\alpha$ EW and Balmer jump, highlighting the crucial role played by medium-bands. The separation could be made even cleaner when adding additional medium bands that help more securely constrain the line strengths of H$\alpha$ and \oiii. However, the effective redshift range probed becomes narrower to capture the specific emission lines with specific medium band filters. 
In Appendix~~\ref{sec:appendix_f480m} we discuss the color-color diagrams and the effective redshift range when utilizing F480M instead of F444W.

To facilitate the identification of Pop~III candidates, we define the red dashed-line regions in the color-color diagrams as selection criteria for Pop~III galaxies. Since this color selection is effective only within specific redshift ranges, we first perform an SED template fitting using \texttt{EAZY} \citep{brammer2008} with the 14 Pop~III templates described in Section~\ref{sec:sed}. We note that this means these two steps of redshift and color-color selections are not strictly independent, but this is essential to ensure the applicability of the color diagrams. 
We set the following redshift-based selection criteria: 
\begin{equation}
\label{eq:zphot}
\begin{split}
    5.0 < z_{\rm phot, PopIII} < 7.5 \land z_{\rm phot, PopIII} - z_{16} < 0.5 \\
    \land z_{84} - z_{\rm phot, PopIII} < 0.5
\end{split}
\end{equation}
where \(z_{\rm phot, PopIII}\) refers to the best-fit photometric redshift derived from \texttt{EAZY} with the Pop~III templates, while \(z_{16}\) and \(z_{84}\) denote the redshift at the 16th and 84th percentile.  
Then, we step forward to the color selection enclosed by the following vertices in the color-color space:
\begin{itemize}
    \item F356W--F277W vs. F410M--F444W (vertices): 
    \begin{equation}
    \begin{split}
        (0.5, 2.5), \, (0.3, 1.0), \, (-0.12, 0.9), \\
        (-0.12, 0.0), \, (-0.8, 0.0), \, (-0.8, 2.5)
    \end{split}
    \end{equation}
    \item  F410M--F200W vs. F356W--F444W (vertices):
    \begin{equation}
    \begin{split}
       (0.6, 1.2), \, (0.4, 0.65), \, (0.55, 0.55), \\ 
       (-0.8, -0.1), \, (-2.5, -0.1), \, (-2.5, 1.2)
    \end{split}
    \end{equation}
    \item F410M--F356W vs. F410M--F444W (vertices):
    \begin{equation}
    \begin{split}
       (1.5, 0.2), \, (0.4, 0.2), \, (-0.2, 0.85), \\ 
       (-2.5, 1.2), \, (-2.5, 2.7), \, (1.5, 2.7), \, (1.5, 1.8)
    \end{split}
    \end{equation}
\end{itemize}

These three color diagrams are based on the same principles: the x-axis colors are sensitive to the weak \oiii\ line and/or the Balmer jump, while the y-axis colors are designed to detect the strong H$\alpha$ line (see also Figure~\ref{fig:sed_comp}).
Due to the clear separation observed in all three color-color diagrams, any one of these criteria alone could suffice for identifying Pop~III candidates. However, applying all three criteria together helps mitigate contamination from noise fluctuations, especially for low S/N sources.
Each diagram places different weights on specific features: for instance, the [F200W -- F410M] color in the second diagram is more directly sensitive to the Balmer jump, the [F356W -- F410M] color in the third diagram better captures the weak \oiii\ line, and the [F277W -- F356W] color in the first diagram is sensitive to both features.
Therefore, combining these three criteria enhances the robustness of the selection process\footnote{ 
In the future, the application of machine learning techniques to perform multi-dimensional separation could further improve the selection process \citep[e.g.,][]{kojima2020}.}. 
Next, we discuss the SED-based selection method after which we turn to evaluate the completeness (Section~\ref{sec:comp}) and contamination rate (Section~\ref{sec:contami}) to ensure reliable selection of Pop~III candidates.

\subsection{SED-based Selection}
\label{sec:sed}

Similar to the high-redshift galaxy search \citep[e.g.,][]{finkelstein2015}, 
another promising method for identifying Pop~III candidates is using SED fitting, 
in addition to the color-based approach (Section~\ref{sec:color}). 
This SED-based approach leverages the full photometric data set rather than relying on specific filter combinations optimized for certain redshift ranges. 
By maximizing the information extracted from all available photometric bands, 
SED fitting may enable a more comprehensive search for robust Pop~III candidates. We extend the concept of robustness of high-redshift candidates based on the difference in the $\chi^{2}$ between the best-fit high-$z$ and forced low-$z$ SEDs to the Pop III use case \citep[e.g.,][]{naidu2022, donnan2023, harikane2023, finkelstein2024}. We use \texttt{EAZY} \citep{brammer2008} to compute the $\chi^{2}$ difference between the best-fit metal-enriched galaxy templates and Pop III templates.

The first template set consists of the \texttt{blue\_sfhz\_13} model subset\footnote{\url{https://github.com/gbrammer/EAZY-photoz/tree/master/templates/sfhz}}, aiming to obtain the best-fit SED as a metal-enriched galaxy. This set incorporates redshift-dependent star formation histories (SFHs) and dust attenuation. Additionally, the linear combinations of log-normal SFHs are further constrained to ensure they do not start earlier than the age of the Universe. The template set is further expanded with a blue galaxy template derived from a high-SNR \jwst\ spectrum of a lensed \(z = 8.50\) galaxy representative of typical nebular conditions in the early universe \citep{carnall2022}. This template extends the color grid coverage for galaxies with extreme emission line equivalent widths, resulting in a total of 14 templates.

The second template set comprises 14 Pop~III templates based on the \zackmodel\ model \citep{zackrisson2011}, aiming to identify the best-fit SED of a Pop~III galaxy. 
These templates are selected from three IMF scenarios of Pop~III.1, Pop~III.2, and Pop~III.Kroupa summarized in Table~\ref{tab:models}. From Pop~III.1 (the most top-heavy IMF), we use three templates with stellar ages \(t_{\rm age} = [0.01, 1.0, 3.6]\) Myr. Pop~III.2 (moderately top-heavy IMF) contributes five templates with \(t_{\rm age} = [0.01, 1.0, 5, 10, 50]\) Myr, and Pop~III.Kroupa (Kroupa IMF) provides six templates with similar stellar ages of \(t_{\rm age} = [0.01, 1.0, 5, 10, 50, 100]\). Note that the difference in age range across the IMF flavors is because no massive stars survive beyond a few Myrs in the top-heavy scenario. In all cases, the gas covering fraction is fixed at \(f_{\rm cov} = 1.0\), ensuring maximal nebular contribution. Although additional Pop~III templates, such as those from \citet{nakajima2022b}, could be included, their contribution to the color-space coverage is minimal (see Figure~\ref{fig:color}). 
To maintain balance with the first template set, we limit the second set to 14 templates, ensuring the same number of models for a fair evaluation of the delta $\chi^{2}$ measurement. 

With these template sets, we run \texttt{EAZY} and obtain the $\chi^{2}$ values for the best-fit metal-enriched galaxy template, \(\chi^2_{\rm galaxy}\), and Pop~III galaxy template set, \(\chi^2_{\rm PopIII}\). By defining  $\Delta \chi^{2}\equiv \chi^{2}_{\rm galaxy} - \chi^{2}_{\rm PopIII}$, we set the following $\chi^{2}$ criteria:
\begin{equation}
    \Delta \chi^{2} \geq\, C_{\rm thresh} \land \chi^{2}(\text{PopIII})< 10
\end{equation}
\[
    {\rm or}
\]
\begin{equation}
\label{eq:sed2}
    \Delta \chi^{2} \geq 30 \land \chi^{2}(\text{PopIII})< 20
\end{equation}
where \(C_{\text{thresh}}\) is a threshold value. 
In high-redshift galaxy studies \citep[e.g.,][]{naidu2022, donnan2023, harikane2023, finkelstein2024}, \(C_{\text{thresh}}\) is typically set to either 4 or 9.\footnote{These thresholds correspond approximately to 2$\sigma$ and 3$\sigma$ significance levels for a degree of freedom (DoF) of 1. However, as SED fitting involves multiple parameters beyond redshift, the actual DoF is \(>1\), and this statistical interpretation should be regarded as an approximation.}  
The choice of $C_{\text{thresh}}$ represents a trade-off between completeness and contamination rates. 
Given the scope of our search for extremely rare objects like Pop~III galaxies,
we prioritize mitigating the contamination rate and adopt $C_{\text{thresh}} = 9$ in the following analyses.  
Since the \(\chi^{2}\) value can naturally increase with the number of filters used in SED fitting, Equation~\ref{eq:sed2} is incorporated to ensure that potential candidates are not overlooked in regions where more extensive filter-sets are available.
Unlike the color-based selection (Section~\ref{sec:color}), the SED-based selection is not restricted to a specific redshift range. However, to ensure a fair comparison between the two methods, we adopt the same redshift criteria defined in Equation~\ref{eq:zphot}. 
In Appendix~\ref{appendix:sed}, we discuss the distributions of the completeness and contamination rate when the SED-based method is applied to a broader redshift range.

Note that we use the photometric redshift estimated with the Pop~III templates ($z_{\rm phot, PopIII}$), rather than galaxy templates, for the redshift criteria above.  
This choice is motivated by the fact that the Lyman-break feature is not always well-constrained at this redshift range, particularly in the faint regime (\(>28~\text{mag}\)), where extremely deep HST/ACS data would be required.  
Without robust constraints from the Lyman-break feature, general galaxy templates may produce high $\chi^{2}$ values due to their inability to reproduce the unique Pop~III signatures (Figure~\ref{fig:sed_comp}), potentially leading to incorrect redshift estimates.  
While Pop~III templates may similarly result in incorrect redshift estimates for general metal-enriched galaxies, the primary goal of this analysis is to identify Pop~III candidates, rather than providing the most accurate redshift estimates for the majority of metal-enriched galaxies.  
Therefore, the photometric redshift refers to the estimate based on the Pop~III templates throughout this paper (otherwise specified), including the following completeness and contamination rate measurements.

\subsection{Completeness}
\label{sec:comp}

\begin{figure*}[t!]
\begin{center}
\includegraphics[trim=0cm 0.1cm 0cm 0cm, clip, angle=0,width=1\textwidth]{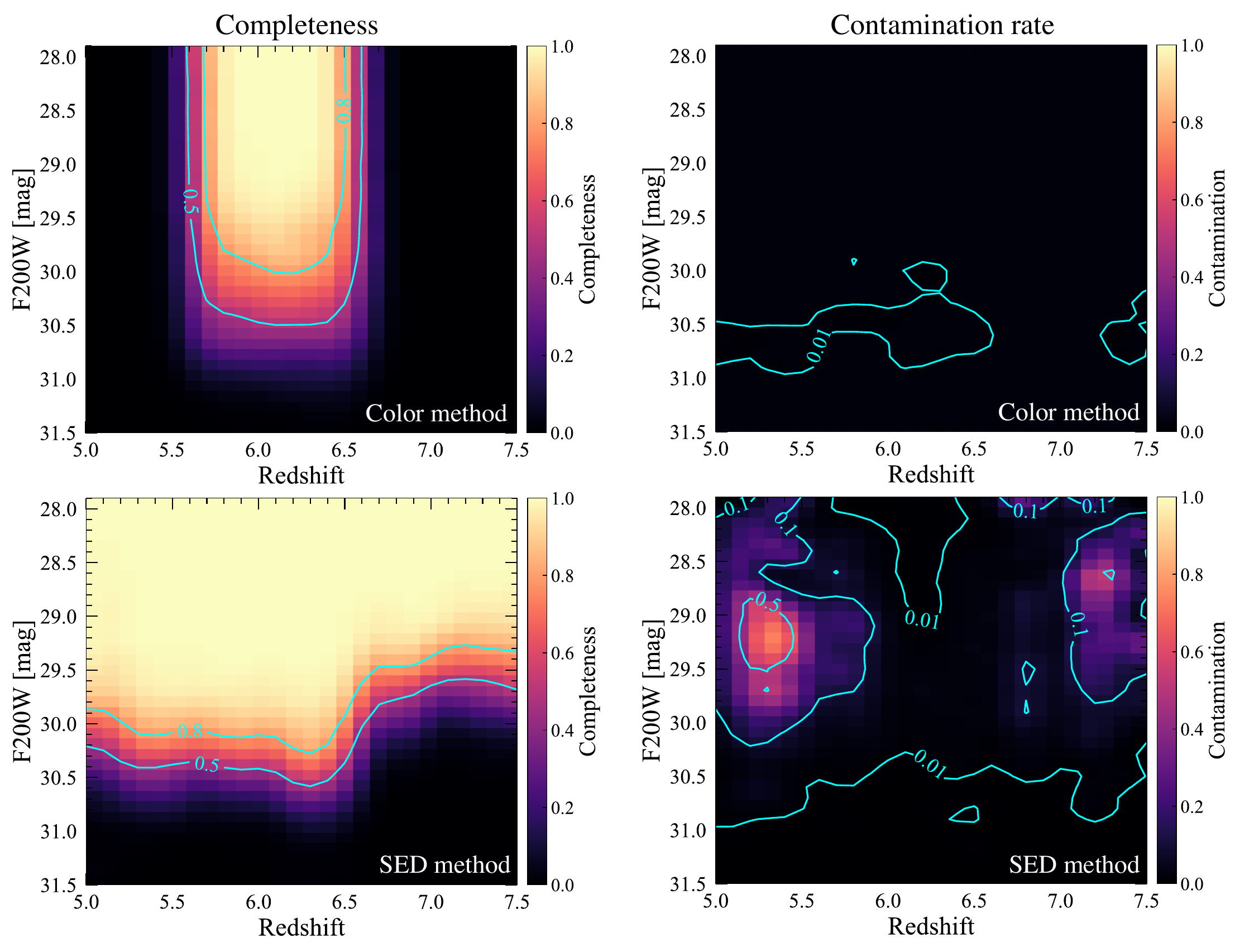}
\end{center}
\vspace{-0.2cm}
 \caption{
\textbf{Completeness (left) and contamination rate (right) of the Pop~III selection based on the color-based (top; Section~\ref{sec:color}) and SED-based (bottom; Section~\ref{sec:sed}) methods.} 
The cyan contours denote 0.5 and 0.8 for completeness and 0.01, 0.1, and 0.5 for contamination rate. 
For completeness, we use a \zackmodel\ Pop~III model (Pop~III.2, $t_{\rm age}=1.0$~Myr, $f_{\rm cov}=1.0$), while confirming similar results with other templates representing young stellar populations (i.e., those with strong nebular emission lines).  
For contamination rate, we utilize the Santa Cruz Semi-Analytic Model (SC SAM) catalog \citep{yung2022}, which includes 1,472,791 metal-enriched mock galaxies spanning $z=0$--10 in a light cone covering $\simeq800$~arcmin$^{2}$. To ensure a fair assessment, we first derive $z_{\rm phot, PopIII}$ using \texttt{EAZY} and bin the galaxies in redshift and observed F200W magnitude grids. 
The completeness and contamination rate are then calculated in each grid through Monte Carlo simulations by adding random noise based on the GLIMPSE data depth and applying the same selection procedures used in the observations.  
The completeness and contamination rates reflect a trade-off between the two methods. 
The color-based method achieves high completeness ($>$50--80\%) within a limited redshift range, accompanied by a low contamination rate, whereas the SED-based method provides high completeness over a broader redshift range but with a relatively higher contamination rate.  
While both completeness and contamination rates are critical, we prioritize achieving a low contamination rate in the search for extremely rare objects like Pop~III galaxies. To this end, we combine the color-based and SED-based methods to maximally mitigate contamination while maintaining robust completeness (Figure~\ref{fig:fiducial}).  
\label{fig:comp_contami}}
\end{figure*}

\begin{figure*}[t!]
\begin{center}
\includegraphics[trim=0cm 0cm 0cm 0cm, clip, angle=0,width=1\textwidth]{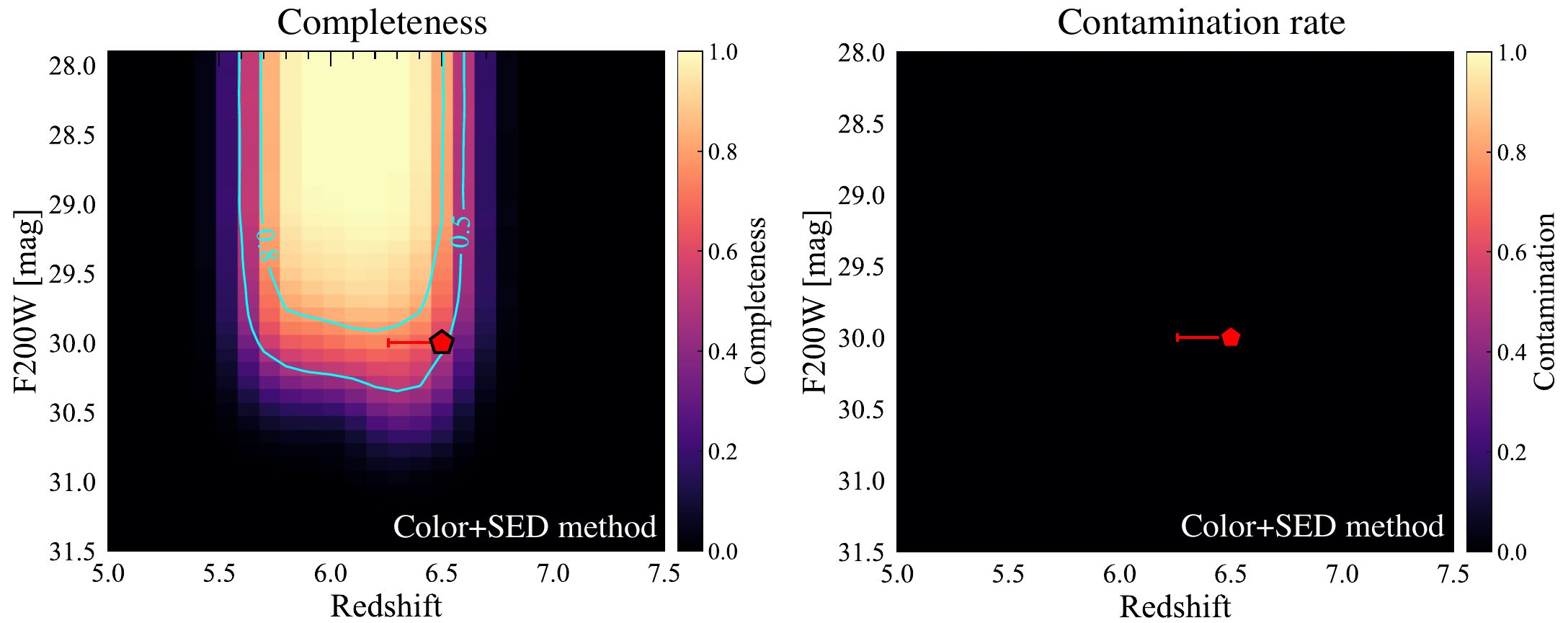}
\end{center}
\vspace{-0.2cm}
 \caption{
\textbf{Completeness and contamination rate for our fiducial Pop~III selection combining the color-based and SED-based methods.} The contamination rate is reduced to zero across the entire parameter space, ensuring a robust search for Pop~III galaxies with NIRCam.  
The red pentagon represents the parameter space of the promising Pop~III candidate \targ\ found in GLIMPSE that satisfies our fiducial selection criteria.  
\label{fig:fiducial}}
\end{figure*}

To evaluate the selection function of our color-based and SED-based methods, we assess the completeness of each approach using Monte Carlo (MC) simulations. 
For both simulations, we adopt the \zackmodel\ Pop~III.2 model \citep{zackrisson2011} with \(t_{\rm age}=1.0\)~Myr and \(f_{\rm cov}=1.0\). 
Note that we verify negligible differences in the final completeness results when repeating the simulations with various Pop~III models, provided \(t_{\rm age} \lesssim 15\)~Myr.  
This outcome is expected, as the key SED features (see Figure~\ref{fig:sed_comp}) begin to diminish as \(t_{\rm age}\) increases (see Figure~\ref{fig:color}). 
NIRCam photometry is calculated for the nine filters used in GLIMPSE, shifting and rescaling the template across redshift and \(m_{\rm F200W}\) grids spanning \(z=[5.0:7.5]\) and \(m_{\rm F200W}=[31.5:28.0]\), respectively. At each grid point, 100 mock NIRCam photometric sets are generated by injecting random noise based on the detection limits of each filter. 
To ensure consistency in redshift criteria across both methods, we apply the \(z_{\rm phot}\) selection defined in Equation~\ref{eq:zphot} by running \texttt{EAZY} with the Pop~III template set for the 100 mock photometric sets at each grid point. For the SED-based method, we also separately run \texttt{EAZY} with the metal-enriched galaxy templates. Following the selection procedures described in Sections~\ref{sec:color} and \ref{sec:sed}, we calculate the completeness for each grid point for both methods.

The left panels of Figure~\ref{fig:comp_contami} show the completeness estimates for the color-based (top panels) and SED-based (bottom panels) methods. 
For the color-based method, we find that the completeness function exhibits steep cutoffs at $z\simeq5.6$ and $z\simeq6.6$. This reflects the dependence of the color-based method on the specific NIRCam filters to capture the unique SED features of Pop~III galaxies, such as bright hydrogen lines and the absence of the \oiii\ line (Figure~\ref{fig:sed_comp}). In other words, at $z \lesssim 5.6$ or $z \gtrsim 6.6$, either the \oiii\ or H$\alpha$ emission line falls outside the F356W and F444W filters, resulting in a loss of these diagnostic features. Within the $z\simeq5.6$--6.6 range, the completeness exceeds \(\sim\)50--80\% for \(m_{\rm F200W} \leq 30.0\), ensuring high completeness within this specific redshift range. 
In contrast, the SED-based method shows high completeness over a broader redshift range, reflecting its ability to leverage the full photometric dataset rather than being restricted to specific filters optimized for a particular redshift. However, the completeness decreases at the faint end (\(m_{\rm F200W} > 30.0\)~mag) for \(z \gtrsim 6.6\), as the H$\alpha$ line moves out of the F444W filter, losing one of the strongest features of Pop~III galaxies in the SED.

\subsection{Contamination Rate}
\label{sec:contami}

Here we investigate the contamination rate for both methods. For this analysis, we utilize the DR3 mock galaxy catalog generated with the Santa Cruz Semi-Analytic Model, as presented in \citet{yung2022}.\footnote{The catalog is available at \url{https://ceers.github.io/sdr3.html\#catalogs}} 
This catalog contains 1,472,791 metal-enriched galaxies across a light cone spanning an area of \(\sim800\)~arcmin\(^2\) and covering the redshift range \(z=0\)--10. The catalog includes photometry for all NIRCam filters, accounting for emission line contributions from the mock galaxy spectra. To simulate realistic observations, we inject random noise into the NIRCam photometry based on the GLIMPSE detection limits (see Section~\ref{sec:glimpse}). We then derive the observed \(m_{\rm F200W}\) magnitudes and estimate \(z_{\rm phot, PopIII}\) using \texttt{EAZY}. The mock galaxies are sorted into redshift and \(m_{\rm F200W}\) grids spanning \(z=[5.0:7.5]\) and \(m_{\rm F200W}=[31.5:28.0]\). Following the selection criteria defined in Sections~\ref{sec:color} and \ref{sec:sed}, we estimate the contamination rate for each grid in both methods. Note that we confirm that most grids contain $>$10--100 galaxies, ensuring reliable statistics.

The right panels of Figure~\ref{fig:comp_contami} show the contamination rate -- the fraction of metal-enriched galaxies that satisfy the Pop III criteria --  estimates for the color-based (top panels) and SED-based (bottom panels) methods.  
For the color-based method, the contamination rate is nearly zero across most of the parameter space, with only a few regions at the faintest regime ($\simeq$~30--31~mag) exhibiting modest contamination rates (\(\sim1\%\)). This is likely due to the noise fluctuation at the faintest regime close to the detection limit, while this modest contamination rate highlights the robustness of the color-based method.  
In contrast, the SED-based method displays contamination rates exceeding \(\sim1\%\) across almost the entire parameter space, with rates exceeding 10--50\% in certain regions.\footnote{
In the contamination rate distribution shown in Figure~\ref{fig:comp_contami}, we note that the redshift is binned based on \(z_{\rm phot, PopIII}\), rather than the true redshift of the mock galaxies, to ensure consistency with the procedure applied to the observed data. Therefore, the high contamination rates of \(>10\)--50\% in specific parameter spaces only apply to sources whose SEDs are initially favored by the Pop~III templates at \(z \simeq 5.0\)--7.5.} 
However, within the specific redshift range \(z\simeq5.8\)--6.6, the contamination rate remains modest (\(<10\%\)). Similar to the color-based method, these are the redshifts where the unique SED features of Pop~III galaxies -- strong Balmer lines and the absence of the \oiii\ line -- are effectively captured by the NIRCam filters.  

Although the SED-based method achieves high completeness over a broad redshift range, there is significant contamination from metal-enriched galaxies. To mitigate this, it is necessary to focus on an optimal redshift range that is very similar to the color-based method. This motivates our choice to apply the SED-based method to decrease contamination by adding it as an additional selection criterion to the color-based method, rather than applying it independently across a wide redshift range.

\subsection{Fiducial Selection Method}
\label{sec:fiducial}

Based on the results of the completeness and contamination tests, we combine the selection criteria of the color-based and SED-based methods (i.e., Pop III candidates must satisfy Eqs.\ref{eq:zphot}--\ref{eq:sed2}).  
Figure~\ref{fig:fiducial} shows the completeness and contamination rates derived from the combined color+SED method.  
The completeness remains largely similar to that of the color-based method due to the broader selection range allowed by the SED-based method.  
However, the contamination rate is reduced to nearly zero across almost the entire parameter space, achieved through the strict selection function provided by the combination of both methods.  
We therefore adopt this combined technique as our fiducial selection method.

\begin{figure*}[t!]
\begin{center}
\includegraphics[trim=0cm 0cm 0cm 0cm, clip, angle=0,width=1.\textwidth]{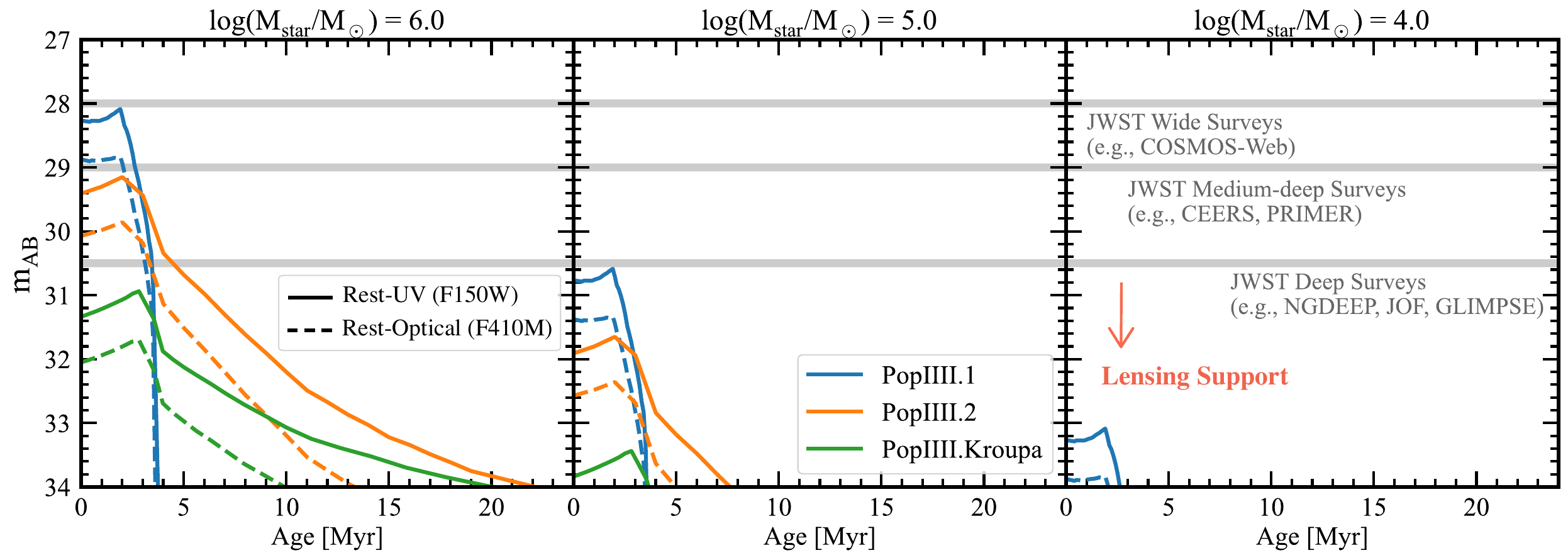}
\end{center}
\vspace{-0.2cm}
 \caption{\textbf{A meaningful Pop III search demands the deepest \jwst\ surveys as well as the assistance of gravitational lensing}. Apparent magnitudes in the rest-frame UV (solid lines; F150W) and optical (dashed lines; F410M) are shown as a function of $t_{\rm age}$ for Pop~III galaxies at $z=6.5$.  
The three colored curves represent the \zackmodel\ Pop~III models with different IMFs: Pop~III.1 (blue; very top-heavy), Pop~III.2 (orange; moderately top-heavy), and Pop~III.Kroupa (green; standard).  
The three $M_{\rm star}$ cases are shown from left to right, with a simple scaling applied to the \zackmodel\ Pop~III models, which originally assume $M_{\rm star}=10^{6}\,M_{\odot}$. Following the arguments outlined in Section~\ref{sec:depth}, we consider $\approx10^{5} M_{\rm{\odot}}$ as our fiducial case -- this makes clear that only the deepest JWST surveys, further aided by lensing, have a reasonable chance at detecting Pop III stars. While the mass-to-light ratio could vary with $M_{\rm star}$ due to differences in nebular parameters (e.g., electron density, ionization parameter), this figure provides a first-order insight into the NIRCam depth required for a meaningful Pop III search (see Section~\ref{sec:depth}). 
The future prospects for the meaningful survey design are also discussed in Section~\ref{sec:future}. 
\label{fig:mobs_age_all}}
\end{figure*}

\subsection{Required Depth for the Pop~III Search}
\label{sec:depth}

Although we define the fiducial selection function in Section~\ref{sec:fiducial}, 
the selection focuses on the SED colors alone. 
It is also important to consider the possible luminosity for the Pop~III galaxies and understand the required data depth for a meaningful Pop~III search.  
The first galaxies are predicted to form within halos where the cooling of primordial gas enables star formation. At $z\sim6$--7, the increasing background of Lyman-Werner radiation likely suppresses molecular hydrogen cooling, making atomic cooling the dominant mechanism. In these cases, a virial temperature of $T_{\rm vir} \sim 10^4\,{\rm K}$, the threshold for efficient atomic cooling, corresponds to a halo mass of $M_{\rm vir} \sim 10^8 \, M_\odot$ \citep[e.g.,][]{bromm2011, kulkarni2021}. In regions yet to be reionized, such metal-free halos are expected to form Pop~III stars upon reaching the atomic cooling limit. Assuming a nominal gas conversion factor $f_{\rm cool}=0.01$ within halos and a time-averaged star formation efficiency $\epsilon_{\star}=0.1$, a single starburst is expected to yield a stellar mass of $M_{\rm star}\simeq 10^{5}\,M_{\odot}$ \citep[see e.g., review by][]{bromm2011}. This estimate may be conservative, as recent \jwst\ studies at $z \geq 6$ suggest $\epsilon_{\star}$ values greater than 0.1. This is supported by the significant abundance of UV-bright galaxies at $z \gtrsim 9$ \citep[e.g.,][]{harikane2023,finkelstein2024}, the early formation of massive galaxies by $z=5$--6 \citep[e.g.,][]{xiao2023, graaff2024,weibel2024}, and the high stellar and gas densities observed in resolved lensed galaxies at $z=6$--10 (e.g., \citealt{adamo2024, fujimoto2024, mirka2025} -- however, see also \citealt{donnan2025} for an alternative viewpoint). Nevertheless, our simple estimate provides a useful baseline.

Figure~\ref{fig:mobs_age_all} illustrates how the apparent magnitude of Pop~III galaxies varies as a function of $t_{\rm age}$.  
We use the \zackmodel\ Pop~III galaxy template ($M_{\rm star}=10^{6}\,M_{\odot}$) with three different IMFs, and also present cases for $10^{5}\,M_{\odot}$ and $10^{4}\,M_{\odot}$ by assuming simple scaling, as per the baseline $M_{\rm star}$ for first galaxies that are dominated by Pop~III stars as described above.  
For reference, the typical 5$\sigma$ depths of \textit{JWST} legacy surveys are shown with horizontal gray lines, corresponding to three survey layers: wide ($\simeq$~28~mag), medium-deep ($\simeq$~29~mag), and deep ($\simeq$~30.5~mag) surveys \citep[e.g.,][]{casey2022, finkelstein2024, donnan2024, bagley2023, robertson2023}.  
In most cases, the Pop~III galaxies are likely to be too faint ($\gtrsim$30.5~mag) for these \jwst\ surveys, except under the most optimistic conditions such as $M_{\rm star}\simeq10^{6}\,M_{\odot}$, top-heavy IMFs, and $t_{\rm age}\lesssim5$~Myr.  
This highlights the immense challenge in successfully identifying Pop~III galaxies, consistent with the absence of robustly confirmed Pop~III sources to date. From Figure \ref{fig:mobs_age_all} it is clear that leveraging gravitational lensing is critical to push past $\approx30.5$ mag, without which the majority of scenarios remain out of reach (involving e.g., slightly older stellar ages $>5$ Myrs). Indeed, the most metal-poor source detected with \textit{JWST} till date is a highly magnified clump at $z=6.6$ ($\mu>100$; \citealt{vanzella2023b}).

\section{Application to Deep JWST Surveys}
\label{sec:obs}

In Section~\ref{sec:method}, we outlined our novel NIRCam selection method for selecting $z \simeq 6$--7 Pop~III galaxies and the required depths for a meaningful search.  
Based on the filters used for the color-color diagrams, which are optimized to capture the key features of the Balmer jump, strong H$\alpha$ emission, and the absence of the \oiii\ line, the NIRCam selection method requires the F200W, F277W, F356W, F410M, and F444W filters (Figure~\ref{fig:color}).  
Also given the preference for deeper data (Figure~\ref{fig:mobs_age_all}), we focus on the following publicly available NIRCam data from \jwst\ legacy surveys: CEERS \citep{finkelstein2024}, PRIMER-UDS/COSMOS \citep{donnan2024}, JADES Origins Field (JOF; \citealt{robertson2023}), UNCOVER+MegaScience \citep{bezanson2022, suess2024}, and the latest Cycle~2 large program, GLIMPSE (Atek et al., in prep.). 
We summarize the data sets used in our analysis in Table~\ref{tab:fields}. 
Collectively, these surveys are representative of the various depth tiers shown in Figure~\ref{fig:mobs_age_all}, enabling a systematic ``wedding cake" search. 
Below, we briefly describe the various surveys and how we incorporate their data.

\begin{deluxetable*}{lp{5cm}ccp{4cm}}
\tablewidth{0.99\textwidth}
\tabletypesize{\footnotesize}
\tablecaption{Summary of datasets used in our systematic Pop~III galaxy search \label{tab:fields}}
\tablehead{
\colhead{Field} & \colhead{NIRCam Surveys} & \colhead{Area [arcmin$^{2}$]} & \colhead{Depth$^{a}$ [AB mag]} & \colhead{Filterset$^{c}$}
}
\startdata
\vspace{-0.3cm}\\
Abell S1063 & GLIMPSE (Atek et al., in prep.; this work; \citealt{kokorev2024}) & 9.4 & 30.2-30.5$^{b}$ & $\geq0.9\mu$m wide filters + F410M + F480M\\
Abell 2744 & UNCOVER \citep{bezanson2022}, MegaScience \citep{suess2024}, ALT \citep{naidu2024}, BEACONS \citep{morishita2024}, \#2883 (PI: Sun), \#2756 (PI: Chen), \#3538 (PI: Iani) & 45 & 29.2-30.2$^{b}$ & All NIRCam wide and medium bands \\
GOODS-S & JOF \citep{robertson2023}, JADES \citep{eisenstein2023a,eisenstein2023b}, FRESCO \citep{oesch2023} & 9.4 & 30.2-31.0 & $\geq0.9\mu$m wide filters + F182M + F210M + F250M + F300M + F335M + F410M \\
EGS & CEERS \citep{finkelstein2024} & 88.1 &  28.6-29.2 & $\geq1.1\mu$m wide filters + F410M \\
UDS & PRIMER \citep{donnan2024} & 170.7 & 27.6-28.4 & $\geq0.9\mu$m wide filters + F410M \\
COSMOS & PRIMER \citep{donnan2024} & 132.2 & 27.6-28.4 & $\geq0.9\mu$m wide filters + F410M \\
\vspace{-0.3cm}
\enddata
\tablenotetext{a}{All depths quoted are 5$\sigma$ depths measured in 0\farcs2 diameter, with the exception of Abell 2744 (0\farcs16 and 0\farcs32 diameter apertures for the SW and LW; \citealt{bezanson2022, suess2024}).}
\tablenotetext{b}{In fields leveraging gravitational lensing (Abell S1063, Abell 2744), intrinsically fainter sources are accessible -- e.g., the effective depth of GLIMPSE in Abell S1063 reaches $>33$ mag in regions with $\mu>10$.}  
\tablenotetext{c}{By selection, coverage in almost all SW+LW wide filters (F090W, F115W, F150W, F200W, F277W, F356W, F444W) and at least one medium-band (F410M) is available in all our studied fields. Abell 2744 is additionally covered by deep F070W imaging as well as by every single medium-band (F140M, F162M, F182M, F210M, F250M, F300M, F335M, F360M, F410M, F430M, F460M, F480M).}
\end{deluxetable*}

\subsection{GLIMPSE}
\label{sec:glimpse}
We performed our own reductions for the GLIMPSE data, and here we briefly explain the survey, observations, data reduction, and its processing. 
The GLIMPSE survey (PID~3293; PIs: H. Atek \& J. Chisholm) is designed to probe the faintest galaxies during the epoch of reionization. Observations were conducted between September 20 and September 28, 2024.  
The target was the well-studied massive lensing cluster Abell~S1063, especially taking advantage of the largest number of highly magnified sources ($\mu>10$), leveraged by one of the most robust lens models, among the 6 Hubble Frontier Field Clusters \citep[e.g.,][]{lotz2017}. 
The program obtained ultra-deep NIRCam imaging across seven broadband at $\geq0.9\mu$m 
and two medium bands of F410M and F480M, with exposure times for each filter ranging from 16 to 39 hours. Specifically, the deepest imaging was performed in the F090W and F115W bands, with total exposure times of 39 hours each, while other filters had exposure times of $\sim$16–23 hours, achieving the almost homogeneous sensitivity at $\sim1$--5$\mu$m. 
The availability of deep HST/ACS imaging ($5\sigma$ depth of $\sim29$~mag) is also beneficial for our study, providing crucial constraints on the Lyman-break features for our targeted populations at $z\sim6$--7.
The orientation of the observations was set to ensure optimal coverage of the lensing cluster. Module B was centered on the high-magnification core of the lensing cluster (magnification factor $\mu\gtrsim3$), while Module B still covered moderately magnified regions ($\mu \sim 1.2$–1.5). 
GLIMPSE reaches depths of 30.2--30.5~AB mag (5$\sigma$, point source, $0\farcs2$-diameter aperture) in broadbands, comparable to the deepest blank-field surveys such as NGDEEP \citep[e.g.,][]{bagley2023} and JOF \citep[e.g.,][]{robertson2023}. By leveraging the magnification provided by Abell S1063, the effective sensitivities achieve unprecedented depths of $\gtrsim 33$ AB mag in the most highly magnified regions ($\mu\gtrsim10$), enabling the detection of galaxies an order of magnitude fainter than previously achievable. 

The raw NIRCam imaging data were processed using the \jwst\ pipeline (v1.9.7) with the \texttt{1293.pmap} context map. Key steps included custom corrections for cosmic rays, $1/f$ noise, and detector artifacts, as well as refined background subtraction to mitigate contamination from bright cluster galaxies. The final mosaics were created using a drizzling algorithm, yielding pixel scales of $0\farcs02$ for the short wavelength channels and $0\farcs04$ for the long wavelength channels.

To account for contamination from intra-cluster light (ICL) and bright cluster galaxies (bCGs), we employed an iterative modeling approach. The ICL was modeled using smooth polynomial functions, while the BCGs were subtracted using \texttt{GALFIT}, employing S\'ersic profile fits. 
These corrections were applied consistently across all exposures and filters. 
After removing ICL and BCG contributions, a secondary local background subtraction was performed to refine the photometry in affected regions. 
This multistep approach ensures accurate measurements of faint sources, particularly near the lensing cluster.

Sources were identified using \texttt{SExtractor}, with detections based on a combined image stack of the F277W, F356W, and F444W bands. To ensure consistent photometry across all bands, point spread functions (PSFs) were empirically constructed by stacking isolated stars in the field. The PSFs of all bands were then matched to the lowest resolution filter, F480M, using convolution kernels. Fluxes were measured in circular apertures with diameters ranging from $0\farcs1$ to $1\farcs2$, and aperture corrections were applied based on the empirically derived growth curves of the matched PSFs. We adopt the $0\farcs2$-diameter aperture-corrected photometry in the following analyses. 
Photometric uncertainties were estimated by placing random apertures in source-free regions of the mosaics, thereby accounting for correlated noise and background variations. 
The final photometric catalog includes all sources with a signal-to-noise ratio (S/N) exceeding $3$ in at least three bands.
Further details of the observation setup, scope, and data processing will be presented in a separate paper (H.~Atek et al. in prep.; see also \citealt{kokorev2024}).

\subsection{UNCOVER+MegaScience}
\label{sec:uncover}

The UNCOVER survey (PID~2561; PIs: I.~Labb\'e \& R.~Bezanson; \citealt{bezanson2022}) and the ``Medium Bands, Mega Science" follow-up program (PID~4111; PI: K.~Suess; \citealt{suess2024}) provide deep NIRCam imaging over the Abell 2744 galaxy cluster, complementing earlier HST observations from the Frontier Fields program \citep{lotz2017}. We utilize the publicly available UNCOVER DR3 \citep{suess2024} that additionally incorporates imaging from various NIRCam surveys including ALT \citep[][]{naidu2024}, BEACONS \citep{morishita2024}, MAGNIF (\#2833, PI: Sun), \#2756 (PI: Chen), and \#3538 (PI: Iani). The DR3 release delivers imaging in all 12 medium bands 
and all 8 broadband filters, 
covering a continuous wavelength range from 0.7 to 5~$\mu$m.
The depths span 29.2--30.2~mag (5$\sigma$, point source) for broadbands with $0\farcs08$-radius and $0\farcs16$-radius apertures for SW and LW filters, respectively. The inclusion of medium-band filters allows precise sampling of key spectral features, such as the Balmer break, strong nebular emission lines (e.g., \oiii\ and H$\alpha$), and the rest-frame optical continuum at our redshifts of interest. This dataset crucially benefits from lensing magnification (mean $\mu\approx2.5$; \citealt{furtak2023a, price2024}) provided by the Abell~2744 cluster. 
The photometric catalog, reduced mosaics, and associated materials are publicly available from the UNCOVER and MegaScience project website\footnote{\url{https://jwst-uncover.github.io/megascience}}. 

\begin{figure*}[t!]
\begin{center}
\vspace{0.2cm}
\includegraphics[trim=0cm 0cm 0cm -0.4cm, clip, angle=0,width=1\textwidth]{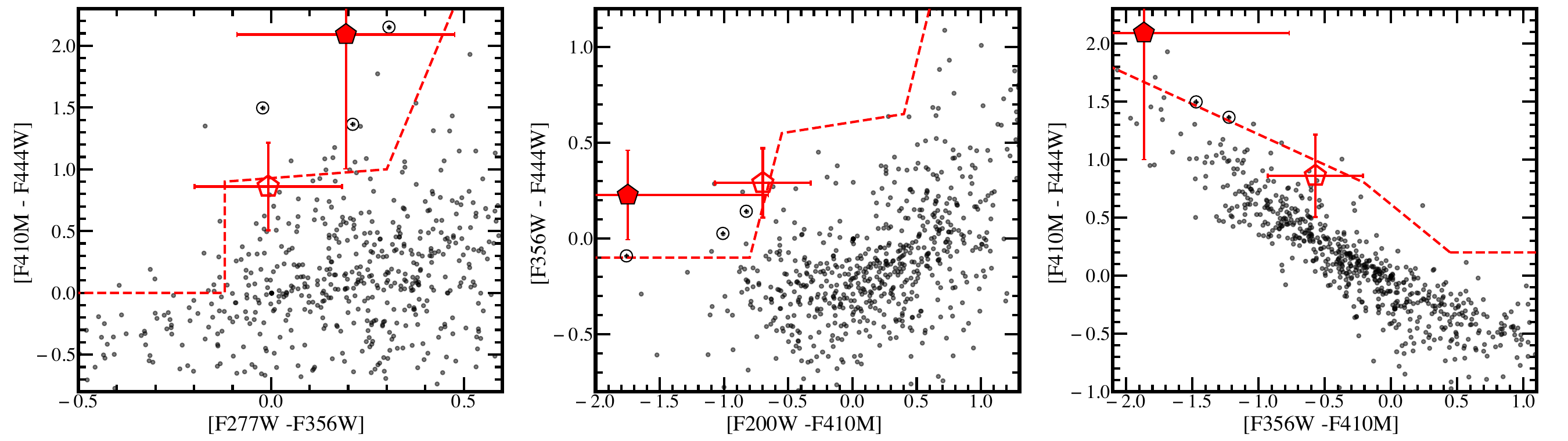}
\end{center}
 \caption{
\textbf{
Same NIRcam color-color diagrams as Figure~\ref{fig:color}, but with actual sources}. 
The red-filled pentagon represents a promising Pop~III candidate, \targ\ at $z \simeq 6.5$ found in GLIMPSE, which meets all three color-color diagrams and the SED criteria.
The open red pentagon denotes a tentative candidate, \targb\ at $z\simeq6.2$, found in JOF (see Appendix~\ref{sec:appendix_jof}). 
The grey dots denote 450 NIRCam sources at $z_{\rm phot}=5.0$--7.5 in GLIMPSE. 
The open black circles present those that meet the three color-color diagrams, while they do not satisfy the SED criteria and thus do not pass our selection. 
The $\sim$1\% contamination rate observed in the color-based method alone (see top right panel of Figure~\ref{fig:comp_contami}), suggests that a few potential contaminations may exist among the 450 NIRCam sources at $z_{\rm phot}=$5.0--7.5. The number of these color-satisfying but non-SED-consistent sources agree with the prediction from the contamination rate estimate, highlighting the importance of combining the color and SED-based methods.  
\label{fig:color_data}}
\end{figure*}

\subsection{CEERS, PRIMER, and JOF}
\label{sec:other_fields}

For the CEERS, PRIMER, and JOF regions, we utilize the photometric catalogs constructed in \citet{weibel2024}.
The reduced images for these datasets are based on the DAWN \jwst\ Archive's v7 imaging products\footnote{\url{https://dawn-cph.github.io/dja/imaging/v7/}}, which were processed using a custom \jwst\ pipeline with standard calibrations, cosmic ray removal, and background subtraction, ensuring consistent quality across fields (see more details in \citealt{valentino2023}). 
The CEERS program \citep[PID~1345;][]{finkelstein2024} offers imaging across the Extended Groth Strip (EGS) field, covering $\sim$100 arcmin$^2$. Observations were conducted in seven NIRCam bands at $1.1\mu$m, including F410M, reaching depths of 28.6--29.2~mag (5$\sigma$, point source, $0\farcs2$-diameter; \citealt{finkelstein2024}). 
The PRIMER survey \citep[PID~1837;][]{donnan2024} is conducted in the COSMOS and UDS fields, with imaging over $\sim$300 arcmin$^2$. Eight NIRCam filters at $0.9\mu$m are used, including F410M, with median depths of 27.6--28.4~mag (5$\sigma$, point source, $0\farcs3$-diameter; \citealt{donnan2024}). 
JOF \citep[PID~1180;][]{robertson2023} targets the GOODS-S region, offering ultra-deep NIRCam imaging in 13 filters at $\geq0.9\mu$m. The JOF dataset achieves depths of up to 30.2--31.0 AB magnitudes (5$\sigma$, point source, $0\farcs2$-diameter; \citealt{robertson2023}). 

The photometry across these fields was systematically extracted. 
Source detection was performed using a weighted combination of F277W, F356W, and F444W images. 
Point spread functions (PSFs) were constructed empirically and matched across all bands, enabling precise photometric measurements. Total fluxes were measured within $0\farcs16$-radius apertures, with aperture corrections derived based on fluxes measured through Kron ellipses and the growth curves of the PSFs.  Photometric uncertainties were estimated by placing random apertures in blank sky regions, accounting for correlated noise.

\section{Results}
\label{sec:result}

\setlength{\tabcolsep}{30pt}
\begin{table}
\begin{center}
\caption{NIRCam photometry of GLIMPSE-16043}
\label{tab:photometry}
\begin{tabular}{lc}
\hline 
\hline
R.A. [deg]  & 342.2123718      \\ 
Dec. [deg]  & $-$44.528751     \\ \hline
F090W [nJy] & $3.17 \pm 0.60$  \\
F115W [nJy] & $4.44 \pm 0.52$  \\
F150W [nJy] & $4.74 \pm 0.61$  \\
F200W [nJy] & $3.54 \pm 0.62$  \\
F277W [nJy] & $3.29 \pm 0.71$  \\
F356W [nJy] & $3.94 \pm 0.74$  \\
F410M [nJy] & $0.71 \pm 1.21$  \\
F444W [nJy] & $4.85 \pm 0.67$  \\
F480M [nJy] & $12.54 \pm 2.47$ \\ 
\hline
\end{tabular}
\end{center}
\tablecomments{
The measurements above are aperture-corrected photometry derived from PSF-matched images in the observed frame (i.e., without correcting for magnification).
}
\end{table}

\begin{figure*}[t!]
\begin{center}
\includegraphics[trim=0cm 0.cm 0cm 0cm, clip, angle=0,width=1.\textwidth]{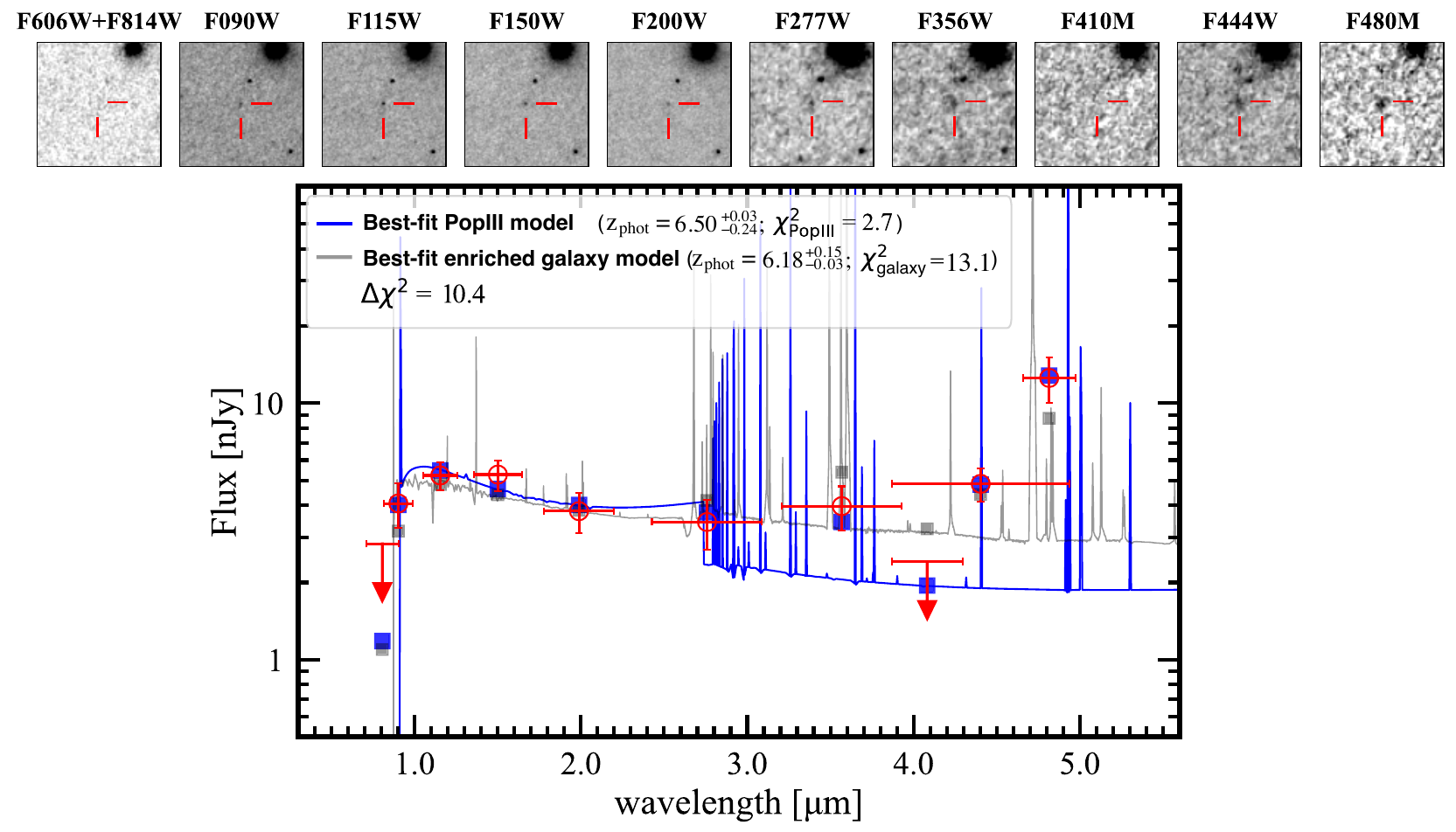}
\end{center}
 \caption{
\textbf{Best-fit SED of \targ\ derived with \texttt{EAZY} using the Pop~III templates (blue curve) and the enriched galaxy templates (grey curve)}, together with the NIRCam and HST $3''\times3''$ cutouts in the top panels. 
The open red circles denote the NIRCam and HST/F814W photometry, and the red arrows indicate 2$\sigma$ upper limits. 
The blue and grey squares present the expected photometry from the best-fit SEDs. 
The photometry favors the best-fit Pop~III templates. 
The complete absence in F410M strongly supports the presence of an extremely strong H$\alpha$ (EW$\approx3000$\AA) and a Balmer jump that are consistent with Pop~III SEDs but are challenging to reproduce along with the other emission line excesses in metal-enriched galaxy SEDs. 
\label{fig:sed_eazy}}
\end{figure*}

\setlength{\tabcolsep}{10pt}
\begin{table*}
\begin{center}
\caption{SED fitting results for \targ\ with different codes}
\vspace{-0.2cm}
\label{tab:sed}
\begin{tabular}{cccccc}
\hline 
\hline
Fitting Type &  Pop~III (\texttt{EAZY}) & Galaxy (\texttt{EAZY}) & Galaxy (\texttt{BEAGLE}) &  Galaxy (\texttt{Prospector})  &  Galaxy (\texttt{BAGPIPES})  \\ \hline
$z_{\rm phot}$ & $6.50^{+0.03}_{-0.24}$ &  6.18$^{+0.15}_{-0.03}$ &   $6.28^{+0.12}_{-0.09}$     & $6.41^{+0.08}_{-0.10}$      & $6.26^{+0.15}_{-0.11}$   \\ 
$\chi^{2}$     &   2.7     & 13.1  & 7.7  & 5.9  & 7.6   \\ 
\hline
\end{tabular}
\end{center}
\end{table*}

\subsection{Pop~III Candidates Found in the Deepest NIRCam Data}
\label{sec:candidate}

We systematically applied our fiducial selection method (Section~\ref{sec:fiducial}) to the NIRCam photometry catalogs from GLIMPSE, UNCOVER, CEERS, PRIMER, and JOF.  
We also performed careful visual inspections of the NIRCam images for all candidates to exclude potential artificial contaminants, such as residual diffraction spikes or extended flux from nearby bright objects.  
As a result, we identified one plausible candidate in GLIMPSE, \targ, which satisfies all selection criteria. 
In Tables~\ref{tab:photometry} and \ref{tab:sed}, we summarize the NIRCam photometry and the $\chi^{2}$ values from different SED fitting methods for \targ.
We also identified a tentative candidate in JOF, \targb, which meets the SED thresholds but falls very close to the borders of the criteria in the color-color diagrams, failing to satisfy all three.  
For completeness, we further investigate and discuss \targb\ in Appendix~\ref{sec:appendix_jof}. However, several suspicious aspects remain in this object, and thus \targb\ should be regarded as a tentative candidate. 

In Figure~\ref{fig:color_data}, we present the NIRCam colors of \targ\ (filled pentagon) and \targb\ (open pentagon) in the color-color diagrams. 
For reference, we also show other NIRCam sources with $z_{\rm phot}=5.0$--7.5 in GLIMPSE (grey dots) and three sources that meet the criteria of the color-based criteria but do not satisfy the SED-based criteria (black open circle). 
\targ\ resides within the color space of Pop~III galaxies and is well separated from the region occupied by metal-enriched galaxies. 
This clear separation is also confirmed in the other three color-color diagrams with F480M (see Appendix~\ref{sec:appendix_f480m}), ensuring its robustness as the candidate. 
\targb\ meets one of the color-color diagrams and falls slightly outside the thresholds in the other two diagrams, while still satisfying the thresholds within the 1$\sigma$ errors. 

Figure~\ref{fig:fiducial} shows the location of \targ\ in the completeness and contamination rate distributions derived from the combined color-based and SED-based selection functions. 
\targ\ is detected with F200W $=$ 30.0~mag (observed frame) and has a photometric redshift of \(z_{\rm phot,PopIII}=6.50^{+0.03}_{-0.24}\). At this parameter space, the completeness is estimated at 50\% and the contamination rate is 0\%, supporting the robustness of the selection. 
Since \targb\ does not meet all the selection criteria of our fiducial method, we rerun the completeness and contamination rate tests optimized to the criteria satisfied by \targb. We confirm that \targb\ also achieves high completeness ($>80\%$) and zero contamination rate at its specific parameter space, while a modest contamination rate ($\sim1\%$) remains in the nearby parameter space, making it a tentative Pop~III candidate. 

Figure~\ref{fig:sed_eazy} shows NIRCam cutouts of \targ. 
The faintness in F410M is visually confirmed, supporting the presence of Pop~III-like key SED features (Figure~\ref{fig:sed_comp}). 
\targ\ appears as a moderately magnified source with a lensing magnification of $\mu=2.9^{+0.1}_{-0.2}$ with radial $\times$ tangential magnifications of $1.28\times2.26$, based on an updated mass model incorporating the GLIMPSE data (L.~Furtak et al., in prep.). After lensing correction, the intrinsic absolute UV magnitude of \targ\ is $M_{\rm UV}=-15.89^{+0.12}_{-0.14}$, indicating an ultra-faint source. 
The successful identification of candidates in these faintest UV regimes aligns with the predicted optimal depths for detecting galaxies dominated by Pop~III stars (Section~\ref{sec:depth}).

Figure~\ref{fig:sed_eazy} also presents the best-fit SEDs for \targ\
and metal-enriched galaxy templates using \texttt{EAZY}, the methodology introduced in Section~\ref{sec:sed}. 
The photometric redshift solutions are securely constrained by the Lyman break in the NIRCam+HST filters at \(\sim0.9\)~$\mu$m and the enhanced flux observed in F444W+F480M, consistent with H$\alpha$ emission at the inferred Lyman-break redshift, while 
possible low-$z$ solutions are further discussed in Section~\ref{sec:lowz}. 
The HST+NIRCam SED is well-fitted by a combination of the young (1--50~Myr) Pop~III.2 models (Section~\ref{sec:sed}) with $\chi^{2}=2.7$ and no dust attenuation, yielding a $Delta\chi^{2}=10.4$ compared to the best-fit \texttt{EAZY} metal-enriched galaxy SED.
This large $\Delta\chi^{2}$ value confirms that \targ\ differs significantly from typical metal-enriched galaxies. 

\begin{figure}
\begin{center}
\includegraphics[trim=0cm 0cm 0cm 0cm, clip, angle=0,width=.45\textwidth]{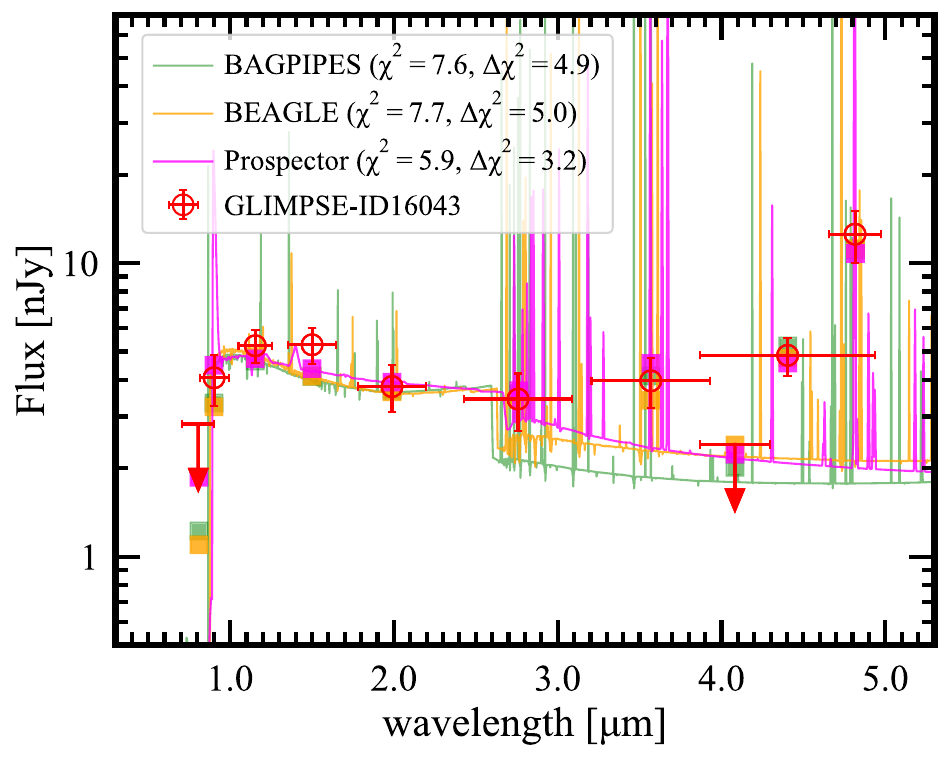}
\vspace{-0.2cm}
\end{center}
 \caption{
\textbf{Best-fit metal-enriched galaxy SEDs generated using \texttt{BAGPIPES} (green curve), \texttt{BEAGLE} (orange curve), and \texttt{Prospector} (magenta curve).}  
The red symbols are the same photometric points as those shown in Figure~\ref{fig:sed_eazy}.  
Despite the greater flexibility offered by these SED fitting codes compared to the Pop~III template fitting with \texttt{EAZY}, the $\Delta\chi^{2}$ values still exceed 3--5 in all cases. Besides, all models reach the lowest $Z_{\rm gas}$ values within their respective grids, \(Z_{\rm gas}/Z_{\odot} \simeq 0.01\), where zero metallicity remains consistent within the uncertainties.  
\label{fig:sed_all}}
\end{figure}

To investigate whether the HST+NIRCam SED is truly challenging for metal-enriched galaxy models beyond \texttt{EAZY}, we further conducted SED fitting using the more flexible SED fitting tools of \texttt{BEAGLE} \citep{chevallard2016}, \texttt{BAGPIPES} \citep{carnall2018}, and \texttt{Prospector} \citep{johnson2021}. 
We assumed parameter ranges typical for high-$z$ star-forming galaxies \citep[e.g.,][]{tang2023, reddy2023b} with the redshift range $z=$0--8. 
The detailed setup of the \texttt{BEAGLE} fits are as described in \citet{endsley2024}, \texttt{Prospector} fits are as described in \citet{naidu2024}, and \texttt{BAGPIPES} fits as described in Chisholm et al., in prep. 
Figure~\ref{fig:sed_all} summarizes the best-fit SEDs obtained from these tools. 
While the $\Delta\chi^{2}$ values decrease compared to \texttt{EAZY}, they generally remain $>3$--5 in all cases, reinforcing a preference for the Pop~III model. 
It is worth mentioning that all best-fit SEDs reach the lowest gas-phase metallicity values in their grids (\(Z_{\rm gas}/Z_{\odot} \simeq 0.01\)), suggesting that even if our candidates are not true Pop~III galaxies, they likely represent extremely metal-poor galaxies. We further discuss alternative scenarios, including the extremely metal-poor galaxy case, in Section~\ref{sec:caveats}.

\subsection{Physical properties}
\label{sec:properties}

We summarize the physical properties of \targ\ in Table~\ref{tab:int_prop}. 
Below, we briefly describe how these properties were derived.

\subsubsection{EW(H$\alpha$), stellar age, and UV continuum slope}
\label{sec:age}

The significant faintness in F410M observed in \targ\ strongly supports the presence of a Balmer jump and strong H$\alpha$ emission. 
By determining the underlying continuum with the best-fit Pop~III SEDs and utilizing the flux excess in F480M and F444W, we infer the rest-frame H$\alpha$ equivalent width (EW) of $2810\pm550{\rm \AA}$. 
These EW(H$\alpha$) values are comparable or even higher than the maximal EW(H$\alpha$) values covered by the \zackmodel\ models \citep{zackrisson2011}, indicating that the majority of the emission is dominated by very young stellar populations ($<5$~Myrs; see e.g., Fig.9 in \citealt{trussler2023}) and/or the presence of very massive stars (VMS; $>100\,M_{\odot}$) in these objects \citep[][]{schaerer2024}. 
From the best-fit Pop~III SEDs, we obtain the mean $t_{\rm age}$\footnote{luminosity-weighted logarithmic mean} of 2.8~Myr, which is consistent with the very young stellar ages implied by the high EW(H$\alpha$) above. 
Using the rest-frame UV filters F115W, F150W, and F200W, we also estimate a UV continuum slope of $\beta = -2.34 \pm 0.36$. 
This value is consistent with the nebular-dominated UV continuum shape predicted for Pop~III SEDs in young stellar ages ($\sim3$\,Myr) with no dust attenuation (see e.g., Fig.~7 in \citealt{trussler2023}), well aligning with the above $t_{\rm age}$ estimate and the observed high EW(H$\alpha$). 

\subsubsection{Stellar mass}
\label{sec:mass}

In the \zackmodel\ models, the stellar component is uniformly set to $M_{\rm star}=10^{6}\,M_{\odot}$. By scaling each template to the best-fit Pop~III SED, we obtain $M_{\rm star}$ values of $3.3\times10^{5}\,M_{\odot}$ after the lens correction. 
We note, however, that the SED shapes of our Pop~III candidates are dominated by the nebular continuum. This indicates that the $M_{\rm star}$ estimates for our candidates strongly depend on the ionizing properties of galaxies, which are regulated by the IMF shape, $t_{\rm age}$, and the nebular conditions such as electron density and ionization parameter. 
For example, if we simply assume the Pop~III.1 IMF and $t_{\rm age}=1$~Myr, given the high EW(H$\alpha$) of $\simeq$~2800${\rm \AA}$, the $M_{\rm star}$ estimates decrease to $0.7\times10^{5}\,M_{\odot}$.  
While a detailed investigation of the ionizing properties including nebular conditions in Pop~III galaxies is beyond the scope of this work, 
these results indicate that a simple order-of-magnitude estimate for $M_{\rm star}$ would be $\approx10^{5}\,M_{\odot}$ for \targ, which is in excellent agreement with the baseline $M_{\rm star}$ expected for newly forming Pop~III galaxies at $z\simeq$~6--7 (Section~\ref{sec:depth}).

\setlength{\tabcolsep}{30pt}
\begin{table}
\begin{center}
\caption{Physical properties of \targ}
\label{tab:int_prop}
\begin{tabular}{lc}
\hline 
\hline
Property              &    Measurement  \\ \hline
$z_{\rm phot}$  &  $6.50^{+0.03}_{-0.24}$   \\
$\mu$           &  $2.9^{+0.1}_{-0.2}$      \\
$\beta$         &  $-2.34\pm0.36$           \\
$M_{\rm UV}$  [mag]    & $-15.89^{+0.12}_{-0.14}$ \\
$M_{\rm star}$ [$M_{\odot}$] &  $\approx10^{5}$ \\
$r_{\rm e}$ [pc]            &   $<40$ \\
$t_{\rm age}$  [Myr]        &   2.8    \\
H$\alpha$ EW [${\rm \AA}$]    &   2810 $\pm$ 550 \\
OIII/H$\beta^{\dagger}$   &  $<0.44$    \\
12+$\log$(O/H)$^{\dagger}$  &  $<6.4$   \\
\hline
\end{tabular}
\end{center}
\tablecomments{
Physical parameters are derived using the best-fit Pop III templates (see Section~\ref{sec:properties} for details). \\
$\dagger$ The 1$\sigma$ upper limit is presented, same as \cite{vanzella2023}.  
}
\end{table}

\subsubsection{Size}
\label{sec:size}

\begin{figure}[t!]
\begin{center}
\includegraphics[trim=0cm 0cm 0cm 0cm, clip, angle=0,width=.5\textwidth]{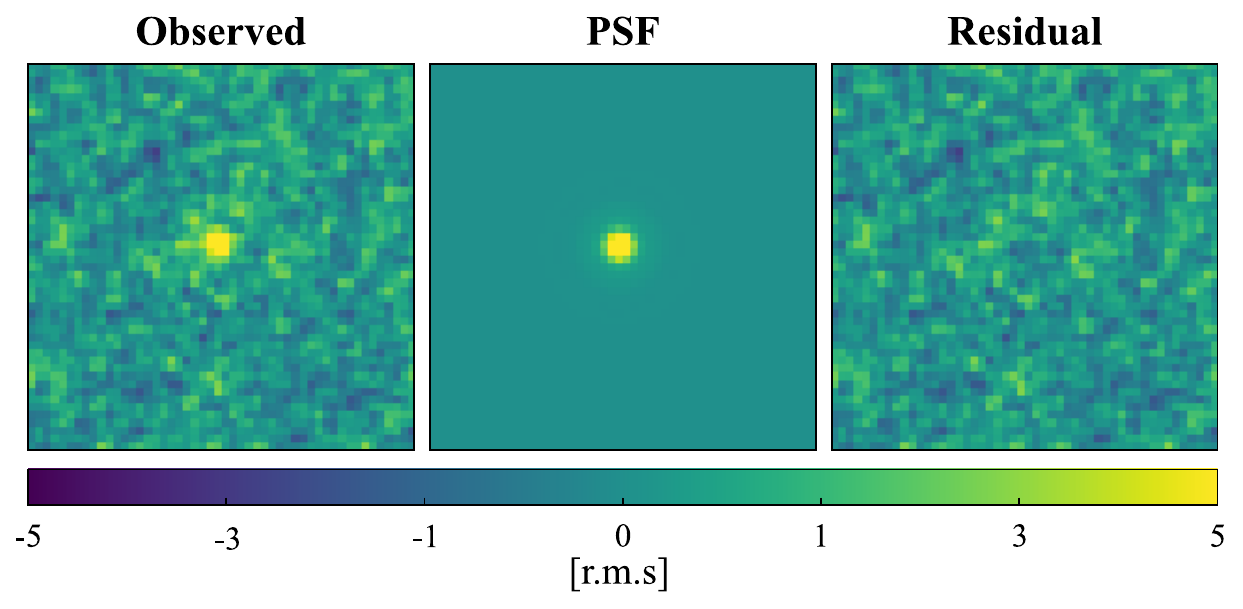}
\end{center}
\vspace{-0.4cm}
 \caption{\textbf{The point-source morphology of the Pop III candidate is consistent with arising from e.g., extremely compact star-clusters.} NIRCam F150W $1''\times1''$ cutout for the observed, PSF, and PSF-subtracted residual image of \targ. 
 The compactness is consistent with the PSF, and we obtain an upper limit of 30~pc for the effective radius after the lens correction.  
\label{fig:galfit}}
\end{figure}

\targ\ exhibits extremely compact morphology.  
Using empirical PSFs in GLIMPSE generated in the same manner as \citet{weibel2024}, 
we perform S\'ersic profile fitting with \texttt{GALFIT} \citep{peng2010} on the high-resolution F150W image.  
Notably, the fitting does not converge, reaching the smallest size in the fitting grid, indicating consistency with a point source. 
Figure~\ref{fig:galfit} presents the NIRCam F150W $1''\times1''$ map around \targ, the PSF in F150W, and the residual map after PSF subtraction. The morphology closely matches the PSF, supporting its compact nature. 
Typically, intrinsic sizes can be reliably constrained down to $\sim$0.5 times the pixel scale \citep[see Appendix~C in][]{messa2022}. Additionally, \cite{ono2023} conducted Monte Carlo simulations and found that the output/input size ratio for the smallest sources (effective radius $r_{\rm e}=$1~pix) with luminosities near the detection limit is approximately $\sim2$ in NIRCam/SW filters \citep[see Fig.~4 \& 5 in][]{ono2023}. By conservatively adopting the possible output/input size ratio of 3, we place an upper limit on the effective radius of \targ\ at $0.5\times$ pixel scale ($=0\farcs02$) $\times3$ $\simeq$40~pc after applying the lensing correction\footnote{
In this system, tangential magnification dominates and is used for lens correction rather than a circularized value.
}. 
Compact star-forming clumps with $r_{\rm e} < 10$--100~pc have been identified in recent \jwst\ studies of strongly lensed galaxies at \(z \gtrsim 6\) within similarly UV-faint and low-mass regimes \citep[e.g.,][]{vanzella2023, adamo2024, mowla2024, fujimoto2024}.  
The comparably small sizes of \targ\ may suggest it is a UV-faint, low-mass compact star-forming region or galaxy potentially dominated by Pop~III stars.  

\subsubsection{Gas-phase Metallicity}
\label{sec:metal}

\begin{figure*}[t!]
\begin{center}
\includegraphics[trim=0cm 0cm 0cm 0cm, clip, angle=0,width=1.\textwidth]{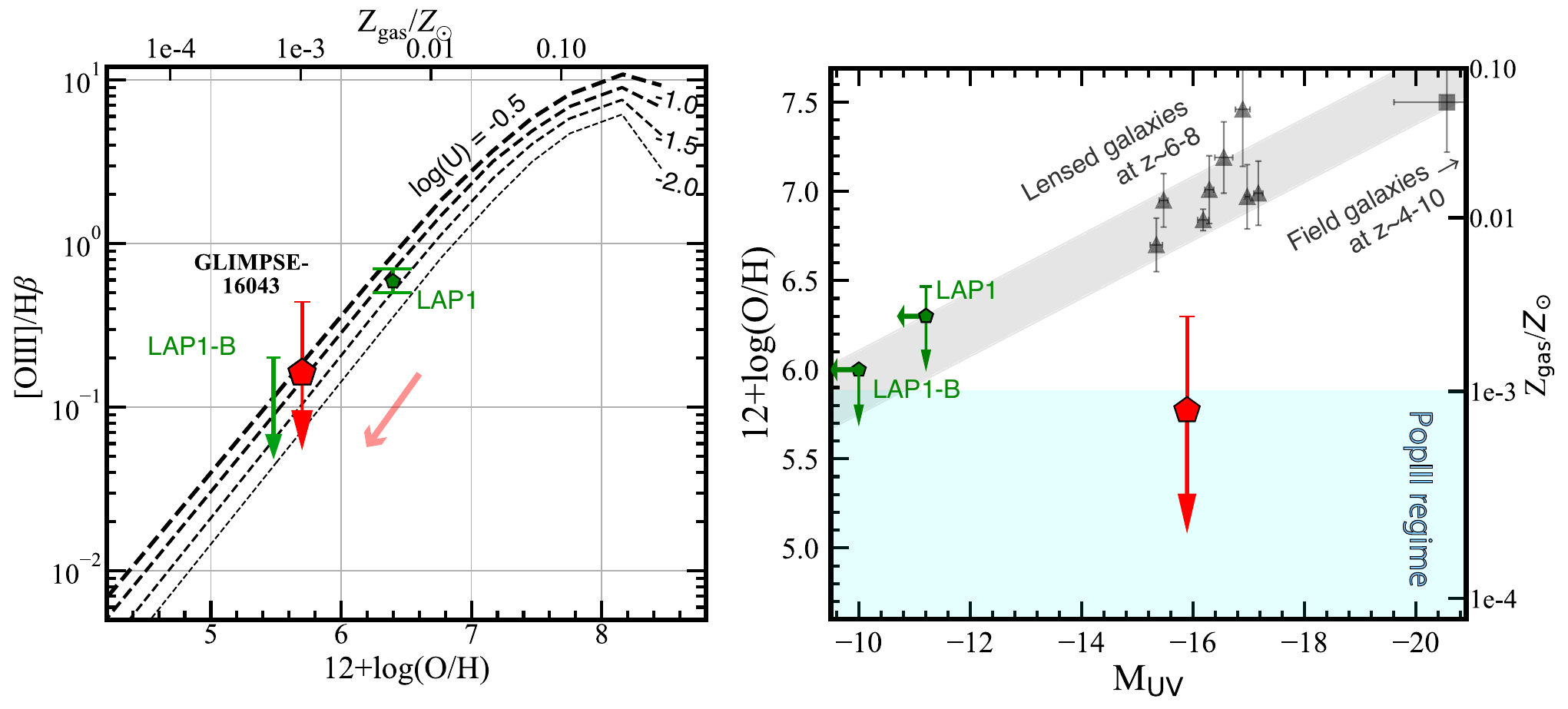}
\end{center}
 \caption{
\textbf{Constraints on the gas-phase metallicity \boldmath $Z_{\rm gas}$}.
\textbf{\textit{Left}:} 
The red pentagon and the upper error bar represent the best-fit estimate and the 1$\sigma$ upper limit on the \oiii/H$\beta$ line ratio, derived from the flux excesses in F480M (for H$\alpha$) and F356W (for \oiii+H$\beta$). In this calculation, the underlying continuum is evaluated using the best-fit Pop~III template, and H$\alpha$/H$\beta$ is fixed at 2.74 (see Section~\ref{sec:metal}). 
The dashed curves indicate photoionization model predictions extrapolated towards the low metallicity regime from \cite{vanzella2023b}, covering an ionization parameter range of [$-0.5$:$-2.0$]. 
The red diagonal arrow shows how the upper limit of 12+$\log$(O/H) shifts according to that of \oiii/H$\beta$.
The green symbols present the very \oiii-weak H$\alpha$ emitters at $z=6.6$, LAP1 and LAP1-B \citep{vanzella2023}
found in a strongly lensed arc with no continuum detection, suggesting their possible origins in nebular emission arising from metal-free pockets within galaxies. 
\textbf{\textit{Right}:} 
Luminosity-Metallicity relation. 
Given the very high EW(H$\alpha$) of $\approx3000{\rm \AA}$, we assume $\log(U)\geq-1.5$ for \targ\ to derive the metallicity from the photoionization model shown in the left panel. The grey symbols denote the recent \jwst\ measurements for lensed \citep{chemerynska2024} and field galaxies \citep{nakajima2023} at $z\simeq$4--10, and the grey shade indicates the 1$\sigma$ region of its best-fit linear relation. 
LAP1 and LAP1-B fall on the best-fit relation. 
In contrast, \targ\ shows significant UV enhancement relative to the best-fit relation, which indicates a unique origin, e.g., due to a top-heavy IMF in a Pop~III galaxy or an extremely metal-poor, low-mass AGN. Under the condition that the dust cooling is minimal, $[{\rm O}/{\rm H}]\simeq-3$ is a critical value for the Pop~III to Pop~II transition via metal cooling \citep[e.g.,][]{bromm2003}, indicating that $[Z_{\rm gas}/Z_{\odot}]< -3$ is the Pop~III regime (cyan shade). 
\label{fig:uv-metal}}
\end{figure*}

In Figure~\ref{fig:uv-metal}, we present constraints on the \oiii/H$\beta$ ratio and the resulting gas-phase metallicity (\(Z_{\rm gas}\)) of \targ, derived from H$\alpha$ and \oiii+H$\beta$ fluxes inferred through flux excesses in the F480M (or F444W) and F356W filters, respectively. 
The best-fit Pop~III SED model is used to estimate the underlying continuum and calculate the flux excesses. 
Although the F480M (or F444W) filter captures both H$\alpha$ and \nii\ lines, the \nii/H$\alpha$ ratio is expected to be \(\lesssim1\%\) in the low-\(Z_{\rm gas}\) regime \citep[e.g.,][]{izotov1997}, making the contribution of \nii\ to the flux in F480M or F444W negligible, if any. 
The excellent agreement of \targ\ with the Pop~III templates in the rest-frame UV suggests nearly zero dust obscuration. 
Consequently, we estimate the H$\beta$ line strength by assuming an intrinsic H$\alpha$/H$\beta$ ratio of 2.74, with Case~B recombination under conditions of an electron temperature of \(2 \times 10^4\)~K and an electron density of \(10^4\)~cm\(^{-3}\) \citep{osterbrock1989}.  
From this analysis, we find the \oiii/H$\beta$ ratio to be consistent with zero, with the 1$\sigma$ upper limit of 0.44, derived from the 1$\sigma$ propagated uncertainties, which are primarily limited by the current NIRCam depth.

The left panel of Figure~\ref{fig:uv-metal} extrapolates photoionization modeling to the lowest \(Z_{\rm gas}\) regime along with a $\log(U)$ grid. Given the observed large EW(H$\alpha$) values in our candidates, we adopt $\log(U) \geq -1.5$ as the most plausible scenario and estimate upper limits of 12+$\log(\mathrm{O/H})$ at $\leq$6.4 ($\leq0.005 Z_{\odot}$), based on the \oiii/H$\beta$ upper limits derived above.  
These constraints represent some of the lowest \(Z_{\rm gas}\) values observed to date at $z > 6$, comparable to LAP1 and LAP1-B, the most metal-poor clumps identified thus far, found within a strongly lensed arc at \(z=6.6\) \citep{vanzella2023b}.  
A critical distinction is that LAP1 and LAP1-B lack continuum detections. 
This suggests that they may represent extremely metal-poor and low-mass ($\lesssim10^{3}$\,$M_{\odot}$) star clusters inside the metal-enriched galaxy or line-emission from pristine nebular gas clouds within the galaxy, rather than star clusters. 
In contrast, our candidates potentially represent the lowest metallicity observed as a single, isolated galaxy at $z>6$.

The right panel of Figure~\ref{fig:uv-metal} compares the Luminosity--Metallicity relation of our candidates with recent \jwst\ measurements of $z\simeq4$--10 galaxies \citep[e.g.,][]{nakajima2023, chemerynska2024} and a linear fit to these observations (black shaded region).  
While LAP1 and LAP1-B fall along the best-fit relation, \targ\ and \targb\ exhibit significant deviations, appearing more than 100 times brighter in the rest-UV than LAP1 and LAP1-B at comparable metallicities.  
This deviation may indicate mechanisms that enhance UV luminosity at a given stellar mass, such as a top-heavy IMF, which is consistent with theoretical predictions for Pop~III galaxies \citep[e.g.,][]{tumlinson2006,fardal2007,dave2008}.  
Alternatively, the deviation could be attributed to significantly lower \(Z_{\rm gas}\) at a given $M_{\rm UV}$, potentially resulting from a recent pristine gas inflow that dilutes the metal abundance and decreases the 12+$\log(\mathrm{O/H})$ value.  
We further explore possible mechanisms behind this deviation from the typical Luminosity--Metallicity relation in Section~\ref{sec:context} and Section~\ref{sec:caveats}.

Although the definition of Pop~III stars is strictly zero metallicity, it is useful to consider a critical metallicity threshold ($Z_{\rm crit}$), above which ``normal'' (low-mass dominated) star formation can occur (i.e., the transition from Pop~III to Pop~II).
This helps us understand how close our candidates may be to this boundary. 
Theoretically, metal-line cooling can set $Z_{\rm crit}$ at about [O/H]~$\sim -3$ \citep{bromm2003},
whereas dust cooling could drive the threshold even lower, to [O/H]~$\sim -5$ or $-6$ \citep{schneider2006}.
However, pair-instability supernovae (PISNe) from very massive primordial stars can produce extremely energetic explosions,
often resulting in significant destruction of newly formed dust grains \citep[e.g.,][]{nozawa2010}.
Furthermore, once the first supernovae occur, metal enrichment within the star-forming regions proceeds rapidly,
so that [O/H]~$\approx -3$ can be reached on short timescales ($\lesssim 10$~Myr; e.g., \citealt{karlsson2008,maio2010}).
Given uncertainties in dust survival and the rapid pace of chemical enrichment,
we adopt [O/H]~$=-3$ as a practical reference for $Z_{\rm crit}$.
In Figure~\ref{fig:uv-metal}, the region below $Z_{\rm crit}$ is shaded in cyan to indicate the Pop~III regime.
Our candidate, with [O/H]~$<-2.3$ 
falls within a factor of a few from $Z_{\rm crit}$. 
Future deep observations may push their upper limits further into the Pop~III regime.

\subsection{Pop~III UVLF at $z \simeq$~6--7}
\label{sec:uvlf}

\begin{figure*}[t!]
\begin{center}
\includegraphics[trim=0cm 0cm 0cm 0cm, clip, angle=0,width=.75\textwidth]{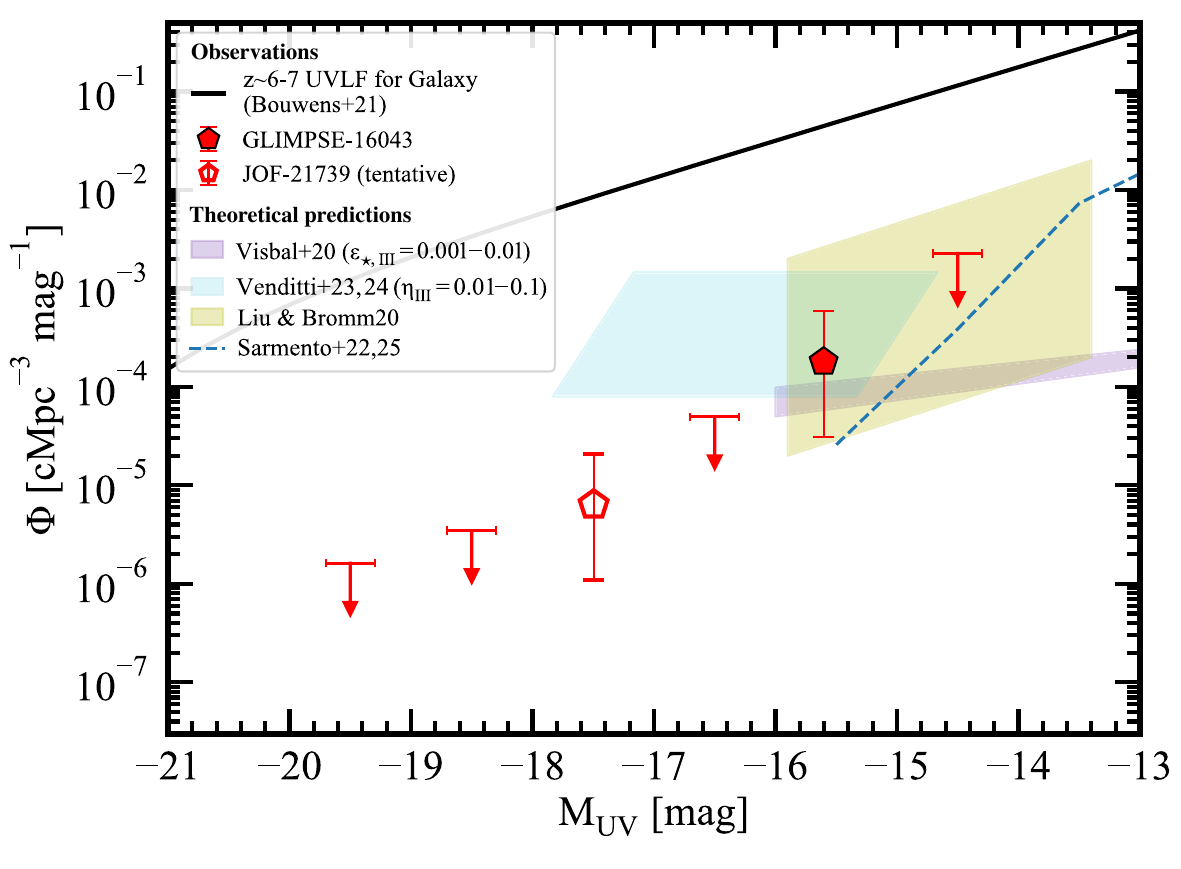}
\end{center}
\vspace{-0.2cm}
 \caption{
\textbf{Pop~III UV Luminosity Function (UVLF) at $z \simeq$~6--7.} 
The red symbols represent the volume density estimates of our Pop~III candidates identified through a systematic search across a total area of $\simeq$~500~arcmin$^{2}$ from the \jwst\ legacy surveys of GLIMPSE, UNCOVER, CEERS, PRIMER, and JOF. 
Filled and open pentagons correspond to \targ\ and \targb, respectively, with error bars reflecting Poisson uncertainties. 
Upper limits are calculated based on Poisson uncertainties at the single-sided confidence level of 84.13\% \citep{gehrels1986}.  
The black curve shows the UVLF of general $z \simeq 6$--7 galaxies for comparison \citep{bouwens2021}, while the colored curves and shades represent theoretical predictions (see details in Appendix~\ref{sec:appendix_model}). 
\targ\ falls well within the range of theoretical predictions in the ultra-faint regime ($M_{\rm UV} \simeq -16$), whereas \targb, if real, may require additional physical mechanisms, such as rapidly rotating stars \citep{liu2024}, to explain its boosted $M_{\rm UV}$ for some models. 
\label{fig:uvlf}}
\end{figure*}

\setlength{\tabcolsep}{20pt}
\begin{deluxetable}{cc}
\tablecaption{Constraints on Pop~III UVLF at $z\sim$6--7}
\tablehead{\colhead{$M_{\rm UV}$} & \colhead{$\Phi$}  \\
   \colhead{[AB mag]}             & [$10^{-5}$ Mpc$^{-3}$~dex$^{-1}$] \\
              (1)                 &    (2)          }
\startdata
$-19.5$ & $<0.16$  \\
$-18.5$ & $<0.35$ \\
$-17.5$ & ($0.64_{-0.53}^{+1.46})\dagger$ \\
$-16.5$ & $<5.0$ \\
$-15.5$ & $18.0_{-14.0}^{+41.4}$ \\
$-14.5$ & $<226.5$ \\
\enddata
\tablecomments{
(1): Absolute UV magnitude.  
(2): Observational constraints obtained from GLIMPSE and other \jwst\ legacy deep field data of UNCOVER \citep{bezanson2022}, PRIMER-UDS/COSMOS \citep{donnan2024}, and CEERS \citep{finkelstein2024}, resulting in a total survey area of $\simeq$500~arcmin$^{2}$ (0.14~deg$^{2}$). 
Errors and upper limits are 1$\sigma$, evaluated with the Poisson uncertainty \citep{gehrels1986}. \\
$\dagger$ This is estimated from the tentative candidate of \targb. If we place an upper limit instead, the 1$\sigma$ upper limit is estimated to be 1.2$\times10^{-5}\,$Mpc$^{-3}$~dex$^{-1}$. 
}
\label{tab:uvlf}
\end{deluxetable}

The presence of late Pop~III star formation and Pop~III galaxies down to $z \simeq 6$ largely depends on several factors, including feedback mechanisms from early galaxies, the efficiency of subsequent chemical mixing, environmental conditions in star-forming regions, and the abundance of the metal-free low-mass halos remaining at $z\simeq6$. 
Conversely, the volume density of Pop~III galaxies at $z\simeq6$ provides critical constraints for simulations of Pop~III formation and evolution in the early universe \citep[e.g.,][]{pallottini2014b, sarmento2018, visbal2020, liu2020, vikaeus2022, venditti2023, venditti2024b}.  
Using the systematic search for Pop~III galaxies with NIRCam and the well-defined selection functions developed in this study (Section~\ref{sec:method}), we provide the first observational constraint on the Pop~III UV luminosity function (UVLF).  
Although spectroscopic confirmation is essential to definitively determine whether these candidates are true Pop~III galaxies, we derive the Pop~III UVLF at \(z \simeq 6\)--7 using our photometric candidates, following the standard approach commonly employed in UVLF studies of high-redshift galaxies.  
The dedicated survey volume, including completeness corrections, also allows us to set meaningful upper limits even if future spectroscopy reveals these candidates to be enriched sources.

First, we evaluate the survey volume as follows.  
Based on the completeness function shown in Figure~\ref{fig:comp_contami}, we focus on the parameter space satisfying the completeness of $\gtrsim50\%$, defining our effective search range as $z = 5.6$--6.6 and observed magnitudes in F200W of $\leq 30.3$~mag.  
Note that the completeness in Figure~\ref{fig:comp_contami} is estimated using GLIMPSE data.  
When considering completeness in other survey fields, we scale the distribution according to the relative data depth in each survey compared to GLIMPSE, while the completeness distribution remains largely consistent across redshifts due to the similar NIRCam filter sets.  
For GLIMPSE, we calculate the effective survey area using the NIRCam field-of-view of 9.4~arcmin\(^2\) and the updated mass model incorporating the GLIMPSE data (L.~Furtak et al., in prep., I. Chermynska et al., in prep.).  
For UNCOVER, the total NIRCam mosaic area is 45~arcmin$^{2}$ \citep{bezanson2022}, and we also calculate the effective survey area using the publicly available v2.0 mass model\footnote{\url{https://jwst-uncover.github.io/DR4.html\#LensingMaps}} \citep{furtak2023a, price2024, suess2024, weaver2023}.  
For CEERS and JOF, we adopt areas of 88.1~arcmin$^{2}$ \citep{finkelstein2024} and 9.4~arcmin$^{2}$ \citep{robertson2023}, respectively.  
For PRIMER, we follow the areas defined by the depths in \cite{donnan2024} and adopt values of 47.8~arcmin$^{2}$, 84.4~arcmin$^{2}$, and 170.7~arcmin$^{2}$ for PRIMER-COSMOS Deep, PRIMER-COSMOS Wide, and PRIMER-UDS, respectively.  
We then bin the effective survey areas according to the absolute UV magnitude ($M_{\rm UV}$) bins, resulting in a total survey area of approximately 500~arcmin$^{2}$ in the brightest bin.  
We convert the survey area into a co-moving survey volume for the \(z = 5.6\)--6.6 slice and apply completeness corrections to our individual detections and upper limits for the final volume density estimates.  
Uncertainties and upper limits are calculated using Poisson errors following the values presented in \cite{gehrels1986}.  
Note that the observed magnitude and intrinsic $M_{\rm UV}$ are not directly correlated due to varying magnifications in lensing fields, and we may calculate the completeness for each magnification at a given observed magnitude. However, our completeness simulations assume point sources (i.e., unaffected by lensing distortions), making the completeness depend solely on the magnitude in the observed frame. We thus adopt an average completeness of 75\% based on the effective survey parameter space defined above in the completeness correction for the upper limit estimates in lensing fields, while we use the observed magnitude and redshift estimate and infer the completeness for our individual candidates.

In Figure~\ref{fig:uvlf}, we show the inferred Pop~III UVLF at $z \simeq 6$--7, derived from our robust candidate \targ\ (red filled pentagon). For completeness, we also plot the tentative candidate \targb\ (red open pentagon). 
Colored curves and shaded regions represent simulation predictions from the literature \citep{visbal2020, liu2020, venditti2023, venditti2024b, sarmento2022}, updated here to match the UVLF format (see Appendix~\ref{sec:appendix_model}).  Our Pop~III UVLF estimates, including the upper limits, are listed in Table~\ref{tab:uvlf}, and the theoretical framework for each Pop~III UVLF simulation prediction is summarized in Appendix~\ref{sec:appendix_model}.

We find that \targ\ aligns remarkably well with several simulation models, despite considerable differences in the assumptions for both galaxy formation and evolution and their implementations of the UV emission (see details in Appendix~\ref{sec:appendix_model}). 
This lends additional support to the scenario that \targ\ is indeed a Pop~III galaxy, independently validated from the perspective of volume density.
On the other hand, most of these simulations exhibit a sharp cutoff beyond $M_{\rm UV} \simeq -16$, corresponding to a mass cutoff around $M_{\rm star} \approx 10^{5}\,M_{\odot}$ due to the limited survey volumes in the simulations and the specific conditions required for Pop~III star formation at $z\sim6$--7 (Section~\ref{sec:depth}). 
Note that this does not immediately rule out the presence of \targb, given the significant uncertainties in the theoretical upper limit of the Pop~III galaxy mass \citep[e.g.,][]{yajima2017b} and in the mass-to-light ratio of Pop~III galaxies. If physical mechanisms enhance UV luminosity at a given stellar mass, the current shapes of the simulated Pop~III UVLFs may extend further into the bright end. Possible mechanisms include rapidly rotating stars undergoing chemically homogeneous evolution, which could increase the brightness by up to $\sim$2 magnitudes \citep{liu2024}, thus making the current Pop~III UVLF predictions consistent also with \targb.

\subsection{Pop~III SFRD at $z \simeq$~6--7}
\label{sec:sfrd}

\begin{figure*}[t!]
\begin{center}
\includegraphics[trim=0cm 0cm 0cm 0cm, clip, angle=0,width=.75\textwidth]{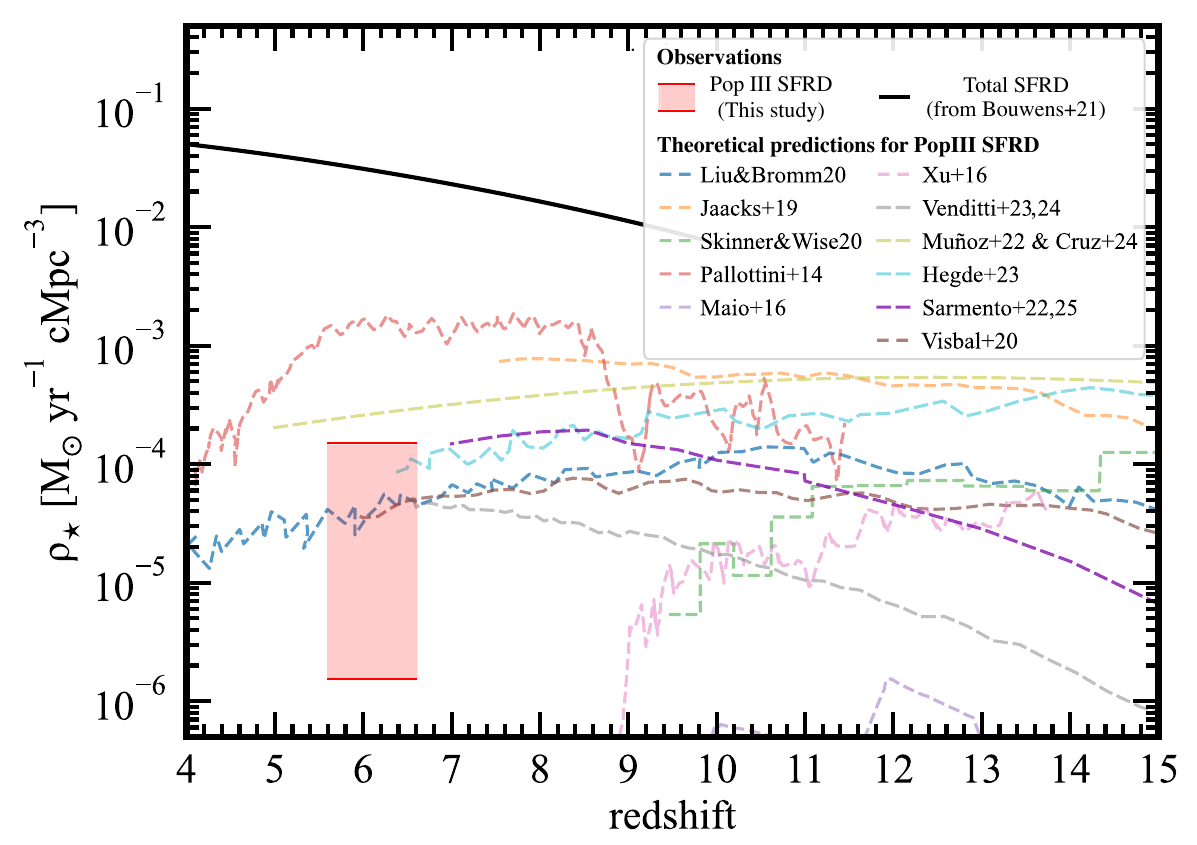}
\end{center}
\vspace{-0.3cm}
 \caption{
\textbf{Cosmic Pop~III star-formation rate density (SFRD) at $z\simeq$~6--7.} 
The red shading and line indicate our conservative lower and upper limit estimates, respectively, derived from our Pop~III UVLF results (see text). 
Colored curves represent theoretical predictions for the Pop~III SFRD from the literature \citep{pallottini2014, maio2016, xu2016, jaacks2019, sarmento2022, skinner2020, visbal2020, liu2020, munoz2022, hegde2023, venditti2023}. 
The black curve shows the total SFRD inferred from the best-fit redshift evolution of the general galaxy UVLFs in \citet{bouwens2021}, integrated down to $M_{\rm UV}=-12$ to match a typical mass limit ($\approx10^{4}\,M_{\odot}$) in the simulations above. 
Our observational constraints fall well within the theoretical range and suggest that the Pop~III galaxies may contribute $\sim0.01$--1\% of the total cosmic SFRD at $z\simeq6$--7.
\label{fig:sfrd}}
\end{figure*}

\setlength{\tabcolsep}{15pt}
\begin{deluxetable}{cc}
\tablecaption{Constraints on Pop~III SFRD at $z$$\sim$6--7}
\tablehead{\colhead{$z$} & \colhead{$\log(\rho_{\star, {\rm PopIII}})$ [$M_{\odot}$~yr$^{-1}$ Mpc$^{-3}$]}  \\
              (1)                 &    (2)          }
\startdata
5.6 -- 6.6 & $[-5.81: -3.82]$  \\
\enddata
\tablecomments{
(1): Redshift range (see Section~\ref{sec:uvlf}).  
(2): Our lower and upper limit estimates by accounting for the single galaxy contribution of \targ\ and integrating the $z=6$ UVLF for general galaxies \citep{bouwens2021} rescaled down to match the data point of \targ\ (see Section~\ref{sec:sfrd}).}
\label{tab:sfrd}
\end{deluxetable}

We also explore possible constraints on the cosmic Pop~III star-formation rate density (SFRD), $\rho_{\star, {\rm PopIII}}$, based on our UVLF results. 
First, for the lower limit, we focus on the single data point of \targ\ to obtain a conservative estimate. We use its lower-limit volume density to derive the corresponding UV luminosity density. 
\cite{schaerer2024} examine the variation of the conversion factor between UV luminosity and SFR, $\kappa_{\rm UV}$\footnote{Defined as per ${\rm SFR(UV) [M_{\odot}~yr^{-1}]} = \kappa_{\rm UV} \times L_{\rm UV} [{\rm erg~s^{-1}~Hz^{-1}}].$
}, 
for different IMFs and stellar populations (including Pop~III) as a function of $t_{\rm age}$ (see Fig.~7 in \citealt{schaerer2024}). 
Based on the best-fit mean $t_{\rm age}=2.8$~Myr and a moderately top-heavy IMF in \targ, we adopt $\kappa_{\rm UV}=0.5\times10^{-28}$ and obtain $\rho_{\star, {\rm PopIII}}=1.5\times10^{-6}$~[$M_{\odot}$~yr$^{-1}$~cMpc$^{-3}$]. 

Second, for the upper limit, we use the best-fit UVLF shape at $z=6$ derived in \cite{bouwens2021}, scaling it to match the upper limit of the UVLF data point of \targ. We then integrate the rescaled $z=6$ UVLF down to $M_{\rm UV}=-12$, which reflects the possible $M_{\rm UV}$ range of Pop~III galaxies down to $\approx10^{4}\,M_{\odot}$. Using the same $\kappa_{\rm UV}$ value, we derive $\rho_{\star, {\rm PopIII}} = 1.5\times10^{-4}$~[$M_{\odot}$~yr$^{-1}$~cMpc$^{-3}$]. 
As discussed in Section~\ref{sec:uvlf}, there is likely a cutoff in the bright end of the Pop~III UVLF based on simulation predictions. Nevertheless, using the same UVLF shape as the general galaxy population provides a conservative upper-limit estimate. We summarize our possible Pop~III SFRD constraints in Table~\ref{tab:sfrd}. 
Note that we focus on \targ\ for our Pop~III SFRD estimates above, because of the less robust nature of \targb. However, the contribution of \targb\ still falls well within our lower and upper limits, given our conservative approach.

In Figure~\ref{fig:sfrd}, we show our possible constraints on the cosmic Pop~III SFRD (red line and shading). For reference, we also show the total SFRD (black curve) by integrating the best-fit redshift evolution of the general galaxy UVLFs from \cite{bouwens2021} down to $M_{\rm UV}=-12$, the same limit used for the Pop~III calculation. For comparison, we include simulation predictions from the literature \citep{pallottini2014, maio2016, xu2016, jaacks2019, sarmento2022, skinner2020, visbal2020, liu2020, munoz2022, hegde2023, venditti2023}. 
Our Pop~III SFRD constraint at $z\simeq6$--7 agrees well with the broad range of simulation predictions. 
Similar to our UVLF results, this SFRD constraint further supports the robust identification of \targ\ as a Pop~III candidate. 
Compared to the total SFRD, our results suggest that Pop~III galaxies may account for $\sim0.01$--1\% of the total cosmic SFRD at $z\simeq6$--7.

We caution that contributions from metal-free pockets/satellites, (i.e., Pop~II+Pop~III mixed objects; candidates -- e.g., \citealt{vanzella2023,wang2024}) are missed in our current Pop~III SFRD estimates, due to the strict requirement for the absence of the \oiii\ line in our selection method  (Section~\ref{sec:method}).
We note, however, that we also applied the same Pop~III selection technique to the clump-based NIRCam photometry for individual clumps in 133 strongly-lesned, resolved galaxies at $z\simeq$5--8 in UNCOVER and GLIMPSE (\citealt{claeyssens2024}, 2025 in prep.), which resulted in null detection. 
Hence, the contribution from the PopII+PopIII mixed objects, even if any, would unlikely change our current SFRD constraints significantly. Concerted efforts will be required to derive a definitive conclusion about this fraction -- for e.g., the development of future \jwst\ lensing cluster surveys.

\subsection{Cosmological Context of \targ}
\label{sec:context}

\begin{figure*}[t!]
\begin{center}
\includegraphics[trim=0cm 0cm 0cm 0cm, clip, angle=0,width=0.7\textwidth]{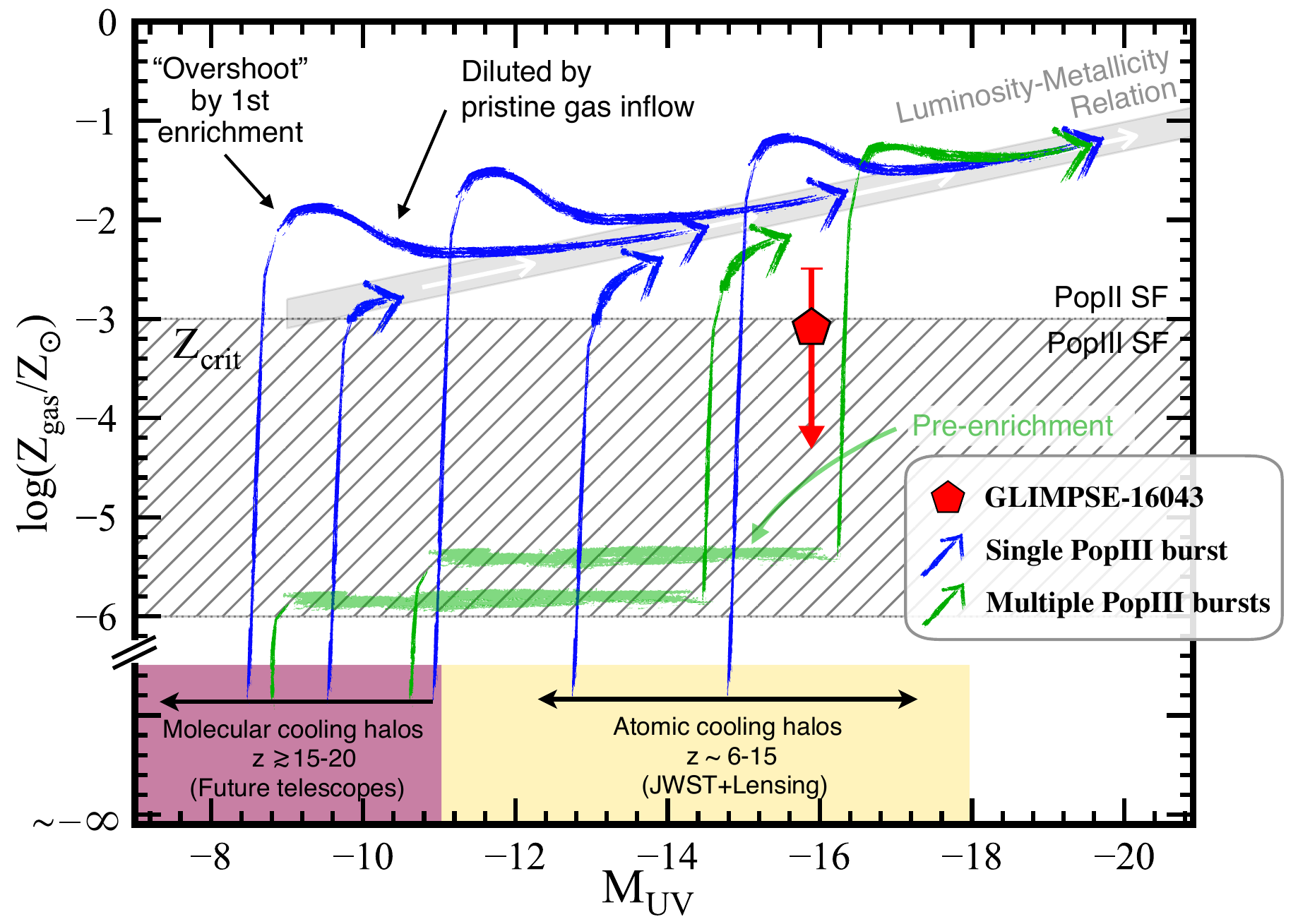}
\end{center}
\vspace{-0.3cm}
\caption{ \textbf{Illustration of {\targ} within the context of early chemical evolution.} The transition from Pop~III to Pop~II star formation (SF) is represented by a hatched region covering $Z_{\rm gas}\sim 10^{-6}-10^{-3}\,Z_{\odot}$, which reflects the range of critical metallicity from different cooling channels \citep[e.g.,][]{schneider2012}. 
Here, we adopt $Z_{\rm crit}=10^{-3}\,Z_\odot$ as a strict criterion for Pop~II star formation with metal-line cooling for definiteness. There are two broad categories of evolutionary pathways (curves with arrows) from the pristine (pure-Pop~III, $Z_{\rm gas}\sim 0$) state towards the extrapolated main sequence relation between $Z_{\rm gas}$ and $M_{\rm UV}$ for Pop~II dominated galaxies (shaded region, see the right panel of Fig.~\ref{fig:uv-metal}). In the first category (blue curve), it takes a single event of metal enrichment from the initial Pop~III starburst to reach $Z\gtrsim Z_{\rm crit}$. In the second category (green curve), the initial starburst with a limited number of massive stars in a molecular-cooling minihalo ($\lesssim 10^7 M_\odot$, $z\gtrsim 15-20$) only enriches the surroundings to a sub-critical metallicity, and one (or a few) subsequent starburst(s) will drive the system above $Z_{\rm crit}$. In both cases, the (last) Pop III enrichment event that achieves $Z\gtrsim Z_{\rm crit}$ can ``overshoot'' the main sequence \citep[e.g.,][]{liu2020,Magg2022,Prgomet2022}, producing a faint but relatively metal-rich (up to $\sim 0.1\ \rm Z_\odot$) galaxy right after SN explosions, which then converges onto the main sequence from above when inflow of fresh pristine gas fuels Pop~II star formation and meanwhile dilutes metals. Such diverse pathways cause a large scatter around the extrapolated main sequence at the faint end, as seen in observations of dwarf galaxies in the Local Volume \citep[e.g.,][]{Simon2019}.
The thickness of each pathway indicates typical timescales of $\sim5$--10\,Myr (thin) and $\sim50$--100\,Myr (thick) at each evolutionary phase. The shorter timescale reflects the rapid metal enrichment by Pop~III SNe, while the longer timescale is primarily governed by the free-fall time, closely related to the `recovery' time for gas to re-collapse after experiencing SN feedback \citep{jeon2014}. 
The magenta and yellow shaded regions represent the $M_{\rm UV}$ ranges associated with broad dark-matter halo mass regimes, where Pop~III star formation is driven by molecular cooling at $z\gtrsim15$--20 and by atomic cooling at later epochs, respectively. 
}
\label{fig:evolution}
\end{figure*}

The proposed selection technique in Section \ref{sec:method} and the Pop~III candidate \targ\ provide a unique opportunity to study the pristine environments of early galaxies. Beyond the low implied \oiii/H$\beta$ ratio, \targ\ has an extreme H$\alpha$ equivalent width, blue continuum slope, and strong Balmer Jump signatures, suggestive of extremely young stellar populations. Further, the relative faint \muv\ suggests that \targ\ resides in a low-mass halo. As such, \targ\ is significantly off the typical Luminosity-Metallicity relation. What do the observed properties of \targ\ suggest about the transformation of the pristine universe into a near-universally metal enriched universe \citep[e.g.,][]{karlsson2013}? More concretely, how do galaxies enrich from the nearly pristine Pop~III  metallicity and onto the well-established Luminosity-Metallicity relation? As galaxies enrich they must exceed the critical metallicity, $Z_{\rm crit}$, where star-formation transitions from the high-mass dominated Pop~III mode to the low-mass dominated Pop~II mode\footnote{The precise value of $Z_{\rm crit}\sim 10^{-6}-10^{-3}\,Z_{\odot}$ is under debate, involving dust- and fine-structure line cooling channels, with a possible dependence on redshift and local environment\citep[e.g.,][]{schneider2012}.}, but this exact transition has so far been observationally under-constrained. 

In broad terms, there are two overall pathways for a star-forming system to achieve this transition: enriching above $Z_{\rm crit}$ in one-step from the first round of  SNe, or from a multi-step process. In general, the single enrichment scenario occurs very quickly (5-10~Myr), consistent with the extreme equivalent widths observed in \targ. In the multiple episode scenario, enrichment from Pop~III stars forming at $z\gtrsim 15$ in minihalos establishes a seed metallicity that fuels subsequent star formation. The level of this pre-enrichment may still be below $Z_{\rm crit} $, depending on the Pop~III SN yields from a small number of massive stars and the mixing and dilution of their ejecta in the surrounding medium \citep[e.g.,][]{jeon2021}. Since the pre-enriched gas is still below $Z_{\rm crit}$, star formation requires subsequent accretion of gas and proceeds quite slowly on the free-fall timescale of 50-100~Myr. Star formation triggered in such pre-enriched, but still $<Z_{\rm crit}$, gas clouds would still lead to a top-heavy IMF similar to the pristine, zero-metallicity case \citep[e.g.,][]{chon2021}. 

We schematically summarize these two pathways from pristine conditions to super-critical enrichment in Figure~\ref{fig:evolution}. We indicate representative enrichment trajectories, for both the one- and multi-step modes, originating in low-mass minihalos. There are two physical regimes for Pop~III star formation: low-mass halos where primordial gas cools via molecular hydrogen, or higher-mass halos where cooling (initially) proceeds via atomic hydrogen line transitions \citep[for comprehensive reviews, see][]{bromm2013,klessen2023}. \targ\ could be an atomic cooling halo that is experiencing a single Pop~III star-forming event, or a system that has experienced previous (multiple-burst) Pop~III pre-enrichment in minihalo progenitor halos. Distinguishing between these scenarios likely requires detailed follow-up observations to constrain the nebular properties (e.g., the cooling function) and stellar properties to test the single enrichment scenario. In particular, the Lyman-Werner (LW) opacity in the rest-frame UV can provide direct observational constraints on the photo-dissociation of molecular gas within \targ\  to estimate whether the LW radiation was sufficiently weak to feed a molecular cooling halo  \citep[e.g.,][]{oh2002,johnson2008}. If confirmed as a Pop~III source, \targ\ would provide an ideal laboratory to reveal the conditions that produced the first stars and could stringently test Pop~III formation models.

\section{Alternatives to the Pop~III Scenario}
\label{sec:caveats}

Although our Pop~III candidates satisfy our stringent selection criteria and show excellent agreement in the volume density with the simulation predictions, which independently strengthen its robustness, it is important to explore all plausible alternative explanations given that we are searching for a very rare population. 
Below, we discuss interpretations for our candidates other than Pop~III. 

\subsection{Extremely Metal-poor Galaxy$?$}
\label{sec:empg}

The first plausible alternative scenario is an extremely metal-poor galaxy with \(Z_{\rm gas}/Z_{\odot}\) meeting the current upper limit of our candidates, but not entirely zero metallicity.  
This possibility naturally arises due to the limitation from the current photometric constraints on the \oiii/H$\beta$ ratio derived from the F356W excess.  
To draw a definitive conclusion, deep follow-up spectroscopy is essential to place more stringent upper limits on the complete absence of \oiii\ and/or to confirm the expected strong He {\sc ii} equivalent width \citep{tumlinson2001,schaerer2002,nakajima2022b}.  

Even if future spectroscopy reveals that our candidates are extremely metal-poor galaxies rather than Pop~III, they still offer exciting new insights.  
First, the current upper limit (\(Z_{\rm gas}/Z_{\odot} < 0.005\)) already reaches the most metal-poor regime explored by \jwst\ at \(z \gtrsim 6\) (Figure~\ref{fig:uv-metal}).  
This constraint is comparable to the values observed in the extremely \oiii-weak clumps in the strongly lensed arc at \(z=6.6\), known as LAP-1 and LAP-1B \citep{vanzella2023}.  
However, the rest-frame UV--optical continuum is undetected in LAP-1 and LAP-1B, leaving open the possibility that these clumps represent nebular emission from metal-poor or metal-free pockets of gas within galaxies rather than star clusters.  
In contrast, our candidates may serve as the first unique laboratory to study extremely metal-poor star clusters at $z > 6$.  
Second, the deviation in the Luminosity-Metallicity relation observed in our targets suggests either pristine gas inflow or enhanced UV flux.  
In the former case, directly detecting gas inflow has been challenging, particularly in ultra-faint, low-mass early galaxies with \(M_{\rm star} \approx 10^{5-6}\,M_{\odot}\).  
Our candidates provide key insights into the gas fueling of abundant low-mass nascent galaxies in the early universe.  
In the latter case, the enhanced UV flux implies a high luminosity-to-mass ratio, offering compelling evidence of a top-heavy IMF in an extremely metal-poor galaxy at \(z > 6\).  
Therefore, the extreme metallicity regime and the deviation in the Luminosity--Metallicity relation observed in our candidates represent in any case fairly unique features, even under the extremely metal-poor galaxy scenario.

\subsection{Extremely Metal-poor, Faint AGN$?$}
\label{sec:agn}

In the Luminosity-Metallicity relation (Figure~\ref{fig:uv-metal}), we find a significant deviation of our candidates from the typical relation, which could be attributed to either pristine gas inflow (i.e., reflecting an offset in $Z_{\rm gas}$) or enhanced UV flux (i.e., reflecting an offset in $M_{\rm UV}$).  
If the latter is true, possible mechanisms include a top-heavy IMF or AGN activity.  
While young star clusters may also exhibit compact morphologies ($\lesssim100$~pc; e.g., \citealt{adamo2024, mowla2024, fujimoto2024}), the point-source morphology of our candidates is also consistent with the AGN scenario.  
Recent \jwst\ studies have reported an abundant population of faint broad-line AGNs \citep[e.g.,][]{kocevski2023, matthee2023, greene2024, naidu2024}, with an estimated abundance of $\simeq5$--10\% among general galaxies at $z=4$--9 down to $M_{\rm UV}=-18$ \citep[e.g.,][]{harikane2023c, maiolino2023c, fujimoto2023c, napolitano2024}.  
Interestingly, the volume density of our Pop~III candidates is almost $\sim1$\% of the general galaxy population at $z \simeq$6--7 (Figure~\ref{fig:uvlf}), consistent with the abundance of faint AGNs, assuming their high abundance extends to fainter UV magnitudes. 
Taken together, these results suggest that the AGN scenario is worth further exploration as an alternative explanation.  
In particular, the UV luminosity of \targ\ ($M_{\rm UV} = -15.9$) lies in an ultra-faint regime, even compared to the recent \jwst\ studies of faint AGNs ($M_{\rm UV} < -18$).  
Assuming a typical Eddington ratio of 0.1--1.0 observed in these \jwst-discovered faint AGNs, this UV luminosity corresponds to a BH mass of \(M_{\rm BH} \simeq 10^{4-5}\,M_{\odot}\).  
This mass range is consistent with seed BHs \citep[e.g.,][]{volonteri2010, inayoshi2020} and intermediate-mass BHs \citep[e.g.,][]{greene2020, natarajan2021}, which are thought to play a key role in the origins of supermassive black holes from the early universe to the present day.  

To explore this possibility, it is crucial to note that the extremely metal-poor conditions are still essential in the material surrounding the AGN, as evidenced by the absence of the \oiii\ line, which is typically very bright in standard SEDs of high-$z$ quasars and AGNs.  
We thus adopt several SED templates specifically constructed for a seed black hole (BH) embedded in metal-poor environments.
The first set is the model presented in \cite{inayoshi2022b}, assuming a seed BH mass of $M_{\rm BH} = 10^{6}\,M_{\odot}$ embedded in a metal-poor ($Z_{\rm gas}/Z_{\odot} = 0.01$) galaxy. This SED includes three primary components: 
(1) radiation from the unresolved nuclear accretion disk surrounding the BH, modeled using a broken power-law spectrum \citep[e.g.,][]{lusso2015}, 
(2) nebular emission lines and continuum reprocessed by the irradiated gas in the surrounding nebular region, calculated via \texttt{CLOUDY} simulations, 
and (3) radiation from the dense accretion disk (resolved in radiation-hydrodynamical simulations within 0.1--100~pc). 
The gas and disk boundary is defined where the electron fraction $x_{\rm e} = 0.9$, marking the transition between the dense accretion disk and the surrounding nebular gas. The viewing angle is set to $60^{\circ}$. 

The second set is the Direct Collapse BH (DCBH) model introduced in \cite{nakajima2022b}. 
The model incorporates the following primary components: (1) an accretion disk emitting a thermal Big Bump component, parameterized by blackbody temperatures $T_{\rm bb} = 5 \times 10^4, 1 \times 10^5, \text{and}\ 2 \times 10^5 \, \text{K}$, (2) a high-energy power-law continuum, described as $f_\nu \propto \nu^\alpha$, with spectral slopes $\alpha = -1.2, -1.6, \text{and} -2.0$, extending into the X-ray regime, and (3) nebular emission by using \texttt{CLOUDY} to simulate gas photoionized by the BH radiation with the plane-parallel geometry, $n_e = 10^3\, \text{cm}^{-3}$, $Z_{\rm gas}/Z_\odot =0$--2, $\log U = [-0.5:-3.0]$. The model does not include dust in the ionized regions, ensuring that emission lines are unaffected by dust absorption or depletion effects, which should not matter when considering extremely metal-poor or metal-free cases.  
For this study, we utilize 9 DCBH templates (three $T_{\rm bb}$ and three $\alpha$) with $Z_{\rm gas} = 0$, $n_e = 1000\, \text{cm}^{-3}$, and $\log(U)=-1.5$, as fiducial cases. 

Incorporating these metal-poor AGN template sets to \texttt{EAZY}, we perform SED fitting to \targ. 
In Appendix~\ref{sec:appendix_agn}, we present the best-fit SEDs with the \cite{inayoshi2022b} and \cite{nakajima2022b} models, separately. 
We find that the best-fit SEDs provide $\chi^{2}$ (and $\Delta\chi^{2}$) $>$40--100. 
These best-fit SEDs reproduce the strong H$\alpha$ emission and absence of \oiii.  
However, the overall red color of the underlying continuum is at odds with the blue continuum and Balmer jump observed in our candidates.  
The red continuum color is mainly attributed to radiation from the BH and accretion disk (see also e.g., \citealt{natarajan2017, volonteri2017}).  
Thus, \(\Delta\chi^{2}\) might be substantially reduced by assuming different surrounding gas conditions in which the nebular continuum dominates the SED.  
This possibility cannot be ruled out given complete unknowns in the surrounding gas conditions for seed BHs in early galaxies.  
Therefore, while the SED features observed in our Pop~III candidates remain challenging to reproduce with the typical metal-poor AGN models assumed in \citet{inayoshi2022b} and \citet{nakajima2022b}, we cannot completely rule out the possibility of the extremely metal-poor AGN which might represent the seed or intermediate BH.

\subsection{Low-$z$ interlopers$?$}
\label{sec:lowz}

\begin{figure*}[t!]
\begin{center}
\includegraphics[trim=0cm 0cm 0cm 0cm, clip, angle=0,width=1.\textwidth]{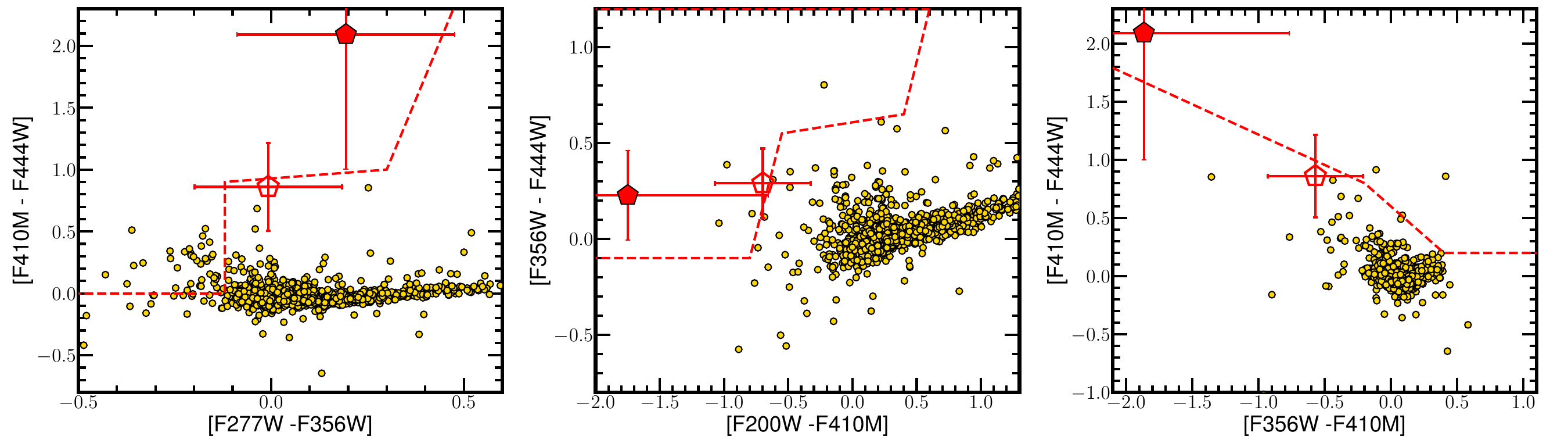}
\end{center}
\vspace{-0.2cm}
 \caption{\textbf{No low-redshift galaxy in the public \jwst\ spectroscopic archive has colors identical to \targ}.  
 Our tests show $z=1.5$ is the key interloper redshift for contaminants where the Pa$\alpha$ line boosts photometry similar to H$\alpha$ (Section~\ref{sec:lowz}). Using 2365 galaxies at $z=0-1.7$ with prism spectra from the DAWN \jwst\ archive, we demonstrate that none of these galaxies when redshifted to $z=1.5$ (golden points) have colors identical to the Pop III candidate (red pentagon). None of these sources satisfy all our color selection criteria (most clearly seen in the third panel). 
 We note, however, the significant error-bars on our photometry, and that the tentative candidate (empty pentagon) is consistent within error-bars with the peripheries of the $z=1.5$ locus.
\label{fig:lowz}}
\end{figure*}

We explore possible lower redshift solutions for \targ\ using \texttt{EAZY} and \texttt{Prospector}. 
We test various scenarios, including brown dwarfs, globular clusters, foreground cluster galaxies, ``Little Red Dot'' AGNs, and dusty emission line objects, which are known to mimic high-redshift galaxies \citep[e.g.,][]{arrabal-halo2023a}. 
To achieve this, we incorporate custom templates for these objects in \texttt{EAZY}, alongside the standard templates included in the \texttt{agn\_blue\_sfhz} template set. 
For example, we add a dusty emission line galaxy template inspired by Fig.~6 of \citet{naidu2022b}. 
We run fits constrained to the redshift range $z = 0$--6. 
All solutions in this range are strongly disfavored, with $\Delta\chi^{2} > 16$, the best-fit being a $z = 1.3$ galaxy $(\Delta\chi^{2} = 16.3)$.

We also run \texttt{Prospector} with a dense sampling of live points to reveal any lower redshift solutions that might not be represented among the \texttt{EAZY} templates. We find the best low-$z$ \texttt{Prospector} solutions to be extreme $z=1.5$ Paschen-$\alpha$ emitters, similar to the best-fit \texttt{EAZY} results with the general galaxy templates (see grey curve in Figure~\ref{fig:sed_eazy}). 
This must occur in a very narrow redshift range ($\Delta z <0.05$) in order to produce the F480M flux excess, thereby mimicking extreme H$\alpha$ emission in the Pop III scenario, 
while the faint F410M remains inexplicable in these fits. Indeed, these hypothetical $z=1.5$ galaxies are strongly disfavored versus the Pop III solution for the GLIMPSE candidate ($\Delta\chi^{2}=7.6$).

Based on lessons learned from the ``Schrodinger's Galaxy" which initially favored a $z \simeq 16$--17 solution in similar $\Delta\chi^{2}$ experiments \citep[e.g.,][]{donnan2023} but was later spectroscopically confirmed as a low-$z$ interloper at $z=4.91$ \citep[e.g.,][]{naidu2022b, arrabal-haro2023b}, we further investigate whether any environmental features might indicate a \(z \sim 1.5\) origin for \targ. 
Using the NIRCam catalog in GLIMPSE and photometric redshift $(z_{\rm phot})$ values derived from general galaxy templates, we identify a luminous source $(\mathrm{F200W}=24.2~\mathrm{mag})$ at $z_{\rm phot} \simeq 1.5$, 
located $4\farcs6$ away from \targ\ (R.A., Dec. = 342.2107, $-44.528362$). 
This proximity places \targ\ within the projected virial radius of this $M_{\star} \approx 10^{9} \, M_{\odot}$ source. 
To assess the statistical likelihood of such an association, we randomly select galaxies from the GLIMPSE catalog and find that 
$\approx 10\%$ of sources have a similarly luminous or more luminous $z_{\rm phot} \approx 1.5$ neighbor within a similar angular separation. Thus, the statistical significance of a physical association between this luminous $z \simeq 1.5$ galaxy and \targ\ is $\approx1$--$2\sigma$.

It is crucial to note that galaxies resembling the theoretical $z=1.5$ SEDs constructed to explain \targ\ with \texttt{Prospector} have never been observed before. Such a source would be among the most powerful starbursts at $z\simeq1.5$, the likes of which are yet to be observed among extremely low-mass, $\approx10^{5}-10^{6} M_{\rm{\odot}}$ dwarf galaxies -- e.g., log(sSFR/yr$^{-1}$)$\approx$-7. This is a stark difference from the fact that $>30\%$ of sources with similar $M_{\rm{UV}}$ as this candidate may be quiescent if we extrapolate trends from higher redshift \citep{endsley2024}. 

Indeed, in Figure~\ref{fig:lowz} we present color-color diagrams like those used in our Pop~III selection (Figure \ref{fig:color}) for a compilation of 2365 galaxies observed by \jwst\ in the NIRSpec/prism mode. Spectra of these sources were homogeneously reduced with the \texttt{msaexp} software \citep{brammer2022b} and distributed in the v3.0 release of the DAWN \jwst\ archive \citep{graaff2024, heintz2024}\footnote{See Acknowledgments for a list of programs that these spectra were collected as a part of.}. 
We select galaxies with the highest quality grade of 3 at $z=0-1.7$, shift them to $z=1.5$, and synthesize NIRCam colors directly from the prism spectra. Note that the color-color diagrams shown in Fig. \ref{fig:color} are focused on LW bands, so we are able to use the lower redshift sources by ignoring the lack of prism coverage at bluer wavelengths $\lesssim0.7-0.8\mu$m. From this exercise it is clear that no source in the public \jwst\ spectroscopic archive has colors that match the color location of \targ.

To summarize -- 
the low-$z$ scenario is still strongly disfavored by the forced low-$z$ SED analysis and the NIRCam color-color comparison with real DJA galaxies shifted at $z=1.5$, while the existence of a $z\simeq1.5$ galaxy at $r\sim4\farcs6$ warrants some caution. 
Importantly, these objects would have to be extraordinary sources in their own right, the likes of which \jwst\ is yet to observe. Spectroscopic follow-up is therefore imperative and the only way forward to confirm the nature of these Pop~III candidates.

\subsection{Two Different Objects Mimicking Colors$?$}
\label{sec:projection}

The chance projection of two objects at different redshifts may alter the observed colors. 
Although our candidates exhibit very compact morphologies, it is still useful to estimate the probability of such a chance projection.  
In GLIMPSE, a total of 59,650 sources are detected in the catalog, corresponding to a surface density of NIRCam objects of 6,345~arcmin$^{-2}$.  
Given the point source morphology of our candidates, we define a projection radius of $0\farcs01$. 
Following the calculation presented in \citet{downes1986}, we estimate the probability of a chance projection to be $\sim$0.06\%.   
This scenario would also require specific combined colors that satisfy all the selection criteria, including the color-color diagrams and SED thresholds.  
As a result, the actual probability is likely reduced by several orders of magnitude.  
We therefore conclude that this scenario is highly unlikely.

\subsection{What is the most plausible alternative solution$?$}
\label{sec:alternative}

We briefly summarize the discussions in Sections~\ref{sec:empg}--\ref{sec:projection}. 
The extremely metal-poor galaxy scenario ($Z_{\rm gas}/Z_{\odot}<0.005$) remains plausible, as it is consistent with the current photometric upper limit on the \oiii/H$\beta$ ratio inferred from the F356W excess (Section~\ref{sec:empg}). 
While the SED features observed in our Pop~III candidates are difficult to reconcile with typical metal-poor AGN models, we cannot exclude the possibility of a metal-poor AGN due to the large uncertainties surrounding seed BH environments in early galaxies (Section~\ref{sec:agn}). 
Contamination by low-$z$ sources is strongly disfavored by the observed SED features, although we cannot rule out rare low-$z$ objects that might exhibit similar SED properties yet remain undiscovered (Section~\ref{sec:lowz}). 
Finally, the chance projection of two distinct sources that collectively produce the observed peculiar colors is highly unlikely (Section~\ref{sec:projection}).

Importantly, however, even under these alternate scenarios, the faintness, the extremely low \oiii/H$\beta$ ratio, and the deviation in the \(M_{\rm UV}\)--\(Z_{\rm gas}\) relation observed in our candidates suggest the identification of either metal-poor seed or intermediate black holes (\(\approx 10^{4-5}\,M_{\odot}\)), a recent pristine gas inflow, or a top-heavy IMF formed in ultra-faint, low-mass early galaxies with \(M_{\rm star} \approx 10^{5-6}\,M_{\odot}\). The low-$z$ interlopers would also be extreme starbursts in ultra-faint galaxies. Regardless of their true origins -- Pop~III or otherwise -- our efficient NIRCam selection method opens a new discovery space in the distant universe. 

\section{Future Prospects}
\label{sec:future}

We have developed an efficient NIRCam-based method for the Pop~III galaxy search (Section~\ref{sec:method}), identified promising candidates in the $\sim500$~arcmin$^{2}$ area of publicly available deep \jwst\ legacy NIRCam data (Sections~\ref{sec:obs} and \ref{sec:result}), and discussed possible alternative explanations (Section~\ref{sec:caveats}).  
Here, we briefly discuss future prospects based on these findings and the current limitations. 

First, deep follow-up spectroscopy is essential to confirm the Pop~III nature of the photometrically selected candidates.  
Although our candidates show excellent agreement with Pop~III galaxy SED models, the presence of very weak \oiii\ remains consistent with the uncertainty in the F356W photometry, which naturally translates to the current upper limits on \oiii/H$\beta$ and subsequent \(Z_{\rm gas}\) estimates.  
This limitation will be the same for all photometric candidates identified in future surveys.  
To confirm the Pop~III origin, an alternative approach is the detection of strong He~{\sc ii}$\lambda$1640 and/or He~{\sc ii}$\lambda$4686 EWs \citep[e.g.,][]{bromm2001,tumlinson2001,schaerer2002,nakajima2022,wang2024}.  
Thus, deep spectroscopy to place more stringent upper limits on \oiii/H$\beta$ and/or detect strong He~{\sc ii} is a logical next step.  
We recommend using NIRSpec Medium or High-resolution gratings rather than the prism mode for such follow-up observations.  
As discussed in Section~\ref{sec:caveats}, ruling out the metal-poor AGN scenario is challenging.  
The use of NIRSpec Medium or High-resolution gratings will enable the detection of broad Balmer lines (e.g., H$\alpha$, H$\beta$) and confirm the complete absence of neighboring \oiii\ lines and/or the large EW of the neighboring He~{\sc ii}$\lambda$4686.

While the presence of a luminous $z\sim1.5$ galaxy at $r\sim4\farcs6$ could be potentially concerning (Section~\ref{sec:lowz}), we must remember that if such environmental considerations were followed as an iron-clad rule, the most distant galaxy presently known at $z_{\rm{spec}}=14.18$, GS-z14-0, \citep{carniani2024,carniani2024b,schouws2024} would never have been selected for follow-up observations. In particular, GS-z14-0's closest neighbor, lying a mere $<1\farcs0$ away, is a luminous $z=3.47$ galaxy with a redshift lining up exactly with the low-$z$ solution for GS-z14-0, with its Balmer break occurring at the same wavelength as the Lyman break of the distant source \citep{carniani2024}. In the pursuit of extraordinary sources, such as Pop~III galaxy candidates, especially in the early years of \jwst, it is perhaps best to obtain follow-up spectra liberally.

Second, observing a large number of lensing clusters is critical for discovering (lensed) Pop III galaxy candidates and enabling time-efficient follow-up spectroscopy.  
Assuming \(M_{\rm star} \approx 10^{5}\,M_{\odot}\) as a typical stellar mass of the first galaxies (Section~\ref{sec:depth}), Figure~\ref{fig:mobs_age_all} shows that a depth even fainter than \(\sim30.5\)~mag is required, exceeding the deepest NIRCam survey depths achieved with more than 100~hours of observation time.  
This highlights the difficulty of discovering such objects in blank field surveys.  
In this context, gravitational lensing is a promising way forward.  
The GLIMPSE program has demonstrated this potential by successfully identifying \targ\ within the survey volume of a single cluster.  
While observing a large number of lensing clusters comes at a cost of shallower depth per cluster, the depth is compensated by collecting numerous high-magnification patches.  
In fact, this ``wide and shallow" strategy is the most time-efficient way to build up the deepest layer of imaging, as demonstrated analytically and by recent large lensing cluster observations \citep{vujeva2023, fujimoto2023b}.  
Moreover, a wide lensing cluster survey mitigates cosmic variance by sampling random lines of sight, which is a crucial advantage in the search for rare galaxy populations.  
Importantly, this approach allows us to probe the intrinsically faintest regimes while the detected sources remain relatively bright in the observed frame, significantly accelerating follow-up spectroscopy.  

Third, incorporating medium-band filters in \jwst\ legacy NIRCam fields is highly beneficial for ensuring robust Pop III selection.  
As demonstrated in Figures~\ref{fig:sed_comp} and \ref{fig:color}, F410M is an essential filter for anchoring the underlying continuum and quantifying the Balmer jump as well as the strengths of H$\alpha$ and \oiii\ lines in conjunction with other broad-band filters at $z\simeq$6--7.  
While the color-color diagrams presented in Figure~\ref{fig:color} rely primarily on broad-band filters, apart from F410M, the addition of other medium-band filters further strengthens constraints on H$\alpha$ and/or \oiii\ line strengths.  
For example, Appendix~\ref{sec:appendix_f480m} presents color-color diagrams using F480M in place of F444W, demonstrating a similar capability for separating metal-enriched galaxies from Pop~III galaxies.  
A disadvantage with the medium-band filters is their narrower validated redshift range for capturing specific emission lines compared to broad-band filters.  
Nevertheless, the NIRCam colors are sensitive to noise fluctuation, which is particularly significant when exploring faint objects.  
Therefore, the independent measurements of key emission line strengths provided by medium-band filters, in addition to the broad-band filters, enhance the reliability of identifications in these challenging regimes.  

Finally, it is important to emphasize that the non-detection of Pop~III galaxies at \(z \simeq 6\)--7 will also be highly constraining, even as deep NIRCam surveys and follow-up spectroscopy continue to expand.  
In Figure~\ref{fig:uvlf}, we find that the volume densities of our Pop~III galaxy candidates are consistent with several simulation predictions.  
This implies that replacing detections with upper limits, which could become significantly more stringent in future surveys, will provide critical constraints on Pop~III formation and evolution models.  
These constraints -- where none exist at the moment -- will revolutionize our understanding of the nature of the first galaxies.

\section{Summary}
\label{sec:summary}

In this paper, we present a novel and efficient NIRCam-based selection method for identifying Pop~III galaxies at \(z \simeq 6\)--7, validated through dedicated completeness and contamination simulations. Leveraging deep NIRCam imaging data from a total area of $\simeq$~500~arcmin$^2$ across several \jwst\ legacy fields, including GLIMPSE, UNCOVER, CEERS, PRIMER, and JOF, we systematically searched for Pop~III galaxy candidates. Below, we summarize the main findings:

\begin{enumerate}
    \item We developed a fiducial selection method combining color-based and SED-based approaches. The color-based method incorporates three color-color diagrams, optimized to capture the key features of Pop~III galaxies at $z=5.6$--6.6, such as the absence of detectable metal lines like \oiii, strong H$\alpha$ emission, and a prominent Balmer jump. The SED-based method relies on the $\chi^{2}$ difference between best-fit SEDs using Pop~III and metal-enriched galaxy templates. Monte Carlo simulations were performed to evaluate completeness and contamination rates. By combining the color and SED-based methods, we confirmed high completeness ($>$50--80\%) and zero contamination across most of the parameter space at $z=5.6$--6.6, above the detection limit of the data. [Figures~\ref{fig:sed_comp}, \ref{fig:color},  \ref{fig:comp_contami}, \ref{fig:fiducial}; Section~\ref{sec:method}]
    
    \item Under realistic assumptions we demonstrate that only the deepest \jwst\ imaging surveys are able to breach the luminosity regime where Pop III-dominated galaxies are most likely to occur ($>30.5$ mag), underscoring the importance of gravitational lensing for meaningful Pop~III galaxy searches. [Figure~\ref{fig:mobs_age_all}; Section~\ref{sec:depth}]
    
    \item We identified one robust Pop~III galaxy candidate, \targ, at $z = 6.50^{+0.03}_{-0.24}$, and one tentative candidate, \targb, at $z = 6.17^{+0.19}_{-0.06}$. \targ\ satisfies all fiducial selection criteria, whereas \targb\ meets the SED-based thresholds but falls near the borders of the color-color diagram criteria, failing to meet all three. [Figures~\ref{fig:sed_eazy}, \ref{fig:sed_all}, \ref{fig:sed_eazy2}, \ref{fig:sed_all2}; Section~\ref{sec:result}, Appendix~\ref{sec:appendix_jof}]

    \item \targ\ is a moderately lensed galaxy ($\mu=2.9^{+0.1}_{-0.2}$), exhibiting textbook features expected of a Pop~III galaxy: strong H$\alpha$ emission with observed rest-frame equivalent width of $2810 \pm 550~{\rm \AA}$, a pronounced Balmer jump, no detectable metal lines (\oiii/H$\beta < 0.44$), and nascent stellar populations with \(t_{\rm age} < 5 \, \text{Myr}\). The inferred stellar mass is \(\sim10^5 \, M_{\odot}\), with intrinsic UV magnitudes of \(M_{\rm UV} = -15.89^{+0.12}_{-0.14}\). It has an extremely compact morphology consistent with a point source, with an upper limit on its effective radius of of $r_{\rm e} < 40$~pc after the lensing correction. [Figures~\ref{fig:galfit},\ref{fig:galfit2}; Section~\ref{sec:properties}, Appendix~\ref{sec:appendix_jof}]

    \item Using the H$\alpha$ and [OIII]+H$\beta$ flux excesses observed in F480M and F356W, we constrain the gas-phase metallicity to 12 + $\log(\text{O/H}) < 6.4$ (\(\simeq0.005\,Z_{\odot}\)). In the $M_{\rm UV}$--$Z_{\rm gas}$ relation, \targ\ significantly deviates from the extrapolated trends of \jwst\ measurements at \(z = 4\)--10, appearing $>10\times$ brighter than expected for its metallicity. This deviation may indicate a top-heavy IMF, pristine gas inflows, or extremely metal-poor AGNs. [Figure~\ref{fig:uv-metal}, Section~\ref{sec:metal}]
    
    \item The tentative candidate, \targb\ at $z = 6.17^{+0.19}_{-0.06}$ found in the JOF field, displays key SED signatures of Pop~III galaxies similar to \targ: faint ($M_{\rm UV} = -17.62^{+0.17}_{-0.15}$), very young ($t_{\rm age} < 5 \,\text{Myr}$), point-source morphology ($r_{\rm e} < 80 \,\text{pc}$), and very low metallicity (12 + $\log(\text{O/H}) < 6.2$, i.e., $Z_{\rm gas} < 0.003$\,$Z_{\odot}$) inferred from the F356W and F444W flux excesses. However, the \texttt{Prospector} fitting analysis yields only a small $\Delta\chi^2=1.2$ between a $z=6.2$ Pop~III and a $z\sim1.5$ Pa$\alpha$ solution. This reaffirms the tentative nature of \targb\ which does not perfectly pass our color-color thresholds. [Figures~\ref{fig:color_data}, \ref{fig:sed_eazy2}, \ref{fig:sed_all2}, \ref{fig:galfit2}; Appendix~\ref{sec:appendix_jof}).

    \item We derived the first observational constraint on the high-redshift Pop~III UV luminosity function (UVLF) and star-formation rate density (SFRD) at $z=5.7$--6.6. The volume density and SFRD of our best candidate, \targ, align well with theoretical predictions, providing independent support for its robustness as a Pop~III galaxy candidate. [Figures~\ref{fig:uvlf}, \ref{fig:sfrd}; Sections~\ref{sec:uvlf}, ~\ref{sec:sfrd}]

    \item While alternative scenarios such as extremely metal-poor galaxies or AGNs cannot be completely ruled out at the moment, even these interpretations make \targ\ and \targb\ exceptional laboratories for studying the first stages of galaxy or black hole formation in the early universe. The low-$z$ scenario of a Paschen-$\alpha$ interloper at $z\approx1.5$ merits caution due to a nearby neighbor at the same redshift, but this would have to be a hitherto unobserved kind of galaxy that is yet to be discovered. Deep, high-resolution follow-up spectroscopy will be critical for confirming the Pop~III nature of these candidates. [Figure~\ref{fig:lowz}; Section~\ref{sec:caveats}]
    
\end{enumerate}

This work paves a clear path for the discovery of the first Pop~III galaxies. Whatever the fate of the present candidates, the methods developed in this study will empower Pop~III galaxy searches throughout the \jwst\ era. The volume density ($\approx10^{-4}$~cMpc$^{-3}$) implied by the finding of our robust candidate is remarkably high, suggesting more than one Pop~III source awaiting discovery per NIRCam pointing. 
The challenge, however, lies in pushing past $\approx30$--31~mag efficiently, while also remaining sensitive to the Pop~III signatures. 

Inspired by these results, we project that a snapshot NIRCam survey of the most magnified regions across a large number of lensing clusters, and deploying carefully chosen medium-bands, could provide significant observational constraints to help distinguish between different Pop~III galaxy simulations, and learn more about the key physical mechanisms of these earliest galaxies and their Pop~III star formation.

Exactly a hundred years ago, our cosmic horizon expanded past the edges of the Milky Way for the first time, with Andromeda and Triangulum marking the boundaries of our place in the Universe  \citep[e.g.,][]{hubble1925}.
As we reflect on the profound discoveries of the last hundred years, it is intriguing to consider how those early surveyors of glass plates would view the prospect that we may soon detect the Universe’s very first stars.

\acknowledgments{
We are grateful to the CEERS, PRIMER, JOF, UNCOVER, and GLIMPSE teams for developing their NIRCam surveys, and to the various \textit{JWST} and \textit{HST} surveys acknowledged in Section~\ref{sec:obs} that enabled our search. We thank Kimihiko Nakajima and Kohei Inayoshi for sharing Pop III and/or AGN templates, Steven Finkelstein for comments on the completeness \& contamination rate simulation, Aaron Yung for SEDs of simulated galaxies, Joel Leja, Ben Johnson, and Sandro Tacchella for advise on SED fitting, and
Takashi Kojima and Hiroto Yanagisawa for discussions.

We made extensive use of the DAWN \jwst\ Archive for various comparisons presented in this paper. Some of the data products presented herein were retrieved from the Dawn \jwst\ Archive (DJA). DJA is an initiative of the Cosmic Dawn Center (DAWN), which is funded by the Danish National Research Foundation under grant DNRF140. The prism spectra used in this paper were observed as part of the following programs, and we are grateful to these teams for helping build the rich spectroscopic legacy of JWST: 1180, 1181, 1210, 1286, 3215 \citep{bunker2023b, deugenio2024}; 1211-1215 \citep{maseda2024}; 1345 \citep{finkelstein2024}; 1433 \citep{hsiao2024a}; 1747 \citep{roberts-borsani2024}; 2028 \citep{wang2024}; 2073 (PI: J. Hennawi); 2198 \citep{barrufet2024}; 2282 \citep{bradley2023}; 2561 \citep{bezanson2022,price2024}; 2565 \citep{nanayakkara2023}; 2750 \citep{arrabal-haro2023b}; 2756 (PI: W.~Chen); 2767 \citep{williams2023}; 3073 \citep{castellano2024}; 4106 (PI: E.~Nelson); 4233 \citep{graaff2024}; 4446 \citep{frye2023}; 4557 (PI: H.~Yan); 6541 (PI: E.~Egami); 6585 (PI: D.~Coulter). 

This work is based on observations made with the NASA/ESA/CSA James Webb Space Telescope. The data were obtained from the Mikulski Archive for Space Telescopes at the Space Telescope Science Institute, which is operated by the Association of Universities for Research in Astronomy, Inc., under NASA contract NAS 5-03127 for JWST. These observations are associated with program \#03293.

This project has received funding from NASA through the NASA Hubble Fellowship grant HST-HF2-51505.001-A awarded by the Space Telescope Science Institute, which is operated by the Association of Universities for Research in Astronomy, Incorporated, under NASA contract NAS5-26555. This work has received funding from the Swiss State Secretariat for Education, Research and Innovation (SERI) under contract number MB22.00072, as well as from the Swiss National Science Foundation (SNSF) through project grant 200020\_207349. The Cosmic Dawn Center (DAWN) is funded by the Danish National Research Foundation under grant DNRF140.
HA and IC acknowledge support from CNES, focused on the JWST mission, and the Programme National Cosmology and Galaxies (PNCG) of CNRS/INSU with INP and IN2P3, co-funded by CEA and CNES.  IC acknowledges funding support from the Initiative Physique des Infinis (IPI), a research training program of the Idex SUPER at Sorbonne Universit\'e. AZ acknowledges support by Grant No. 2020750 from the United States-Israel Binational Science Foundation (BSF) and Grant No. 2109066 from the United States National Science Foundation (NSF); and by the Israel Science Foundation Grant No. 864/23. PN acknowledges support from the Gordon and Betty Moore Foundation and the John Templeton Foundation that fund the Black Hole Initiative (BHI) at Harvard University where she serves as one of the PIs. 
}

\facilities{\textit{JWST}}

\software{{\sc Source Extractor} \citep{bertin1996},  {\sc astropy} \citep{astropy2013}.
}

\appendix

\section{SED models}
\label{sec:appendix_sed}

To develop the color-color diagrams for the color-based selection (Section~\ref{sec:color}), we use several Pop~III SED models in the literature and generate Pop~II SED models using a public SED code. Below, we briefly describe the models we use in our analysis. 

\subsection{Pop~III models}

\subsubsection{\cite{zackrisson2011}}
The stellar component is based on single stellar population of Pop~III stars from \citet{schaerer2002}, \citet{schaerer2003}, and \citet{raiter2010} with a total mass of $10^{6}\,M_{\odot}$, while the nebular emission is computed using the photoionization code \texttt{Cloudy} \citep{ferland1998, ferland2013, ferland2017}.
\zackmodel\ supports three types of initial mass functions (IMFs) for Pop~III stars:  
a very top-heavy IMF using the \cite{schaerer2002} stellar SSP with a power-law IMF with a slope of $\alpha= 2.35$ across the mass range 50--500 $M_{\odot}$; 
a moderately top-heavy IMF assuming a log-normal distribution with characteristic mass at $10M_{\odot}$ and 1~$M_{\odot}$, but with wings extending from 1 to 500~$M_{\odot}$; 
and the Kroupa IMF covering a broad stellar mass range of 0.1--100~$M_{\odot}$. 
These three models are dubbed Pop~III.1, Pop~III.2, and Pop~III. Kroupa, generally reflecting the characteristic mass of stars $M_{\star,\rm IMF}$ to be $\sim$100, 10, and 1~$M_{\odot}$, respectively.
The templates span a wide range of stellar ages (\(t_{\rm age}\)) from 0.01 to 100 Myr. 
For this study, we specifically adopt ages of \([0.01, 0.1, 1, 5, 10, 30] \, \mathrm{Myr}\) when available for all three IMF scenarios, with the gas covering fraction of \(f_{\rm cov} = 1.0\), ensuring maximal contributions from nebular emission. 
The \texttt{Cloudy} calculation is performed with the electron density of $n_{\rm e}=100$~cm$^{-3}$ in the \zackmodel\ model.  

\subsubsection{\cite{nakajima2022b}}

With \texttt{Cloudy}-based calculations, \citet{nakajima2022b} explore a range of physical conditions for Pop~III stellar populations and their surrounding gas, using the SEDs from \citet{schaerer2003} and \citet{raiter2010}. These models assume a Salpeter IMF, while varying the mass range: 1–100~\(M_\odot\), 1–500~\(M_\odot\), and 50–500~\(M_\odot\).
In this paper, we refer to these three models as Pop~III.Sal1-100, Pop~III.Sal50-500, and Pop~III.Sal1-500, respectively.  
The wider mass ranges, particularly the top-heavy 50–500~\(M_\odot\) IMF, represent environments where high-mass stars dominate, as expected in metal-free conditions with limited cooling mechanisms. For our analysis, we use models with a gas electron density of \(n_e = 10^3 \, \mathrm{cm^{-3}}\) and an ionization parameter of \(\log U = -1.5\). These conditions reflect typical values inferred in star-forming regions at high redshift \citep[e.g.,][]{sanders2023, isobe2023b, reddy2023b}, ensuring consistency with observed properties of high-energy ionizing sources. 

\subsection{Pop~II models}

\subsubsection{BAGPIPES}
The mock SEDs are generated with the default setup of \texttt{BAGPIPES}, which assumes a \citet{kroupa2002} IMF and the \citet{bruzual2003} stellar population library.
For this study, we explore a parameter space motivated by recent \jwst\ observations of early galaxies, covering gas-phase metallicities of \(Z_{\rm gas}/Z_\odot = [0.01, 0.05, 0.10, 0.20]\), ionization parameters \(\log U\) of \([-3, -2, -1]\), a fixed electron density of \(n_e = 10^{3} \, \mathrm{cm^{-3}}\), zero dust attenuation, and redshifts \(z = 5.0\) to \(7.5\) in steps of \(\Delta z = 0.1\). 
These conditions are also optimal to encompass the diversity of metal-enriched young galaxies in the early universe, which may have NIRCam colors close to the Pop~III galaxies. Photometric data for the \jwst/NIRCam filters are generated directly from these SEDs, including contributions from both stellar continuum and nebular emission lines.

\section{Color-Color Selection with F480M}
\label{sec:appendix_f480m}

\begin{figure*}[t!]
\begin{center}
\includegraphics[trim=0cm 0cm 0cm 0cm, clip, angle=0,width=1\textwidth]{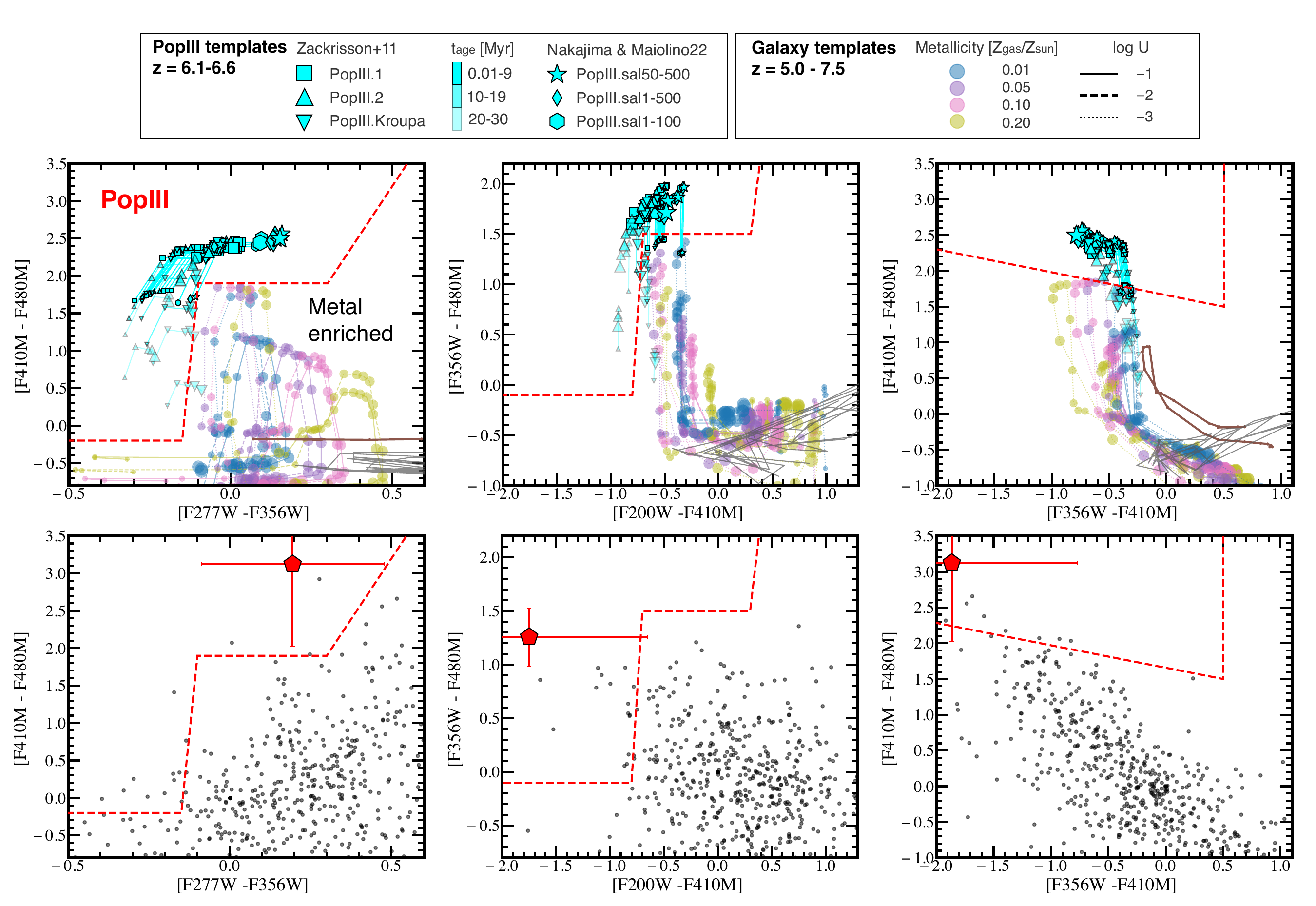}
\vspace{-0.4cm}
\end{center}
 \caption{
 Same as Figure~\ref{fig:color} (top panels) and Figure~\ref{fig:color_data} (bottom panels), but replaced F444W with F480M which also efficiently isolates the extreme H$\alpha$ emission for Pop III galaxies, but a narrower redshift range of $z=6.1$--6.6.  
 Among the NIRCam sources at $z_{\rm phot}=5.0$--7.5 in GLIMPSE (grey dots), no sources meet the three color-color diagrams with F480M, except for \targ.  
\label{fig:appendix_color}}
\end{figure*}

As discussed in Section~\ref{sec:color}, the selection criteria for Pop~III galaxies primarily rely on color-color diagrams constructed using NIRCam broad-band filters (F200W, F277W, F356W, F444W) and the medium-band filter F410M. However, other medium-band filters such as F480M can also play a critical role in refining the selection function by providing additional constraints on the key emission line of H$\alpha$ at similar redshifts, although the available redshift range is even narrowed naturally due to its narrower filter response, compared to F444W. 

In Figure~\ref{fig:appendix_color}, we present color-color diagrams incorporating F480M, replacing F444W. We define the color selection enclosed by the following vertices in the color-color space:
\begin{itemize}
    \item F277W--F356W vs. F410M--F480M (vertices): 
    \begin{equation}
    \begin{split}
        (-1.5, 4.5), \, (-1.5, -0.2), \, (-0.15, -0.2), \\
        (-0.1, 1.9), \, (0.3, 1.9), \, (0.3, 1.9), \, (0.7, 4.5)
    \end{split}
    \end{equation}
    \item  F200W--F410M vs. F356W--F480M (vertices):
    \begin{equation}
    \begin{split}
       (-2.5, 3.2), \, (-2.5, 0.1), \, (-0.8, -0.1), \\ 
       (-0.7, 1.5), \, (0.3, 1.5), \, (0.5, 3.2)
    \end{split}
    \end{equation}
    \item F356W--F410M vs. F410M--F480M (vertices):
    \begin{equation}
    \begin{split}
       (0.5, 3.5), \, (0.5, 1.5), \, (-2.0, 2.3), \\ 
       (-2.0, 3.5)
    \end{split}
    \end{equation}
\end{itemize}
The x-axis colors remain sensitive to the Balmer jump and weak \oiii\ emission, while the y-axis colors are designed to isolate strong H$\alpha$ emission. 
We find that \targ\ also meets all three color-color diagrams, strengthening its robustness as the Pop~III candidate. 
These diagrams demonstrate that F480M also effectively captures the unique SED features of Pop~III galaxies, providing an additional support for distinguishing them from metal-enriched galaxies.

\section{SED-Based Selection at \(z = 3\)--8}
\label{appendix:sed}

\begin{figure}[t!]
\begin{center}
\includegraphics[trim=0cm 0cm 0cm 0cm, clip, angle=0,width=0.5\textwidth]{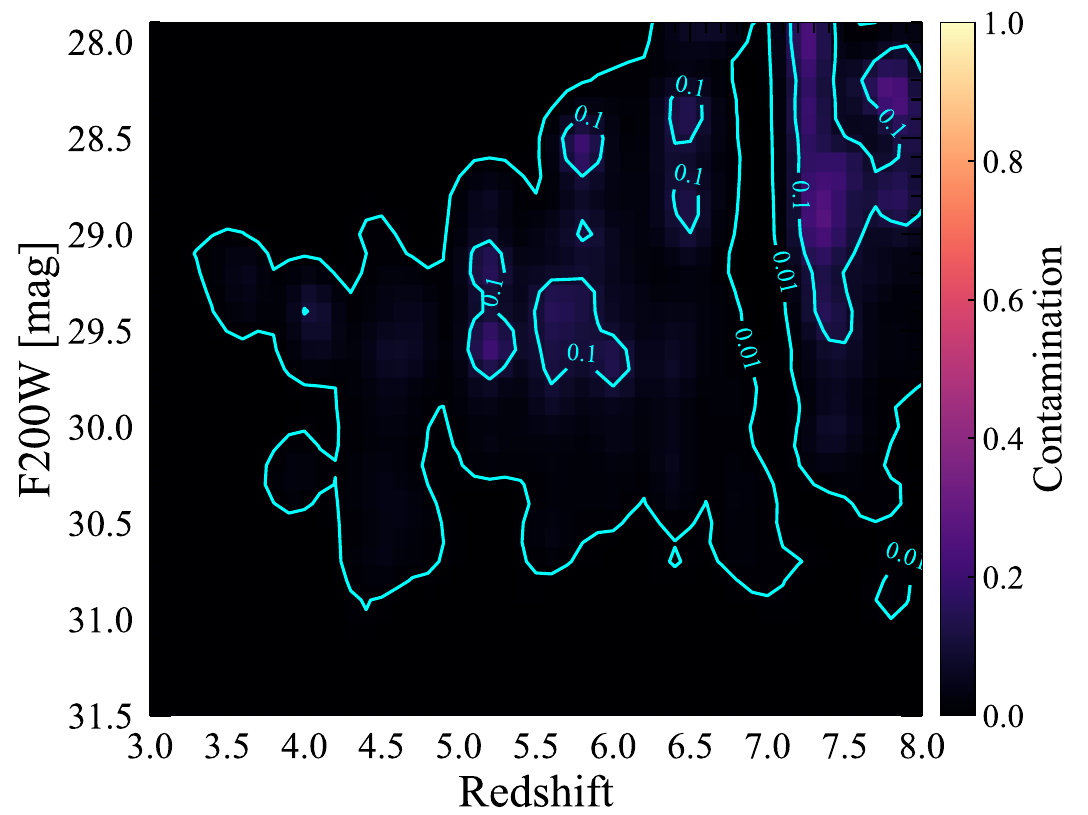}
\end{center}
\caption{
Same as the right panel of Figure~\ref{fig:comp_contami}, but for the SED-based method extended to a broad redshift range (\(z = 3\)--8) using all 20 NIRCam broad-band and medium-band filters. 
Compared to the contamination rate of the SED-based method shown in Figure~\ref{fig:comp_contami}, the inclusion of additional photometric data points reduces the contamination rate across much of the parameter space. 
However, contamination rates exceeding \(1\%\) persist in most regions, presenting challenges for identifying rare populations such as Pop~III galaxies. 
\label{fig:appendix_contami_wide}}
\end{figure}

In Section~\ref{sec:sed}, we introduced the SED-based method optimized for $z\sim$6--7, aligning it with the color-based method available at \(z = 5.6\)--6.6. The strength of the SED-based method, however, is its ability to fully leverage all available photometric data, making it particularly valuable for broader redshift ranges. Notably, the contamination rate in the SED-based selection may be reduced as the number of photometric data points increases. To explore this potential, we extend the SED-based selection to a wider redshift range of \(z = 3\)--8, assuming a case that all 20 NIRCam broad-band and medium-band filters are available, as in surveys like UNCOVER+MegaScience \citep{bezanson2022, suess2024}.

In Figure~\ref{fig:appendix_contami_wide}, we present the contamination rate of the SED-based method using mock galaxies from the Santa Cruz Semi-Analytic Model catalog \citep{yung2022}, in the same manner as Section~\ref{sec:contami}, but now extended to the broader redshift range and full filter set. Compared to the filter set in Section~\ref{sec:sed}, the contamination rate generally decreases due to the increased photometric coverage. Nevertheless, we observe that most of the parameter space still exhibits contamination rates exceeding \(1\%\), which poses challenges for identifying rare populations like Pop~III galaxies. 
Interestingly, there exists a favorable parameter space at \(z \simeq 3\)--5 and magnitudes \(\lesssim 29.0\), where the contamination rate shows almost entirely zero. However, careful simulation studies are necessary to verify whether Pop~III galaxies brighter than 29~mag can plausibly exist at \(z \simeq 3\)--5, making this range a promising but uncertain avenue for future exploration.

\section{A Tentative candidate found in JOF}
\label{sec:appendix_jof}

\begin{figure*}[t!]
\begin{center}
\includegraphics[trim=0cm 0.1cm 0cm 0cm, clip, angle=0,width=1.\textwidth]{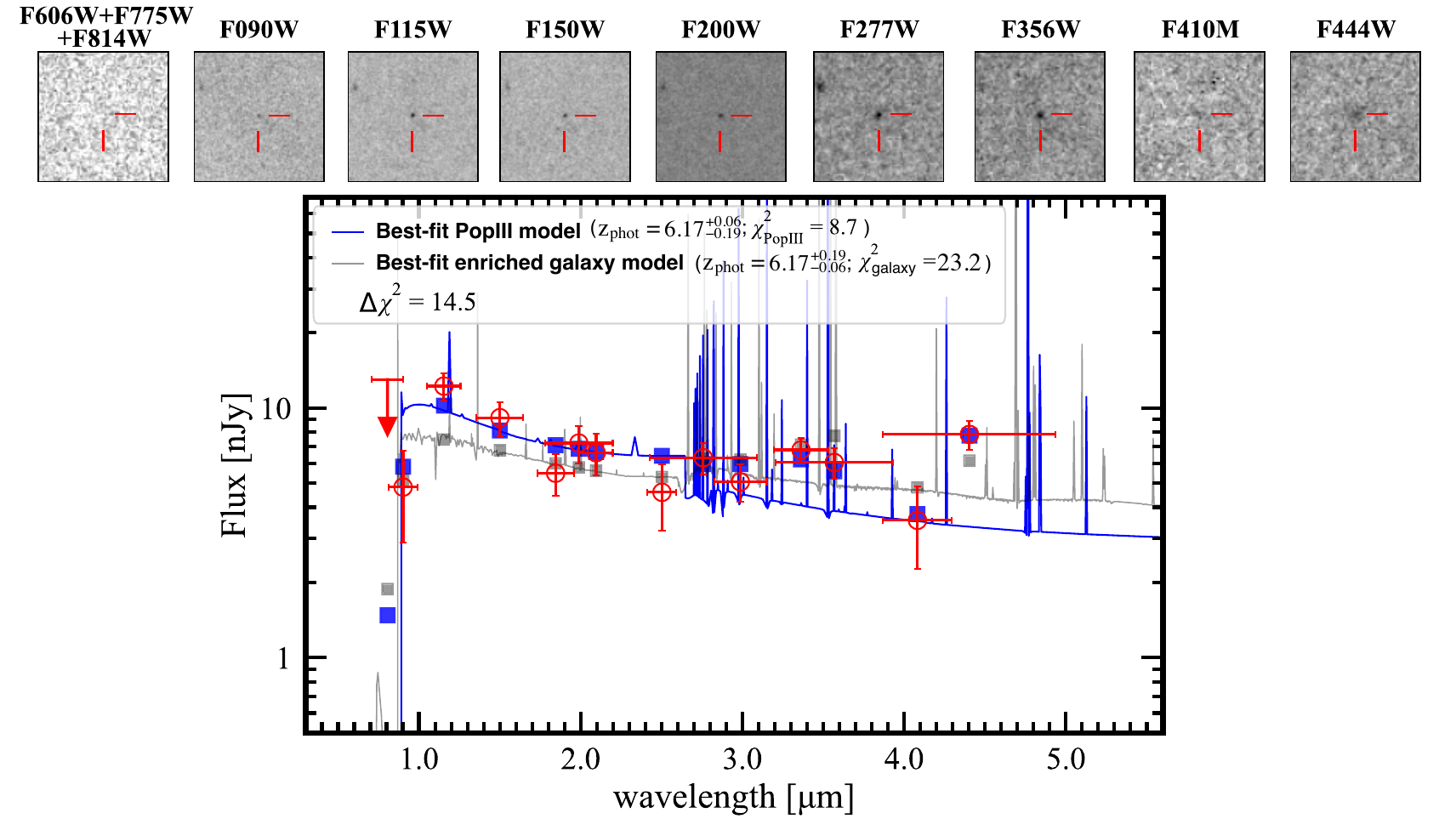}
\end{center}
\vspace{-0.2cm}
 \caption{
Same as Figure~\ref{fig:sed_eazy}, but for the tentative candidate \targb. 
The faint F410M is also confirmed in \targb, strongly supporting the presence of a Balmer jump and extremely strong H$\alpha$ (EW$\approx3600$\AA) that are consistent with Pop~III SEDs but are challenging to reproduce along with the other emission line excesses in metal-enriched galaxy SEDs. 
\label{fig:sed_eazy2}}
\end{figure*}

\begin{figure}
\begin{center}
\includegraphics[trim=0cm 0.1cm 0cm 0cm, clip, angle=0,width=.5\textwidth]{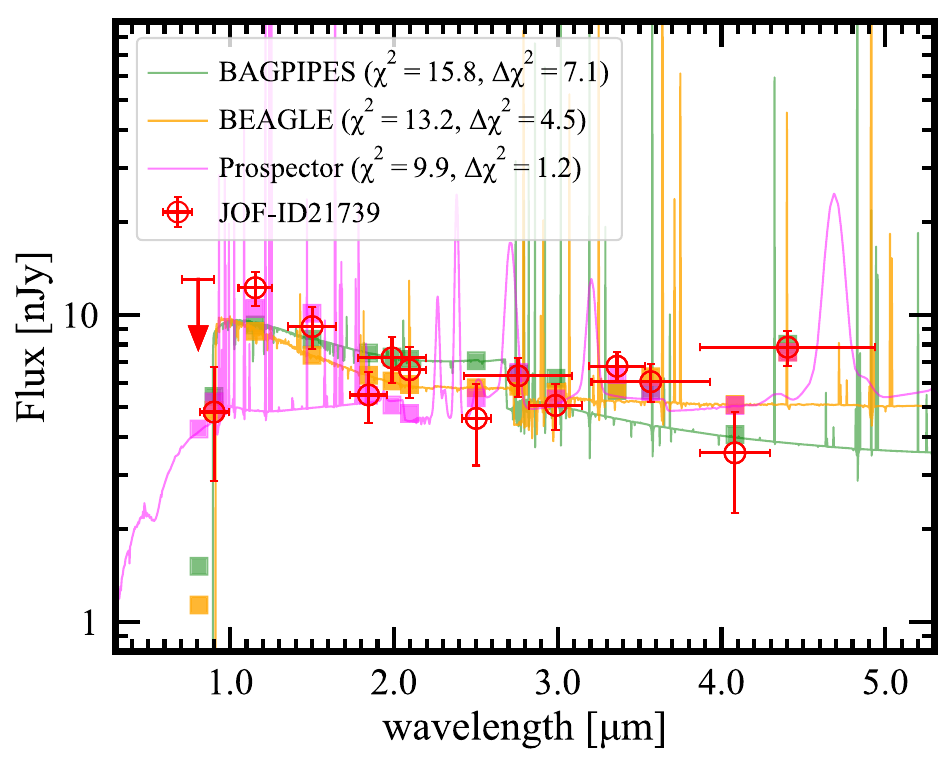}
\end{center}
\vspace{-0.2cm}
 \caption{
Same as Figure~\ref{fig:sed_all}, but for the tentative candidate \targb. 
The minimal $\Delta\chi^{2}$ of $\sim1$ obtained in \texttt{Prospector} reaffirms the tentative nature of this candidate. 
\label{fig:sed_all2}}
\end{figure}

\setlength{\tabcolsep}{30pt}
\begin{table}
\begin{center}
\caption{NIRCam photometry of JOF-21739}
\label{tab:phot_jof}
\vspace{-0.2cm}
\begin{tabular}{lc}
\hline 
\hline
R.A. [deg]  & 53.0346126       \\ 
Dec. [deg]  & $-$27.893660     \\ \hline
F090W [nJy] & $3.76 \pm 1.50$  \\
F115W [nJy] & $10.40 \pm 1.22$ \\
F150W [nJy] & $8.22 \pm 1.21$  \\
F182M [nJy] & $6.07 \pm 0.93$  \\
F200W [nJy] & $6.75 \pm 1.11$  \\
F210M [nJy] & $6.21 \pm 1.13$  \\
F250M [nJy] & $4.37 \pm 1.28$  \\
F277W [nJy] & $6.04 \pm 0.81$  \\
F300M [nJy] & $4.86 \pm 0.80$  \\
F335M [nJy] & $6.69 \pm 0.67$  \\
F356W [nJy] & $6.00 \pm 0.79$  \\
F410M [nJy] & $3.54 \pm 1.27$  \\
F444W [nJy] & $4.85 \pm 0.67$  \\
\hline
\end{tabular}
\end{center}
\end{table}

As introduced in Section~\ref{sec:candidate}, we find a tentative candidate, \targb, in the JOF field \citep{robertson2023}, which meets the SED criteria, while the NIRCam colors are marginal around the thresholds. 
Figure~\ref{fig:sed_eazy2} shows the NIRCam and HST cutouts and the best-fit SEDs with the Pop~III and metal-enriched galaxy templates using \texttt{EAZY} for \targb. 
The faintness in F410M is confirmed also in \targb, supporting the presence of Pop~III-like key SED features (Figure~\ref{fig:sed_comp}). 
\targb\ is also detected as a very UV-faint source with $M_{\rm UV} = -17.62^{+0.17}_{-0.15}$ and is well reproduced by a combination of young ($0.01$--$50$~Myr) Pop~III.1 and Pop~III.2 models with no dust attenuation, resulting in $\chi^{2} = 8.7$ and $\Delta\chi^{2} = 14.5$ compared to the best-fit metal-enriched galaxy SED. 
Note that, although the $\chi^{2}$ and subsequent $\Delta\chi^{2}$ values are generally larger for \targb\ compared to \targ, this difference reflects the availability of more NIRCam filters in the JOF field than in GLIMPSE, used in the SED fitting. 

When we run other flexible SED fitting tools of \texttt{BAGPIPES}, \texttt{BEAGLE}, and \texttt{Prospector}, for \targb\ in the same manner as \targ, we obtain $\Delta\chi^{2} = 1.2$ with a low-redshift solution at $z \sim 1.5$ with \texttt{Prospector}. 
In fact, Figure~\ref{fig:lowz} shows that the NIRCam color locations of \targb\ overlap with the color distributions of spec-$z$-confirmed $z \sim 1.5$ galaxies, supporting that the low-$z$ solution serves as a reasonable alternative solution for this candidate. 
These results reaffirm the less robust nature of \targb\ as a Pop~III candidate.

\begin{figure}
\begin{center}
\includegraphics[trim=0cm 0cm 0cm 0cm, clip, angle=0,width=.5\textwidth]{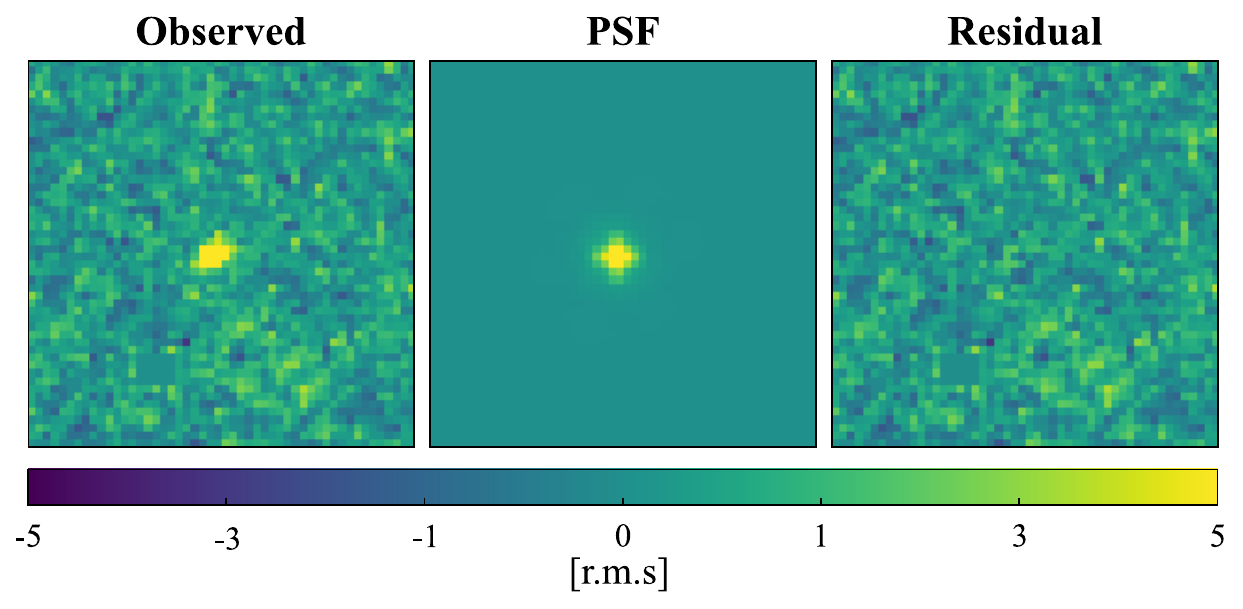}
\end{center}
\vspace{-0.4cm}
 \caption{
 Same as Figure~\ref{fig:galfit}, but for the tentative candidate of \targb\ whose morphology is also consistent with the PSF. 
\label{fig:galfit2}}
\end{figure}

Nevertheless, we infer the basic physical properties of \targb\ as a Pop~III candidate in the same manner as \targ\, as follows. 
We estimate an H$\alpha$ EW of $3600 \pm 430 \, {\rm \AA}$, a mean $t_{\rm age}$ of 0.3~Myr, $\beta=-2.79\pm0.05$, and a stellar mass of $\approx 10^{5-6} \, M_{\odot}$. 
\targb\ also exhibits a compact morphology, and the S\'ersic profile fitting with \texttt{GALFIT} on the high-resolution F150W image does not converge, reaching the smallest size in the fitting grid. 
In Figure~\ref{fig:galfit2}, we show the F150W $1''\times1''$ cutout around \targb, together with the PSF and the residual maps, demonstrating that the morphology agrees well with the PSF.  
We place an upper limit of $\simeq$~80~pc for the effective radius, in the same manner as \targ\ (Section~\ref{sec:size}). 
The NIRCam coordinates, photometry, difference in SED codes, and physical properties are summarized in Table~\ref{tab:phot_jof}, Table~\ref{tab:sed2}, and Table~\ref{tab:prop_jof}, respectively.

\setlength{\tabcolsep}{10pt}
\begin{table*}
\begin{center}
\caption{SED fitting results for \targb\ with different codes}
\vspace{-0.2cm}
\label{tab:sed2}
\begin{tabular}{cccccc}
\hline 
\hline
Fitting Type &  Pop~III (\texttt{EAZY}) & Galaxy (\texttt{EAZY}) & Galaxy (\texttt{BEAGLE}) &  Galaxy (\texttt{Prospector})  &  Galaxy (\texttt{BAGPIPES})  \\ \hline
$z_{\rm phot}$ & $6.17_{-0.19}^{+0.06}$ &   $6.17_{-0.06}^{+0.19}$  &  $6.41^{+0.14}_{-0.18}$     &   $1.51^{+0.03}_{-0.02}$   &  $6.37^{+0.16}_{-0.18}$  \\ 
$\chi^{2}$     &   8.7  & 23.2  & 13.2  & 9.9  & 15.8   \\ 
\hline
\end{tabular}
\end{center}
\vspace{-0.2cm}
\tablecomments{
Although \targb\ meets our condition of $\Delta\chi^{2}>9$ with \texttt{EAZY}, its NIRCam colors do not meet all the color-selection criteria, thereby we classify it as a tentative Pop~III candidate.  
}
\end{table*}

\setlength{\tabcolsep}{30pt}
\begin{table}[t!]
\begin{center}
\caption{Physical properties of JOF-21739}
\label{tab:prop_jof}
\vspace{-0.2cm}
\begin{tabular}{lc}
\hline 
\hline
ID          & JOF-21739   \\ \hline
$z_{\rm phot}$  & $6.17^{+0.19}_{-0.06}$    \\
$\beta$         & $-2.79\pm0.05$ \\
$M_{\rm UV}$  [mag]    &  $-17.62^{+0.17}_{-0.15}$   \\
$M_{\rm star}$ [$M_{\odot}$] & $\approx10^{5-6}$ \\
$r_{\rm e}$ [pc]       & $<80$ \\
$t_{\rm age}$  [Myr]        & 0.3 \\
H$\alpha$ EW [${\rm \AA}$]    & 3600 $\pm$ 430 \\
OIII/H$\beta$   & $<0.32$  \\
12+$\log$(O/H)  & $<6.2$  \\
\hline
\end{tabular}
\end{center}
\tablecomments{
Physical parameters are derived using the best-fit Pop III templates (see Section~\ref{sec:properties} for details). 
The lens correction is applied. \\
$\dagger$ The 1$\sigma$ upper limit is presented, same as \cite{vanzella2023}.  
}
\end{table}

\section{Theoretical frameworks for Pop~III UVLF}
\label{sec:appendix_model}

\subsection{Visbal et al. (2020) model}
\label{sec:visbal2020}
We utilized a modified version of the semi-analytic model from \cite{visbal2020} to estimate the Pop III UVFL. This model uses halo merger trees from cosmological N-body simulations to follow the formation of Pop III and metal-enriched star formation including the 3-dimensional variations in Lyman-Werner feedback, reionization, and metal enrichment of the IGM. The main change from \cite{visbal2020} is that the minimum halo mass in which Pop III star formation occurs is now calibrated from the simulations of \cite{kulkarni2021}. The fitting formulae from \cite{kulkarni2021} include the simultaneous impact of the Lyman-Werner intensity, redshift, and the baryon-dark matter streaming velocity \citep{tseliakhovich2010}. 
The model is applied to 10 different 3~Mpc across simulation boxes with dark matter particle resolution of $8000~M_\odot$. This resolution enables us to follow early metal enrichment from Pop III star formation in low-mass minihalos.

In order to estimate the Pop III UVLF, we determine the number of Pop~III galaxies that form in our simulation boxes from $z = 5.9 - 6.8$ (Pop III star formation is assumed to occur in instantaneous bursts). We convert Pop III stellar mass to UV luminosity with the results from \cite{raiter2010, schaerer2002} (assuming the IMF in ``Model C''). Additionally, we make the simplifying assumption that the Pop III galaxies are bright for a lifetime of 3 Myr and then abruptly become very faint. We note that our 10 simulation boxes are assigned a range of baryon-dark matter streaming velocities (from 0-3 times the typical value of 30 km/s at recombination), with the relative abundance of various relative velocities taken into account.

We compare two model parameterizations to the observational constraints in Figure~\ref{fig:uvlf}. We note that in both models Lyman-Werner feedback results in all Pop III galaxies found forming in halo masses near the atomic cooling threshold, which are forming stars for the first time. For the first model, we assume the fiducial parameters given in \cite{visbal2020} (see Table~1). Here the star formation efficiency $\epsilon_{\star,\rm III}$ (defined as the fraction of gas in a halo that forms Pop III stars during the burst) is $0.001$. This leads to Pop III stellar masses of ${\sim}10^4~M_\odot$, which are substantially fainter than \targ. For the second model, we adopt the same fiducial parameters except for the star formation efficiency, which is increased to $0.01$. This leads to Pop III galaxies with stellar mass of ${\sim}10^5 ~ M_\odot$  and a Pop~III UVLF that is in remarkable agreement with the constraints from \targ. We note that from the \cite{visbal2020} model, we expect a second population of Pop~III galaxies with luminosities similar to JOF-21739 for an efficiency of $0.01$, however the effective volume of our 10 N-body simulations is not sufficient to probe their abundance. This second population is expected to form in halos more massive than the atomic cooling limit at a virial mass where gas photoheated by reionization can collapse gravitationally.

\subsection{Liu \& Bromm et al. (2020) model}
\label{sec:liu2020}
Starting from the rate density of Pop~III star-forming halos at $z\simeq 6.5$, as found in the simulations of \citet{liu2020}, we derive approximate values for the Pop~III UV luminosity function for sources with stellar mass $\sim 10^5$ and $\sim 10^4$\,$M_{\odot}$, as follows.  
The emergence rate of Pop~III star-forming halos is regulated by the photoheating feedback of reionization and turbulent metal mixing. The latter can be modeled as a diffusion process with coefficient $D_{\rm mix}=\beta_{\rm mix}(v_{\rm vir}R_{\rm vir}/3)$, where $\beta_{\rm mix}\lesssim1$ is an efficiency parameter, and $v_{\rm vir}$, $R_{\rm vir}$ are the virial velocity and radius of the host halo, where the Pop~III pocket is located. Without enhanced photoheating feedback and metal mixing, corresponding to late reionization (or strong shielding) and $\beta_{\rm mix}=0$ (original simulation results), the simulations predict a rate density of $\dot{n}\sim 6\times 10^{-8}$\,yr$^{-1}$\,cMpc$^{-3}$. If reionization is nearly complete at $z\simeq 6.5$ (preventing star formation in small halos $\lesssim 4\times 10^8\ M_\odot$) and metal mixing is highly efficient (with $\beta_{\rm mix}=1$), this value can be reduced by up to a factor of $\sim 100$ \citep[see fig.~13 in][]{liu2020}. 

If Pop~III clusters are visible for $t\sim$3\ Myr (see Fig.~\ref{fig:mobs_age_all}), the corresponding number density of Pop~III hosting halos is $\dot{n}t\sim 0.2$\ cMpc$^{-3}$. 
The question now is which fraction of these halos host massive Pop~III clusters with $\sim 10^5$\,$M_{\odot}$ (which is the Pop~III stellar mass inferred for GLIMPSE-16043, corresponding to $M_{\rm UV}\simeq -15.9$, according to the analysis above). It is challenging to answer this question because the formation efficiency and mass distribution of Pop~III star clusters are determined by the sub-parsec-scale properties of Pop~III star-forming clouds and the balance between cloud collapse and protostellar feedback \citep[e.g.,][]{liu_feedback2024}, which are mostly unresolved in cosmological simulations and treated with sub-grid models. 
In the \citet{liu2020} simulations, such massive clusters do not exist, possibly because of the limited volume of the computational box and idealized sub-grid models for star formation and stellar feedback, with an average mass of Pop~III stars per halo of $\sim 1,000 \,M_{\odot}$. 
For simplicity, we fix the overall formation efficiency of Pop~III stars to that found in the simulations and group \textit{all} Pop~III stars into 10$^5\,M_{\odot}$ clusters to estimate the fraction effectively as $\sim 0.01$. The resulting maximum number density of $10^5$\,$M_{\odot}$ Pop~III clusters is $2\times 10^{-3}$\,cMpc$^{-3}$, which we take as an approximation for the upper limit of UVLF $\Phi$ at $M_{\rm UV}\simeq -15.9$, assuming a magnitude spread of order unity (see Fig.~\ref{fig:mobs_age_all}). Similarly, we estimate the $\Phi$ value corresponding to a Pop~III stellar mass of $\sim10^{4}\,M_{\odot}$, by assuming an effective host halo fraction of $\sim 0.1$, and a resulting UV luminosity that is about 2.5 magnitudes fainter ($M_{\rm UV}\simeq -13.4$). 
Given these upper limits and considering the theoretical uncertainty from photoheating feedback and metal mixing, we represent the UVLF estimates based on \citet{liu2020} with the shaded yellow region in Fig.~\ref{fig:uvlf}. As can be seen, there is reasonable agreement, also demonstrating the point that the empirical UVLF can serve to constrain various physical processes governing Pop~III star formation during reionization.

\subsection{Sarmento et al. (2022, 2025) model}
\label{sec:sarmento2025}
We used a customized version of Ramses-RT \citep{teyssier2002, rosdahl2013, rosdahl2015a}, a cosmological adaptive mesh refinement (AMR) simulation to track Pop III star formation in a 24~Mpc $h^{-1}$ box. We set the maximum refinement level such that the best average resolution is 91.6~$h^{-1}$ co-moving pc (cpc) in the densest, refined regions. 

Our version of Ramses-RT tracks the unpolluted fraction of gas in each simulation cell using a subgrid estimate of turbulence. This allows our team to track Population III star formation in otherwise polluted cells \citep{sarmento2017} by estimating the fraction of unpolluted gas as a function of time. This model thereby improves the estimate of Pop III stellar mass at each epoch. 

The rest-frame UV ($1500\AA$) luminosities of our star particles' (SPs') are based on a set of simple stellar population SED models parameterized by the SPs' ages, metallicities, and masses. The SEDs are based on \textit{STARBURST 99} \citep{starburst99} for Pop II SPs.  Pop III SPs use \cite{raiter2010} and \cite{schaerer2003} SEDs.  Pop II stars with $Z_{\star} > Z_{\rm crit} = 10^{-5}Z_\odot$ are modeled on a \cite{salpeter1955} IMF with masses between 0.8 and 100 $M_\odot$. Pop III SPs with $Z_{\star} \leq Z_{\rm crit}$ are based on the \cite{raiter2010,schaerer2003} SEDs for a zero-metallicity population and utilize a log-normal IMF centered on a characteristic mass of $60M_{\odot}$ with $\sigma=1.0$ and a mass range $1\, M_{\odot} \le M_\star \le 500\, M_{\odot}$. The SEDs model instantaneous bursts across the age range of SPs in the simulation for both types of stars. 

While the data for this study is contained in a yet to be published work, the methodology is outlined in detail in \cite{sarmento2018}.

\subsection{Venditti et al. (2023, 2024) model}
\label{sec:venditti2024}

We adopted the suite of \texttt{dustyGadget} \citep{graziani2020} 
cosmological simulations introduced in \citet{dicesare2023}, 
which consists of eight volumes with a comoving side of $50 h^{-1}$~cMpc, a total number of $2 \times 672^3$ particles, and a mass resolution for dark matter/gas particles of $3.53 \times 10^7 h^{-1} ~ \rm M_\odot$/$5.56 \times 10^6 h^{-1} ~ \rm M_\odot$ each, evolved from $z \simeq 100$ down to $z \simeq 4$. These are the largest simulations available that include a model for Pop~III star formation and feedback: Pop III particles representing stellar populations with a mass of $\sim 2 \times 10^6 ~ \rm M_\odot$ are formed when the star-forming condition is met in a gas below the critical metallicity $Z_\mathrm{crit} = 10^{-4} ~ \rm Z_\odot$; a Salpeter-like IMF within the mass range $[100, 500] ~ \rm M_\odot$ is assumed, following mass-dependent yields from stars in the PISN range $[140, 260] ~ \rm M_\odot$. The simulations have demonstrated good agreement with available model predictions and observations of the cosmic star-formation-rate/stellar-mass density evolution and with important scaling relations \citep{dicesare2023}. The main findings on the statistics and physical properties of Pop~III-forming environments have been detailed in \citet{venditti2023}, while their detectability through the HeII line at 1640~\AA\ and through PISNe has also been discussed in \citet{venditti2024a, venditti2024b}, respectively. 

The UVLF estimates in Figure~\ref{fig:uvlf} have been obtained by associating each Pop~III particle with the specific luminosity at 1500~\AA\ of the closest Yggdrasil model, i.e. the closest-age SED from the instantaneous burst, Pop~III.1 database, assuming a Salpeter-like IMF in the range $[50,500] ~ \rm M_\odot$ and zero metallicity, and including nebular corrections with a covering fraction $f_\mathrm{cov} = 1$. 
As the Yggdrasil database only presents photo-ionization calculations for $10^6 ~ \rm M_\odot$ stellar populations, results have been scaled by the total stellar mass of each particle. We also take into account an efficiency factor $\eta_\mathrm{III} = 0.01 - 0.1$ with respect to our Pop~III mass resolution element $M_\mathrm{III,res} \sim 2 \times 10^6 ~ \rm M_\odot$ (such that $M_\mathrm{III} = \eta_\mathrm{III} M_\mathrm{III,res}$), as described in \citet{venditti2024a, venditti2024b}. Note that, while these assumptions are not entirely consistent with the adopted feedback model, they represent a reasonable compromise between consistency with the underlying simulation assumptions and a more realistic emission model. 
We also caution that the results of photo-ionization simulations are not easily scalable by stellar mass in principle, as changes in the normalization of the radiation input into the nebula may result in unpredictable changes in the output spectral shape. However, we emphasize that the scope of this paper is to provide a first-order comparison, 
while a more in-depth study of the predicted emission of Pop~III-hosting \texttt{dustyGadget} galaxies will be provided in a future paper (Venditti et al. in prep.), by performing \textit{ad hoc} \texttt{Cloudy} simulations.

All stellar particles from one of the most star-forming cubes in the simulations (U12) at $z \simeq 6.7$ have been considered here, without taking into account any halo classification. Particularly, they also include objects below the nominal stellar-mass-resolution threshold $M_\star \simeq 10^{7.5} ~ \rm M_\odot$ adopted in \citet{venditti2023}, corresponding to galaxies resolved with less than $\sim 20$ stellar particles. 
Pop~III-hosting halos above this threshold always host Pop~II stellar populations at the same time (with Pop~III stars either found at the periphery of the central galaxy or in metal-free satellites), while Pop~III-only systems are all in the poorly-resolved regime. Tailored zoom-in simulations will shed further light on this halo population.

\section{Metal-poor AGN}
\label{sec:appendix_agn}

\begin{figure}[h]
\begin{center}
\includegraphics[trim=0cm 0cm 0cm 0cm, clip, angle=0,width=0.5\textwidth]{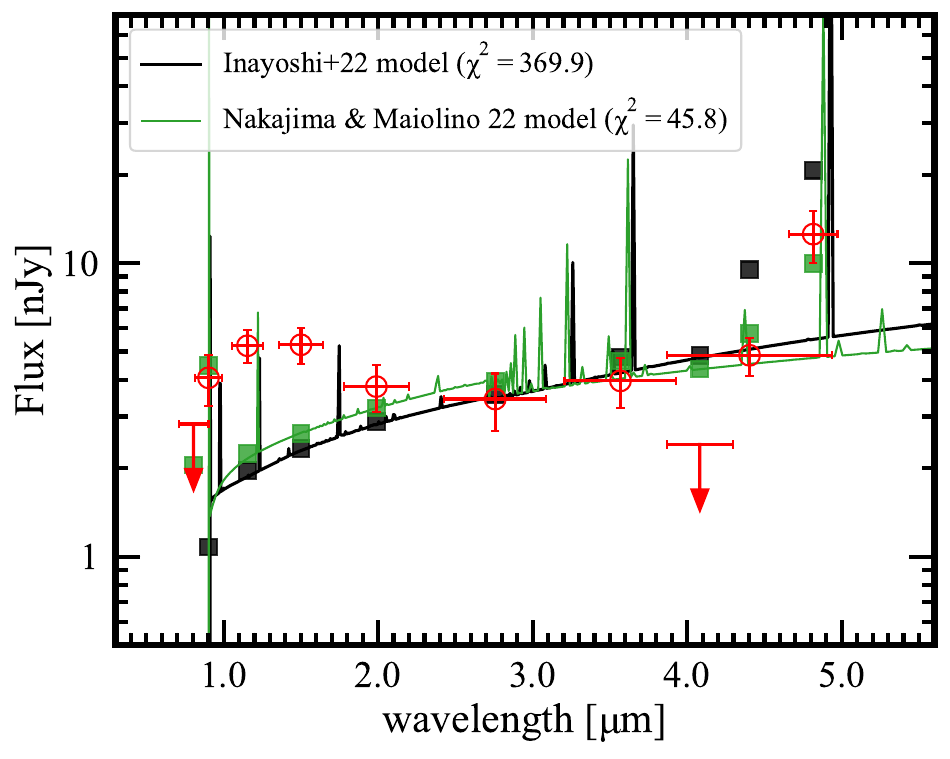}
\end{center}
 \caption{
Best-fit SED with seed BH models presented in \cite{inayoshi2022b} and \cite{nakajima2022b}. 
The red symbols represent the NIRCam and HST photometry of \targ\ in the same manner as Figure~\ref{fig:sed_eazy}. 
The large deviations are observed in the rest-frame UV continuum shape, resulting in the significant $\Delta\chi^{2}$ values and suggesting that \targ\ is less likely explained by the typical metal-poor AGN models in literature. However, the deviations might be mitigated by incorporating alternative surrounding gas conditions where nebular continuum emission dominates the SED, and we cannot rule out the possibility of the metal-poor AGN scenario. 
\label{fig:appendix_agn}}
\end{figure}

As discussed in Section~\ref{sec:agn}, the compact morphology and the deviation in the $M_{\rm UV}$--$Z_{\rm gas}$ relation suggest a possible AGN origin for our candidates. This scenario is particularly intriguing, as our candidates may represent seed or intermediate-mass BHs with $M_{\rm BH} \simeq 10^{4-5}\,M_{\odot}$, given their ultra-faint luminosities. 
In Figure~\ref{fig:appendix_agn}, we present the best-fit SEDs for \targ\ obtained using \texttt{EAZY} and the metal-poor AGN templates introduced by \cite{inayoshi2022b} and \cite{nakajima2022b}. 
We find that the best-fit AGN SEDs yield significantly higher $\chi^{2}$ values (40--100) and subsequent $\Delta\chi^{2}$ values compared to the best-fit SED with Pop~III templates ($\chi^{2}=2.7$; see Figure~\ref{fig:sed_eazy}). 
The primary discrepancy lies in the continuum shape: the AGN templates exhibit a red continuum, which contrasts with the blue continuum observed in our candidates. This difference might be mitigated by assuming alternative surrounding gas conditions where nebular continuum emission dominates the SED.
Therefore, although the current results disfavor the AGN scenario, we cannot entirely rule out the possibility of extremely metal-poor AGN origins for our candidates. Further studies that explore AGN models incorporating nebular-dominated emission will be essential for a comprehensive assessment of this scenario.

\bibliographystyle{apj}
\bibliography{apj-jour,reference}

\end{document}